\documentclass[conference]{IEEEtran}
\IEEEoverridecommandlockouts 

\usepackage[utf8]{inputenc}   
\usepackage[T1]{fontenc}      
\usepackage{amsmath,amssymb,amsfonts} 
\usepackage{bm}               
\usepackage{dsfont}           
\usepackage{nicefrac}         
\usepackage{microtype,soul}   
\usepackage{xcolor}           
\usepackage{textcomp}         

\usepackage{graphicx}         
\graphicspath{{./figures/}}  
\usepackage{booktabs}         
\usepackage{multirow}         
\usepackage{tabularx}         
\usepackage[caption=false,font=normalsize,labelfont=sf,textfont=sf]{subfig}

\usepackage{algorithm}        
\usepackage{algorithmic}      

\usepackage{cite}             
\usepackage{url}              
\usepackage[colorlinks=true,citecolor=blue,breaklinks=true]{hyperref}

\newtheorem{theorem}{Theorem}[section]

\newtheorem{definition}[theorem]{Definition}
\newtheorem{lemma}[theorem]{Lemma}
\newtheorem{proposition}[theorem]{Proposition}

\newenvironment{restatement}[1]{%
  \par\noindent\textbf{#1.}\itshape\quad
}{\par}



\renewcommand{\texttt}[1]{{\fontfamily{lmtt}\selectfont#1}}

\usepackage{refcount}         
\usepackage{tikz}             
\def\BibTeX{{\rm B\kern-.05em{\sc i\kern-.025em b}\kern-.08em
    T\kern-.1667em\lower.7ex\hbox{E}\kern-.125emX}}
\begin{document}

\title{Exploiting Aggregate Feedback for Decentralized Equilibrium Learning in Markov Games}

\author{\IEEEauthorblockN{1\textsuperscript{st} Siying~Huang}
\IEEEauthorblockA{\textit{The School of Mathematical Sciences} \\
\textit{UCAS}\\
Beijing, China \\
huangsiying@amss.ac.cn}
\and
\IEEEauthorblockN{2\textsuperscript{nd} Yifen~Mu}
\IEEEauthorblockA{\textit{SKLMS} \\
\textit{AMSS, CAS}\\
Beijing, China \\
mu@amss.ac.cn}
\and
\IEEEauthorblockN{3\textsuperscript{rd} Ge~Chen}
\IEEEauthorblockA{\textit{SKLMS} \\
\textit{AMSS, CAS}\\
Beijing, China \\
chenge@amss.ac.cn}
\and
\IEEEauthorblockN{4\textsuperscript{th} Qingshan~Jia}
\IEEEauthorblockA{\textit{Department of Automation} \\
\textit{Tsinghua University}\\
Beijing, China \\
jiaqs@tsinghua.edu.cn}
}

\maketitle

\begin{abstract}
Markov games (MGs) provide a general framework for multi-agent sequential decision-making, where efficient equilibrium learning remains a central challenge. Many practical MGs exhibit an aggregative reward structure, giving rise to aggregative Markov games (AMGs). In such systems, observable aggregate feedback can provide richer information for policy learning than selected-action bandit feedback alone. However, existing decentralized equilibrium learning algorithms for general-sum MGs do not explicitly exploit this additional information. To address this gap, we develop a decentralized adaptive stage-based V-learning algorithm built upon an aggregation-predictive policy-learning oracle. The algorithm leverages aggregate feedback at two timescales: for fast-timescale policy learning, it combines \emph{backward-looking evaluation} of all actions' immediate losses using realized aggregates with \emph{forward-looking prediction} from historical aggregates; 
for slow-timescale value learning, aggregate
fluctuations guide the adaptive adjustment
of stage lengths for value updates. We prove a high-probability no-regret guarantee for the proposed oracle and derive tighter stage-wise individual-gap bounds than prior methods. Combining these results with the adaptive stage schedule, we establish polynomial sample complexity guarantees for learning (coarse) correlated equilibria. Finally, we apply the proposed framework to demand response in renewable-integrated smart grids. Numerical results demonstrate improved equilibrium learning performance and strategic stability over benchmark methods, while maintaining competitive user rewards and favorable grid-operation performance.

\end{abstract}


\section{Introduction}
Many real-world sequential decision problems
involve strategic interactions among multiple
agents in a dynamic environment.
Markov games (MGs)~\cite{shapley1953stochastic}
provide a mathematical framework for such
problems, with applications in smart grids
~\cite{zhang2026adaptive,wutac2026nonstationary},
autonomous driving~\cite{kiran2021deep},
and other complex systems.

However, equilibrium computation in MGs is notoriously difficult: agents' decisions affect not only their immediate rewards but also future state transitions, and each agent must contend with both the strategic decisions of others and the evolving game environment ~\cite{giannou2022convergence}. In fact, recent work has proved that computing an approximate Markov perfect equilibrium in general-sum MGs is PPAD-complete ~\cite{deng2023complexity}. This leads us to consider a more tractable alternative, coarse correlated equilibrium (CCE), which can be computed efficiently through no-regret learning in normal-form games~\cite{papadimitriou2008computing,hart2000simple}.
However, the situation is more subtle in MGs: even computing approximate stationary Markov CCEs is PPAD-hard ~\cite{daskalakis2023complexity}.

Given this computational hardness, learning algorithms provide a natural way to approach equilibria in such complex games, while only stationary non-Markov and nonstationary Markov CCEs can be learned efficiently for MGs~\cite{daskalakis2023complexity}. This line of work has attracted increasing research interest in recent years.
In particular, we are interested in decentralized learning algorithms, since agents in practice often have access only to local information and centralized coordination can be costly.

There have been recent advances in decentralized
CCE learning for MGs, including V-learning and
stage-based V-learning algorithms with polynomial
sample complexity guarantees
~\cite{jin2021v,mao2022provably,mao2022improving}.
However, these algorithms may still require a large number of interactions with the environment, and their equilibrium guarantees
often rely on output policies constructed from historical learning policies rather than the
last iterate.  
On the other hand, many practical MGs possess additional structure and information that these algorithms do not fully exploit, leaving opportunities for improving learning efficiency. A representative example is the aggregative structure, where an agent's reward depends on its own action and an aggregate of other agents' actions. Such aggregative Markov games (AMGs)  arise in electricity demand response ~\cite{ye2016game,li2019stackelberg,li2023demand}, communication networks, and market competition ~\cite{nocke2018multiproduct}. In AMGs, agents can obtain richer information for policy learning from aggregate feedback than from selected-action bandit feedback alone.

This idea has been explored in the literature, primarily for static aggregative games, where aggregate information is typically obtained via a central operator or communication networks~\cite{grammatico2017dynamic, paccagnan2019tac,koshal2016distributed, linearly2023aggregative}.
However, these methods cannot be directly extended to AMGs, since the aggregative structure of immediate rewards does not necessarily carry over to the value functions in MGs. More recently, Ghaemi et al. investigated Nash equilibrium learning in continuous-space network AMG, but their guarantees rely on additional structural assumptions~\cite{ghaemi2024expected}.

This paper aims to exploit aggregate feedback to improve decentralized equilibrium learning in general-sum AMGs while retaining theoretical guarantees, which remains largely unexplored. Our main contributions are as follows.
\begin{itemize}
    \item We develop a decentralized
    adaptive stage-based V-learning
    algorithm built upon an
    aggregation-predictive hybrid-feedback
    Exp3-IX oracle.
    The algorithm exploits aggregate
    information at two timescales:
    on the fast timescale for policy
    learning, realized aggregate
    feedback enables
    \emph{backward-looking evaluation}
    of all candidate actions'
    immediate losses, while historical
    feedback supports
    \emph{forward-looking prediction};
    on the slow timescale for value
    learning, aggregate fluctuations
    guide adaptive stage lengths
    for optimistic value updates.

    \item We prove a high-probability no-regret guarantee for the proposed aggregation-predictive oracle and derive tighter stage-wise individual-gap bounds than prior methods. To be specific, these individual-gap bounds preserve the standard worst-case order without requiring accurate predictions, while yielding tighter stage-wise upper bounds than standard bandit Exp3-IX under favorable prediction conditions. Combining these guarantees with the adaptive stage schedule, we establish an episode complexity of $\widetilde O(SA_{\max}T^5/\epsilon^2)$ for learning an $\epsilon$-CCE. The framework further extends to CE learning, with complexity $\widetilde O(SA_{\max}^2T^5/\epsilon^2)$.

    \item We apply the proposed framework
to renewable-integrated smart grids,
a natural AMG arising from demand response,
and demonstrate improved strategic stability
and favorable grid-operation performance.
Specifically, the learned policies achieve
smaller deviation gaps and lower training
gap-AUC than stage-based CCE learning baselines.
Compared with general-purpose MARL methods
and representative demand-response rules,
they exhibit smaller incentives for
unilateral deviations while maintaining
competitive user rewards, high renewable
utilization, and moderate electricity-price
fluctuations.
\end{itemize}

The rest of this article is organized as follows:
Section~\ref{sec:preliminaries}
introduces the game model and
equilibrium concepts.
Section~\ref{sec:cce_ce}
presents the proposed algorithm
and theoretical guarantees for learning CCE.
Section~\ref{sec:ce} further extends this framework to CE learning.
Section~\ref{sec:simulations}
reports the smart-grid application
and numerical results, and
Section~\ref{sec:conclusions}
concludes the paper.

\section{Preliminaries}\label{sec:preliminaries}

\noindent\textbf{Markov game (MG).} An $N$-player episodic Markov game~\cite{jin2021v,mao2022improving} is defined by a tuple: 
\begin{equation}\label{MG1}
\{\mathcal{T},\mathcal{N},\!\{\mathcal{S}^t\}_{t=1}^T,\!\{\mathcal A_i^t(\cdot)\}_{i\in\mathcal N,t\in\mathcal T},\!\{r_i^t\}_{i\in\mathcal N,t\in\mathcal T},\!\{p^t\}_{t\in\mathcal T}\},
\end{equation}
where  i) $\mathcal{T} =\{1,2,\dots,T\}$ is the set of finite time steps in each episode; ii) $\mathcal{N} = \{1,2,\dots,N\}$ is the set of agents; iii) $\mathcal{S}^t$ is the finite state space at time $t$, where \( s^t \in \mathcal{S}^t \) denotes the state of the system at time \( t \);
 iv) $\mathcal A_i^t(s^t)$ denotes the finite action space available to agent $i$ at time $t$ for state $s^t\in\mathcal S^t$, where \( a_i^t \in \mathcal A_i^t(s^t) \) is the action taken by agent \( i \).
      The joint action space at time $t$ is $\mathcal A^t(s^t)=
\prod_{i\in\mathcal N}\mathcal A_i^t(s^t)$, with joint action
$\bm a^t=(a_1^t,\ldots,a_N^t)$.
    v) $r_i^t(s^t,\bm{a}^t) \in [0,1]$ is the immediate payoff  (reward) of agent $i$ at time $t$; and vi) $p^t(\cdot\mid s^t,\bm{a}^t)\in\Delta(\mathcal S^{t+1})$ is the state transition kernel at time $t$, where $\Delta(\mathcal S^{t+1})$ denotes the probability simplex over $\mathcal S^{t+1}$. For convenience, we let $\mathcal S^{T+1}$
consist of a single terminal state.

Notably, this formulation \eqref{MG1} allows time-varying state spaces $\mathcal{S}^t$ and stage-dependent action spaces $\mathcal A_i^t(s^t)$, which captures finite-horizon problems with evolving constraints.
For notational convenience, we define
$S := \max_{t\in\mathcal T} |\mathcal S^t|$, $A_i := \max_{t\in\mathcal T}\max_{s^t\in\mathcal S^t} |\mathcal A_i^t(s^t)|, \forall i\in\mathcal N$,
and
$A_{\max} := \max_{i\in\mathcal N} A_i$.

\vspace{.8em}
\noindent\textbf{Aggregative Markov game (AMG).}
We consider a class of MGs
with aggregative immediate rewards,
referred to as \emph{aggregative
Markov games} (AMGs).
Given the
state, each agent's immediate reward depends only on
its own action and an aggregate of
the other agents' actions:
\[
r_i^t(s^t,\bm a^t)
=r_i^t(s^t,a_i^t,d_{-i}^t),
\qquad
d_{-i}^t=\sigma_{-i}(\bm a_{-i}^t),
\]
where $\bm a_{-i}^t$ denotes
the opponents' action profile,
and $d_{-i}^t$ is its aggregate.
We also denote the aggregate of
all agents' actions by
$d^t=\sigma(\bm a^t)$.
For additive aggregation,
$d^t=\sum_j a_j^t$ and
$d_{-i}^t=d^t-a_i^t$.
Notably, we only require immediate rewards
to have an aggregative structure, with no
additional restrictions on the transition kernel.

Such structures arise in many large-scale systems where strategic interactions are mediated through aggregate quantities, such as total electricity demand in smart grids and traffic loads in congestion networks. 
In these systems, the aggregate quantities can often be observed or recovered from system-level feedback. 
Accordingly, we consider AMGs with \textit{aggregate feedback}. Specifically, each agent $i$ knows its local aggregative reward function $r_i^t(s,a_i,d_{-i})$. After each interaction with others, agent $i$ can observe or recover the overall aggregate $d^t$, and the opponents' aggregate $d_{-i}^t$ can be obtained via a known mapping $d_{-i}^t =\phi_i(s^t,a_i^t,d^t)$.  
For instance, in electricity demand response, aggregate demand may be publicly available or inferred from electricity prices, as illustrated in our smart-grid application in Section~\ref{sec:smart-grid}. Such aggregate feedback provides information about aggregate interactions without revealing individual opponents' actions, and similar feedback structures have been adopted in equilibrium seeking for aggregative games ~\cite{grammatico2017dynamic,paccagnan2019tac, belgioioso2023semi}.

\vspace{.8em}
\noindent\textbf{Policy and value function.}
A (Markov) policy for agent $i$ is $\pi_i=(\pi_i^1,\ldots,\pi_i^T)$, where each step-$t$ decision rule $\pi_i^t(\cdot\mid s^t)\in\Delta(\mathcal A_i^t(s^t))$ specifies its action distribution at the step-state pair $(t,s^t)$.
Let $\Pi_i$ be the set of
agent $i$'s Markov policies
and $\Pi=\prod_{i\in\mathcal N}\Pi_i$.
A joint policy (or policy profile)
$\bm\pi=(\pi_1,\ldots,\pi_N)\in\Pi$
induces a distribution over
state-action trajectories.

For a policy profile $\bm\pi$, and for any $t\in\mathcal{T}$, $s \in \mathcal{S}^t$, and $\bm{a} \in \mathcal{A}^t(s^t)$, the value function and the state-action value function (or $Q$-function) for agent $i$ are defined as: 
\begin{align}\label{eqn:value}
V_{t,i}^{\bm\pi} (s) &:= \mathbb{E}_{\bm\pi} \bigg[ \sum_{t'=t}^{T}r^{t'}_i(s^{t'},\bm{a}^{t'})\mid s^t = s \bigg],\\ Q_{t,i}^{\bm\pi} (s,\bm{a}) &:= \mathbb{E}_{\bm\pi} \bigg[ \sum_{t'=t}^{T}r^{t'}_i (s^{t'},\bm{a}^{t'})\mid s^t = s, \bm{a}^t = \bm{a} \bigg]. 
\end{align}

\noindent\textbf{Best response and Nash equilibrium.}
 For agent $i$, a policy $\pi_i^{\star}$ is a \emph{best response} to $\pi_{-i}$ for a given initial state $s^1$ if $V_{1,i}^{\pi_i^{\star}, \pi_{-i}}(s^1) = \sup_{\pi_i} V_{1,i}^{\pi_{i }, \pi_{-i}}(s^1)$. A policy profile $\bm\pi = (\pi_{i}, \pi_{-i })\in\Pi $ is a \emph{Nash equilibrium} (NE) if $\pi_{i }$ is a best response to $\pi_{-i}$ for all $i\in\mathcal{N}$.

\vspace{.8em}
\noindent\textbf{Correlated policy.}
More generally, a (non-Markov) \emph{correlated policy} is defined as \( \bm\pi = (\pi^1(\cdot),\ldots,\pi^T(\cdot)) \), where each step-$t$ decision rule satisfies
$\pi^t(z,h^t)\in\Delta(\mathcal A^t(s^t))$,
with
$h^t=(s^1,\bm a^1,\ldots,
s^{t-1},\bm a^{t-1},s^t)$
denoting the history up to step $t$,
and $z\in\mathcal R$ representing
shared randomness.
Thus, $\pi^t$ specifies a distribution
over the joint action space
$\mathcal A^t(s^t)$.
Shared randomness $z$ (e.g., a common random
seed) allows agents to coordinate their action
selection and generate correlated joint actions.
Let $\pi_i=(\pi_i^1(\cdot),\ldots,\pi_i^T(\cdot))$
and $\pi_{-i}=(\pi_{-i}^1(\cdot),\ldots,
\pi_{-i}^T(\cdot))$ denote the corresponding
marginal policies of $\bm\pi$, whose step-$t$
outputs take values in
$\Delta(\mathcal A_i^t(s^t))$ and
$\Delta(\mathcal A_{-i}^t(s^t))$, respectively.

For non-Markov correlated policies, the value function at $t=1$ is defined analogously to~\eqref{eqn:value}. A best response $\pi_i^{\star}$ with respect to the non-Markov policies $\pi_{-i}$ is a policy (independent of the randomness of  $\pi_{-i}$) that maximizes agent $i$'s value at step 1, i.e., $V_{1,i}^{\pi_i^{ \star}, \pi_{-i}}(s^1) = \sup_{\pi_i} V_{1,i}^{\pi_{i }, \pi_{-i}}(s^1)$, and is not necessarily Markov. 

\vspace{.8em}
\noindent\textbf{(Coarse) correlated equilibrium.}
Given the PPAD-hardness of computing NE
even in normal-form games
~\cite{daskalakis2009complexity}, two widely used relaxations are \textit{correlated equilibrium} (CE) and \textit{coarse correlated equilibrium} (CCE). Both concepts allow correlation among the agents' actions through a common random signal, while differing in the class of allowed deviations. A CCE ensures no agent has the incentive to deviate from a correlated policy $\bm\pi$ by playing a different independent policy. 

\begin{definition}\label{def:CCE}
	(CCE). A correlated policy $\bm\pi$ is an $\epsilon$-approximate coarse correlated equilibrium for any initial state $s^1\in \mathcal{S}^1$ if
	$
	V_{1,i}^{\pi_i^{ \star}, \pi_{-i}}(s^1) - V_{1,i}^{\bm\pi}(s^1) \leq \epsilon,\forall i\in\mathcal{N}.
	$
\end{definition}

CE imposes a stronger incentive condition
by allowing an agent to modify
its recommended actions.

\begin{definition}
	(Strategy modification). A strategy modification for agent $i$ is a collection $\psi_i = \{\psi_{t,i}^{s^t}:t\in\mathcal{T},s^t\in\mathcal{S}^t\}$, where each $\psi_{t,i}^{s^t}:\mathcal{A}_i(s^t)\rightarrow \mathcal{A}_i(s^t)$ maps a recommended action to another action. 
\end{definition}

Given a strategy modification $\psi_i$, for any policy $\bm\pi$, step $t$ and state $s^t$, if $\bm\pi$ selects the joint action $\bm{a}^t = (a_{1}^t,$ $\dots,a_{N}^t)$, then the modified policy $\psi_i \diamond \bm\pi$ will select $(a_{1}^t,\dots,a_{i-1}^t, \psi_{t,i}^{s^t}(a_{i}^t),$ $a_{i+1}^t,\dots,a_{N}^t)$. 
Let $\Psi_i$ denote the set of all possible strategy modifications for agent $i$. A CE ensures no agent has the incentive to deviate from a correlated policy $\bm\pi$ by using any strategy modification.  

\begin{definition}
	(CE). A correlated policy $\bm\pi$ is an $\epsilon$-approximate correlated equilibrium for initial state $s^1$ if 
$\sup_{\psi_i\in \Psi_i} V_{1,i}^{\psi_i\diamond \bm\pi}(s_1) - V_{1,i}^{\bm\pi}(s_1) \leq \epsilon,\forall i\in\mathcal{N}$.
\end{definition}

In finite normal-form general-sum games,
NE is a special case of CE, and CE is
a special case of CCE.
Moreover, the set of CCEs is always
nonempty and convex~\cite{nisan2007algorithmic}.

\vspace{.8em}
\noindent\textbf{Decentralized learning.}
We consider decentralized learning
over $K$ episodes in an unknown
environment, with initial states
drawn from a fixed distribution
$\rho\in\Delta(\mathcal S^1)$.
At each step $t\in\mathcal T$,
agents observe the public state
$s^t\in\mathcal S^t$ and simultaneously
select their actions
$a_i^t\in\mathcal A_i^t(s^t)$.
After the joint action, each agent
$i$ receives its private immediate reward
$r_i^t(s^t,\bm a^t)$, observes the next
state $s^{t+1}$, and obtains the
aggregate feedback $d^t$, from which
$d_{-i}^t$ can be recovered.
However, agents do not observe opponents'
individual actions, rewards, or
policies, and no inter-agent
communication is required
during learning.
 
\begin{algorithm*}[!tbp]
\caption{Adaptive Stage-Based V-Learning
with Aggregate Feedback (Agent $i$)}\label{alg:sbv}
\begin{algorithmic}[1]
    \STATE \textbf{Initialize:} let $H_t=T-t+1$ and $\overline V_{T+1,i}\gets0$.
    For all feasible $t,s,a$, set
    $\overline V_{t,i}(s),\widehat V_{t,i}(s)\gets H_t$,
    $\widetilde L^t(s)\gets T$,
    $\widetilde C^t(s),\widetilde r_i^t(s),\widetilde v_i^t(s),\widetilde E_i^t(s)\gets0$,
    $\mathcal D^t(s)\gets\emptyset$,
    $Z_{t,i}(s,a)\gets0$,
    $M_{t,i}(s,a)\gets1/(2H_t)$,
    and $\pi_i^t(a\mid s)\gets1/|\mathcal A_i^t(s)|$.
    \STATE Initialize $(\eta,\beta)$ for each $(t,s)$ using \eqref{eq:main-pred-parameters}, evaluated at the planned length $\widetilde L^t(s)$ and $A=|\mathcal A_i^t(s)|$.
    \FOR{episode $k\gets1$ to $K$}
        \STATE Record
    $\pi_i^{t,k}(\cdot\mid s)
    \gets\pi_i^t(\cdot\mid s)$
    for all $t\in\mathcal T$
    and $s\in\mathcal S^t$.
        \STATE Receive initial state $s^1$.
        \FOR{step $t\gets1$ to $T$}
            \STATE $\widetilde c:=\widetilde C^t(s^t)\gets\widetilde C^t(s^t)+1$.
            \STATE Take action $a_i^t\sim\pi_i^t(\cdot\mid s^t)$; observe $r_i^t,s^{t+1}$,
            obtain $d^t$, and recover $d_{-i}^t=\phi_i(a_i^t,d^t)$.\label{line:6}
            \STATE Append $d^t$ to the current-stage sequence $\mathcal D^t(s^t)$.\label{line:7}
            \STATE $\widetilde r_i^t(s^t)\gets\widetilde r_i^t(s^t)+r_i^t$,
            $\widetilde v_i^t(s^t)\gets\widetilde v_i^t(s^t)+\overline V_{t+1,i}(s^{t+1})$.
            \STATE $f(a)\gets[1-r_i^t(s^t,a,d_{-i}^t)]/H_t$ for all $a$;
            $g\gets[H_t-1-\overline V_{t+1,i}(s^{t+1})]/H_t$.
            \STATE $\displaystyle Z_{t,i}(s^t,a)\gets Z_{t,i}(s^t,a)+f(a)+
            \frac{\mathbf1\{a=a_i^t\}\,g}{\pi_i^t(a\mid s^t)+\beta}$, for all $a$.\label{line:10}
            \STATE $\widetilde E_i^t(s^t)\gets\widetilde E_i^t(s^t)+\max_a|f(a)-M_{t,i}(s^t,a)|^2$.\label{pred-error-update}
            \STATE $\displaystyle M_{t,i}(s^t,a)\gets
            \frac{(\widetilde c-1)M_{t,i}(s^t,a)+f(a)}{\widetilde c}$, for all $a$.\label{line:forecast-update}
            \IF{$\widetilde C^t(s^t)=\widetilde L^t(s^t)$}\label{line:12}
                \STATE $\widehat V_{t,i}(s^t)\gets
                [\widetilde r_i^t(s^t)+\widetilde v_i^t(s^t)]/\widetilde c+b_{\widetilde c}$,
                where $b_{\widetilde c}=b^{\mathrm{pred}}_{t,i}(\widetilde c,\widetilde E_i^t(s^t);s^t)$ \eqref{eq:pred-bonus}.\label{line:14}
                \STATE $\overline V_{t,i}(s^t)\gets\min\{\widehat V_{t,i}(s^t),H_t\}$.
                \STATE $\lambda(s^t)\gets f_{\mathrm{MAD}}(\mathcal D^t(s^t))$ (Algorithm~\ref{alg:fluctMAD}).\label{line:16}
                \STATE $\widetilde L^t(s^t)\gets
                \max\{\widetilde L^t(s^t)+1,\,
                \lfloor\lambda(s^t)(1+1/T)\widetilde L^t(s^t)\rfloor\}$.\label{line:17}
                \STATE Recompute $(\eta,\beta)$ using \eqref{eq:main-pred-parameters} for the updated stage length.
                \STATE $\widetilde C^t(s^t),\widetilde r_i^t(s^t),\widetilde v_i^t(s^t),\widetilde E_i^t(s^t)\gets0$;
                $\mathcal D^t(s^t)\gets\emptyset$.\label{line:18}
                \STATE $Z_{t,i}(s^t,a)\gets0$,
                $M_{t,i}(s^t,a)\gets1/(2H_t)$, for all $a$.\label{line:19}
            \ENDIF
            \STATE $\displaystyle \pi_i^t(a\mid s^t)\gets
            \frac{\exp\{-\eta[Z_{t,i}(s^t,a)+M_{t,i}(s^t,a)]\}}
            {\sum_{a'}\exp\{-\eta[Z_{t,i}(s^t,a')+M_{t,i}(s^t,a')]\}}$, for all $a$.\label{line:11}
        \ENDFOR
    \ENDFOR
    \STATE \textbf{Output:}
The historical policy sequence
$\{\pi_i^{t,k}\}_{t\in\mathcal T,\,
k\in[K]}$.
\end{algorithmic}
\end{algorithm*}

\section{Decentralized CCE learning in AMGs}\label{sec:cce_ce}

This section presents the main contributions
of this paper.
We develop a decentralized learning algorithm
for general-sum AMGs that exploits aggregate
feedback to improve stage-wise individual-gap
bounds while retaining polynomial sample
complexity guarantees for learning an
approximate CCE.
\subsection{Algorithmic Design Ideas}
In decentralized Markov games, from the perspective of any individual agent, the environment reduces to a single-agent Markov decision process (MDP) within a single episode. Across multiple episodes, however, the environment perceived by each agent becomes non-stationary, making it highly challenging for agents to simultaneously estimate value functions and optimize policies.

A natural way to address this difficulty
is a \emph{two-timescale} design, an
established approach in decentralized
multi-agent learning
~\cite{sayin2021decentralized,
sayin2022fictitious}, where value estimation
evolves on a slower timescale while policy
optimization proceeds on a faster one.
To implement this idea, inspired by
stage-based V-learning~\cite{mao2022improving},
we partition visits to each step-state
pair $(t,s)$ into \emph{stages}:
policies are updated at each visit,
while value estimates are updated
only at stage boundaries.

Our key design is to exploit aggregate feedback at both timescales:
\begin{itemize}
    \item \textbf{Fast timescale
    (policy learning):}
    Observed aggregate feedback
    enables \emph{backward-looking
    evaluation} of immediate losses
    for all candidate actions,
    rather than only the selected
    action. Historical aggregate
    feedback further enables
    \emph{forward-looking prediction}
    of immediate losses to guide
    no-regret policy updates.

    \item \textbf{Slow timescale
    (value learning):}
    Fluctuations in the observed
    aggregate guide \textit{adaptive stage
    lengths}, allowing more samples
    to be collected before value
    updates when aggregate
    variability is high.
\end{itemize}

\subsection{Algorithm Description}

\begin{figure*}[!t]
    \centering
    \includegraphics[width=\textwidth]{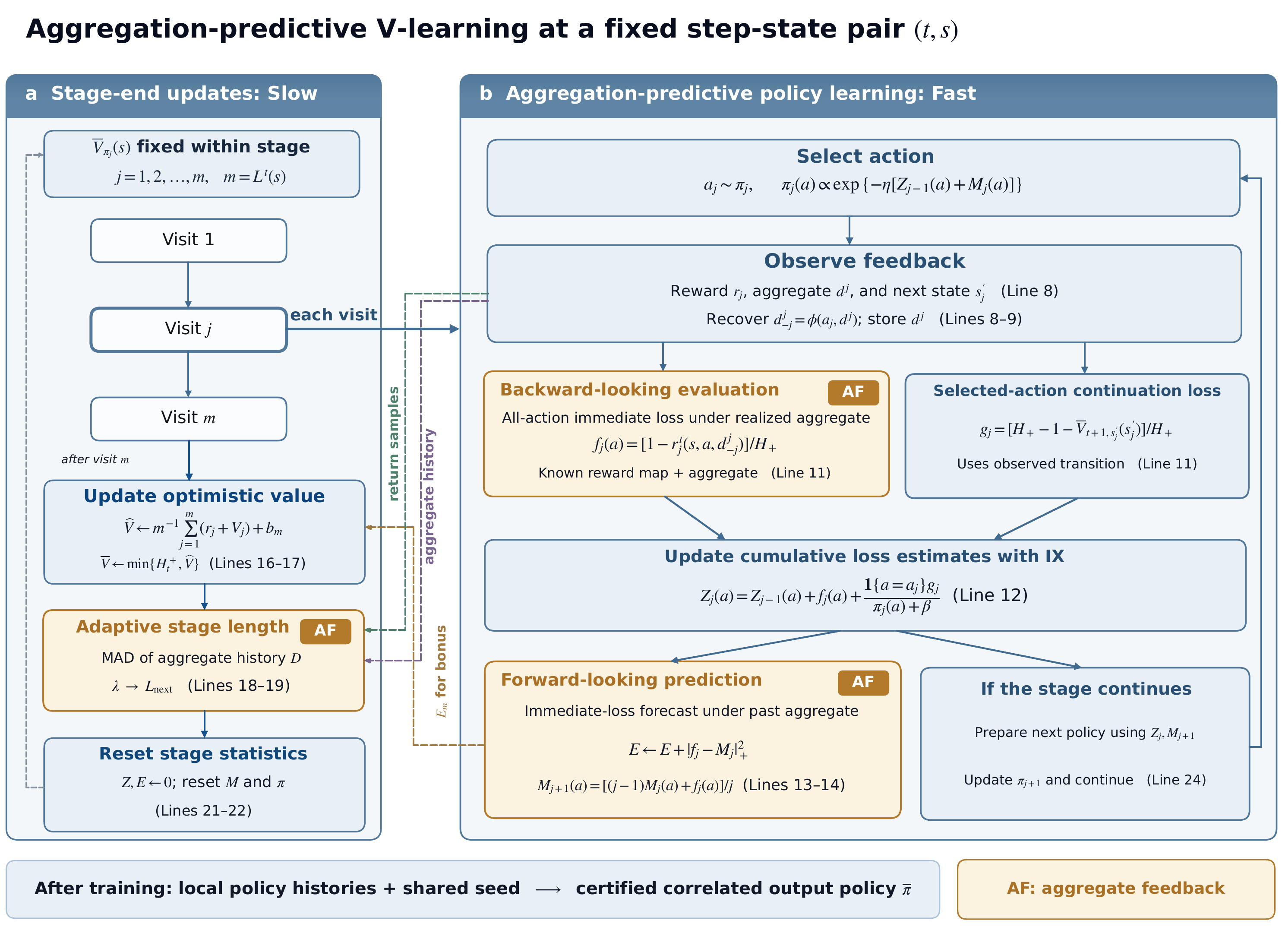}
    \caption{Two-timescale mechanism of the
proposed learning algorithm.}
    \label{fig:algorithm_mechanism}
\end{figure*}

Following the above design, 
Algorithm~\ref{alg:sbv} presents the proposed
adaptive stage-based V-learning algorithm
with aggregate feedback, with its two-timescale mechanism illustrated in Fig.~\ref{fig:algorithm_mechanism}.

\noindent\textbf{Slow timescale:
Adaptive stages and optimistic
value estimation.}
For each step-state pair $(t,s^t)$, the agent partitions its visits into \textit{stages}, 
with an initial stage length $\widetilde L^t(s)=T$.
At the end of each stage, a fluctuation coefficient $\lambda(s^t)\in[\lambda_{\min},1]$ is computed from the aggregate sequence $\mathcal D^t(s^t)$ observed within the stage for the \textit{adaptive stage-length update}:
\begin{equation*}
\widetilde L^t(s^t)\gets
\max\left\{
\widetilde L^t(s^t)+1,\,
\left\lfloor
\lambda(s^t)\left(1+\frac1T\right)
\widetilde L^t(s^t)
\right\rfloor
\right\}.
\end{equation*}
This update retains an increasing
stage-length schedule inspired by
the near-geometric design
\cite{zhang2020almost}, while adapting its
growth to aggregate variability:
larger fluctuations yield a larger
$\lambda(s^t)$, allowing more samples to be
collected in the next stage before value updates.
In practice, $\lambda(s^t)$ can be constructed
from standard dispersion measures, such as
the mean absolute deviation (MAD)
~\cite{geary1935ratio}, adopted in
Algorithm~\ref{alg:fluctMAD}, or the
coefficient of variation (CV)
~\cite{hendricks1936sampling}.

\begin{algorithm}[!tbp]
\caption{Adaptive Fluctuation Coefficient $f_{\text{MAD}}(\mathcal{D}^t(s^t))$ via MAD}\label{alg:fluctMAD}
\begin{algorithmic}[1]
    \STATE \textbf{Input:} Positive aggregate quantity sequence $\mathcal D^t(s^t)=\{d^{t,1},\dots,d^{t,n}\}$, minimum fluctuation coefficient $\lambda_{\text{min}} \in [\frac{T+1/2}{T+1},1]$, empirical upper bound $MAD_{\text{max}}$
    \STATE \textbf{Output:} Adaptive fluctuation coefficient $\lambda(s^t) \in [\lambda_{\text{min}}, 1]$

    \IF{$|\mathcal{D}^t(s^t)| < 2$}
        \STATE $\lambda(s^t) \gets 1$; \texttt{// Default: no fluctuation}\;
        \RETURN $\lambda(s^t)$
    \ENDIF

    \STATE $\bar{d} \gets \frac{1}{|\mathcal{D}^t(s^t)|}\sum_{d^{t,j} \in \mathcal{D}^t(s^t)} d^{t,j}$;

    \STATE $MAD \gets \frac{1}{|\mathcal{D}^t(s^t)|}\sum_{d^{t,j} \in \mathcal{D}^t(s^t)} |d^{t,j} - \bar{d}|$;

    \STATE $\gamma \gets \min\left(\frac{MAD}{MAD_{\text{max}}}, 1\right)$; 
    \STATE $\lambda(s^t) \gets \lambda_{\text{min}} + (1 - \lambda_{\text{min}}) \cdot \gamma$;
    \RETURN $\lambda(s^t)$
\end{algorithmic}
\end{algorithm}

To implement slow-timescale value learning,
value estimates are updated only at the end
of each stage. Specifically, when the visitation count $\widetilde C^t(s^t)$ reaches the planned stage length $\widetilde L^t(s)=\tilde c$, agent $i$ updates its optimistic value estimate:
\begin{equation}\label{eq:description-value}
\begin{aligned}
\widehat V_{t,i}(s^t)
&\gets
\frac{\widetilde r_i^t(s^t)
+\widetilde v_i^t(s^t)}{\tilde c}+b_{\tilde c},\\
\overline V_{t,i}(s^t)
&\gets \min\{\widehat V_{t,i}(s^t),H_t\},
\end{aligned}
\end{equation}
where $H_t = T-t+1$ is the remaining horizon, and $\widetilde r_i^t(s^t)$ and $\widetilde v_i^t(s^t)$ accumulate the observed rewards and estimated continuation values over the stage, respectively.
The stage-end bonus $b_{\tilde c}$ accounts for the stage-wise policy-learning regret and the sampling error in value estimation:
\begin{equation}\label{eq:pred-bonus}
b_{\tilde c}:=
\begin{cases}
\min\left\{H_t,\, H_t\sqrt{2\iota/\tilde c}\right\}, & A=1,\\[2mm]
\min\left\{H_t,\, U_{\tilde c}\!\left(\widetilde E_i^t(s^t); s^t\right)\right\}, & A\ge2,
\end{cases}
\end{equation}
with
\begin{align*}
U_{\tilde c}(x; s^t)
:=& \frac{H_t}{\tilde c}\Bigg[
\frac{C_A}{\eta}+2\eta x
+\frac{4\eta A\tilde c(H_t-1)^2}{H_t^2} +\iota \\& +(1+\frac1{2H_t})\sqrt{2\tilde c\iota}
\Bigg].
\end{align*}
Here $A=|\mathcal A_i^t(s)|$, and
$\widetilde E_i^t(s)$ denotes the
cumulative squared immediate-loss
prediction error within the stage,
computed by the fast-timescale
policy learner described below. The parameters are set as
\begin{equation}\label{eq:main-pred-parameters}
\begin{aligned}
\iota&=\log\frac{16NSA_{\max}TK}{p},\qquad
C_A=\log A+\frac{\iota}{4},\\
\eta&=\min\left\{H_t,\,
H_t\sqrt{\frac{C_A}{\widetilde L^t(s^t)[2+4A(H_t-1)^2]}}\right\},\\
\beta&=\frac{2\eta(H_t-1)}{H_t}.
\end{aligned}
\end{equation}
The learning parameters $(\eta,\beta)$ are fixed at stage initiation according to the planned stage length.
After the stage-end value and stage-length
updates, the stage-specific variables 
are reset as specified in
Line~\ref{line:18}.

\noindent\textbf{Fast timescale:
Aggregation-predictive policy learning.}
At each visit to $(t,s^t)$ within a stage,
agent $i$ selects an action according
to its current policy
$\pi_i^t(\cdot\mid s^t)$ and updates
the policy using our aggregation-predictive
hybrid-feedback Exp3-IX oracle
(Lines~\ref{line:6}--\ref{line:forecast-update}
and~\ref{line:11}).
Unlike standard bandit Exp3-IX,
the proposed oracle exploits aggregate
feedback for full-action immediate-loss
evaluation and prediction, as detailed below.

\emph{Hybrid-feedback loss estimation.}
After selecting
$a_i^t\sim\pi_i^t(\cdot\mid s^t)$,
agent $i$ observes its reward
and the next state $s^{t+1}$,
and obtains the aggregate feedback
$d^t$, from which $d_{-i}^t$
can be recovered.
Using its known reward function,
the agent then computes the normalized
immediate loss of every candidate action:
\begin{equation*}
f(a)=
\frac{1-r_i^t(s^t,a,d_{-i}^t)}{H_t},
\qquad \forall a\in\mathcal A_i^t(s^t).
\end{equation*}
This enables \emph{backward-looking evaluation}
of all candidate actions under the realized
aggregate of the opponents' actions.
By contrast, the normalized
continuation loss
\begin{equation*}
g=
\frac{H_t-1-
\overline V_{t+1,i}(s^{t+1})}{H_t}
\end{equation*}
is available only for the selected
action, since counterfactual next
states are not observed.

The agent then combines these two loss components
to update the cumulative loss estimate
$Z_{t,i}(s^t,a)$:
\begin{equation}\label{eq:Z}
Z_{t,i}(s^t,a)\gets
Z_{t,i}(s^t,a)+f(a)
+\frac{\mathbf1\{a=a_i^t\}g}
{\pi_i^t(a\mid s^t)+\beta}.
\end{equation}
Here, the immediate losses of all
actions are incorporated directly,
while the continuation loss is
estimated from selected-action
feedback using implicit exploration
(IX)~\cite{neu2015explore}.

\emph{Aggregation-based loss prediction.}
Beyond backward-looking loss evaluation,
the agent further uses historical aggregate
feedback for \emph{forward-looking prediction}
of immediate losses.
Let $\tilde c$ denote the current within-stage
visit count.
By collecting the sequence
$\{d_{-i}^{t,j}\}_{j=1}^{\tilde c}$
of opponents' aggregates, the agent can
predict their aggregate behavior without
estimating their individual policies.
As a natural predictor, we use the empirical
distribution
$\widehat\mu_{\tilde c}(d_{-i})
=\frac1{\tilde c}\sum_{j=1}^{\tilde c}
\mathbf1\{d_{-i}^{t,j}=d_{-i}\}$.
Under this distribution, the predicted
immediate loss of action $a$ for the next
visit is given by
\begin{equation*}
\begin{aligned}
M_{t,i}(s^t,a)
&=\mathbb E_{d_{-i}\sim\widehat\mu_{\tilde c}}
\left[\frac{1-r_i^t(s^t,a,d_{-i})}{H_t}\right] \\
&=\frac1{\tilde c}\sum_{j=1}^{\tilde c}
\frac{1-r_i^t(s^t,a,d_{-i}^{t,j})}{H_t}.
\end{aligned}
\end{equation*}
Thus, starting from $M_{t,i}(s^t,a)=1/(2H_t)$,
the agent updates the prediction through
a running average:
\begin{equation}\label{eq:M}
M_{t,i}(s^t,a)\gets
\frac{(\tilde c-1)M_{t,i}(s^t,a)+f(a)}
{\tilde c}.
\end{equation}
In addition, the squared prediction error
$\|f-M_{t,i}(s^t,\cdot)\|_\infty^2$,
evaluated before the forecast update,
is accumulated in $\widetilde E_i^t(s^t)$
for the stage-end bonus~\eqref{eq:pred-bonus}.

\emph{Predictive policy update.}
The agent combines the cumulative loss estimate
$Z_{t,i}$~\eqref{eq:Z} and the immediate-loss forecast
$M_{t,i}$~\eqref{eq:M} through optimistic exponential
weights~\cite{rakhlin2013predictable}:
\begin{equation*}
\pi_i^t(a\mid s^t)\gets
\frac{
\exp\{-\eta[
Z_{t,i}(s^t,a)+M_{t,i}(s^t,a)
]\}
}{
\sum_{a'}
\exp\{-\eta[
Z_{t,i}(s^t,a')+M_{t,i}(s^t,a')
]\}
}.
\end{equation*}
At each stage boundary, $Z_{t,i}$ and
$M_{t,i}$ are reinitialized
(Lines~\ref{line:18}--\ref{line:19}),
and the next stage begins with a uniform policy.

\noindent\textbf{Output policy.}
Finally, the local policy histories are used
to construct the correlated output
policy $\bar{\bm\pi}$ through
Algorithm~\ref{alg:certify},
following~\cite{bai2020near}.
Here $\pi_i^{t,k}(\cdot\mid s)$ denotes
the policy prescribed before the step-$t$
action of episode $k$ at state $s$.
A shared random seed coordinates
historical policy selection during
execution of the output policy,
while no inter-agent communication
is required during training.

\begin{algorithm}[!tbp]
\caption{Construction of the Output Policy $\bar{\bm{\pi}}$}\label{alg:certify}
\begin{algorithmic}[1]
    \STATE \textbf{Input:} The distribution trajectory specified by Algorithm~\ref{alg:sbv}: $\{\pi_{i}^{t,k}:i\in\mathcal{N},t\in \mathcal{T},k\in[K]\}$;

    \STATE Uniformly sample $k$ from $[K]$;
    \FOR{step $t\gets 1$ to $T$}
        \STATE Receive $s^{t}$;
        \IF{$\tilde c^{t,k}(s^t)=0$}
            \STATE Execute fixed uniform local policies for the remaining episode and stop this construction.
        \ENDIF
        \STATE Uniformly sample $j$ from $\{1,2,\dots, \tilde{c}^{t,k}(s^{t})\}$; \COMMENT{ For a state $s^t$, $\tilde{c}^{t,k}$ denotes the number of visits to the state $s^t$ (at the $t$-th step)  in the stage right before the current stage}\;
        \STATE Set $k \gets \tilde{l}_{j}^{t,k}$; \COMMENT{ $\tilde{l}_{j}^{t,k}$ is the index of the episode that this state was visited the $j$-th time among the total $\tilde{c}^{t,k}$ times} \;
        \STATE Take joint action $\bm{a}^{t}\sim \times_{i=1}^N\pi_{i}^{t,k}(\cdot \mid s^{t})$;
    \ENDFOR
\end{algorithmic}
\end{algorithm}

\subsection{Theoretical Results}

We establish the theoretical guarantees of
Algorithm~\ref{alg:sbv} in two parts.
We first analyze the regret of the
aggregation-predictive hybrid-feedback
Exp3-IX oracle and derive improved bounds on
the stage-wise individual gap.
We then show how these stage-wise guarantees,
together with the adaptive stage-length
schedule, lead to a global CCE guarantee. 

\subsubsection{Within-Stage No-Regret Oracle}
We begin by establishing a high-probability external-regret guarantee for the aggregation-predictive hybrid-feedback Exp3-IX oracle employed in Algorithm~\ref{alg:sbv} (Lines~\ref{line:6}--\ref{line:forecast-update} and~\ref{line:11}).
At a fixed state, within-stage learning under decentralized interactions with other agents naturally leads to an adversarial multi-armed bandit problem with hybrid feedback, formulated as follows.

Consider $A$ actions and $\tilde c$ rounds.
At round $j$, the learner selects
$a_j\sim p_j$ and observes the
immediate-loss vector $f_j\in[0,F]^A$
for all actions, together with a
continuation loss $g_j\in[0,G]$
only for the selected action, where
$F+G\le1$.
For Algorithm~\ref{alg:sbv},
$F=1/H_t$ and $G=(H_t-1)/H_t$.
Let $\mathcal F_{j-1}$ denote the
pre-round history.
We assume that $f_j$ is conditionally
independent of the current action draw,
and that the forecast
$M_j\in[0,F]^A$ is
$\mathcal F_{j-1}$-measurable.

The oracle updates
\begin{align}
p_j(a)
&\propto
\exp\{-\eta[Z_{j-1}(a)+M_j(a)]\},
\label{eq:pred-oracle-policy}\\
Z_j(a)
&=Z_{j-1}(a)+f_j(a)
+\frac{\mathbf1\{a_j=a\}g_j}
{p_j(a)+\beta},
\label{eq:pred-oracle-loss}
\end{align}
where $Z_0=0$, $\eta>0$,
$\eta F\le1$, and $\beta=2\eta G$.
Define the conditional expected loss as
\begin{equation}\label{eq:pred-expected-loss}
\ell_j(a)=
\mathbb E[
f_j(a)+g_j
\mid\mathcal F_{j-1},a_j=a
].
\end{equation}
For Algorithm~\ref{alg:sbv}, this
corresponds to the normalized
Bellman loss defined in
\eqref{eq:stage-expected-loss}.

\begin{proposition}
[Regret guarantee for aggregation-predictive oracle]
\label{prop:pred-regret}
For the hybrid-feedback adversarial
bandit problem described above,
suppose the learner follows
\eqref{eq:pred-oracle-policy}--
\eqref{eq:pred-oracle-loss}.
Then, for any $\tilde c\ge1$ and
$\delta\in(0,1)$, with probability
at least $1-\delta$,
\begin{equation}\label{eq:pred-regret-prop}
\begin{aligned}
\mathcal R_{\tilde c}
:={}&
\sum_{j=1}^{\tilde c}\langle p_j,\ell_j\rangle
-\min_a\sum_{j=1}^{\tilde c}\ell_j(a)\\
\le{}&
\frac{\log A+L_\delta/4}{\eta}
+2\eta E_{\tilde c}
+4\eta A\tilde c G^2\\
&+L_\delta
+\frac{2F+G}{2} \sqrt{2\tilde c L_\delta},
\end{aligned}
\end{equation}
where $L_\delta=\log(4A/\delta)$ and
$E_{\tilde c}=\sum_{j=1}^{\tilde c}
\|f_j-M_j\|_\infty^2$. For fixed $A$, $F$, $G$, and $\delta$,
the bound further implies
$\mathcal R_{\tilde c}=O(\sqrt{\tilde c})$
when $\eta=\Theta(\tilde c^{-1/2})$.
\end{proposition}
The proof is given in Appendix~\ref{app:cce}.

Proposition~\ref{prop:pred-regret} establishes a prediction-dependent no-regret guarantee with sublinear regret for our oracle. Unlike conventional bandit algorithms, it exploits aggregate feedback to evaluate all immediate-action losses while relying on bandit feedback for continuation losses. 
\subsubsection{Stage-wise Individual Gap}
We next apply the regret guarantee to
the stage-wise
individual gap in
Algorithm~\ref{alg:sbv}. The stage-wise
individual gap \(\Gamma_{\tilde c}\) measures the difference between the average expected value of the best unilateral deviation policy and the average realized value of the agent’s policy within the stage. 
For a state $(t,s)$, let $\tilde c$
denote the length of its previous
completed stage, and let $\tilde l_j$
be the episode index of its $j$-th visit
in that stage.
Combining Proposition~\ref{prop:pred-regret}
with martingale concentration yields
the following result.
\begin{lemma}[Stage-wise individual gap]\label{lemma:bonus}
With probability at least
$1-p/2$, it holds for all agents $i \in \mathcal{N},$ episodes $k \in [K],$  steps $t \in \mathcal{T},$  and states $s^t\in\mathcal S^t$ with $\tilde c>0$,
\begin{align}
 &\Gamma_{\tilde c} := \max_{\pi_i^t}\frac1{\tilde c}\sum_{j=1}^{\tilde c}
 \mathbb D_{\pi_i^t\times\pi_{-i}^{t,\tilde l_j}}
 (r_i^t+\mathbb P_t\overline V_{t+1,i}^{\tilde l_j})(s)
 \notag\\
 &\quad-\frac1{\tilde c}\sum_{j=1}^{\tilde c}
 \left[r_i^t(s,\bm a^{t,\tilde l_j})
 +\overline V_{t+1,i}^{\tilde l_j}(s^{t+1,\tilde l_j})\right]
 \le b_{\tilde c}.\label{eq:pred-lemma-bound}
\end{align}
\end{lemma}
The proof is provided in Appendix~\ref{app:cce}.

The bound of stage-wise individual gap in Lemma~\ref{lemma:bonus} depends explicitly on the cumulative prediction error $E_{\tilde c}$, which varies with the choice of forecast rule $M_j$. 
The following proposition establishes its
worst-case guarantee, quantifies the
improvement over standard bandit Exp3-IX,
and gives a further bound under
accurate continuation prediction.
\begin{theorem}[Prediction-dependent bounds on the stage-wise individual gap] \label{thm:pred-comparison} 
Consider the stage-based learning framework
in Algorithm~\ref{alg:sbv}.
For an agent $i$ and a completed stage
at $(t,s)$ of length $\tilde c$, let
$A=|\mathcal A_i^t(s)|\ge2$ and
$H_t=T-t+1$.
For any $p\in(0,1)$, let $\iota$ be defined
in \eqref{eq:main-pred-parameters}.
The following results hold.
\begin{enumerate}
\renewcommand{\labelenumi}{\roman{enumi}).}
\item \textbf{Worst-case guarantee.} For any bounded, history-based prediction
rule $M_j\in[0,1/H_t]^A$ used in the
aggregation-predictive oracle, let
$\Gamma_{\tilde c}^{M}$ denote the
corresponding stage-wise individual gap,
defined as in Lemma~\ref{lemma:bonus}.
Then, with probability at least $1-p/2$,
  \begin{equation}\label{eq:pred-worst-gap} 
    \begin{aligned} \Gamma_{\tilde c}^{M} \le \min\left\{ H_t,\, U_{\tilde c}(E_{\tilde c}^{M};s^t) \right\} \le 6H_t\sqrt{\frac{A\iota}{\tilde c}}, \end{aligned} \end{equation} 
where $E_{\tilde c}^{M} =\sum_{j=1}^{\tilde c} \|f_j-M_j\|_\infty^2$. In particular, Algorithm~\ref{alg:sbv} satisfies this bound with $E_{\tilde c}^{M}=\widetilde E_i^t(s)$ and $\Gamma_{\tilde c}^{M}=\Gamma_{\tilde c}$ in Lemma~\ref{lemma:bonus}.  

\item \textbf{Improvement over standard bandit Exp3-IX.}
Replacing the proposed oracle with standard
bandit Exp3-IX yields a larger stage-wise
upper bound. Specifically, for
Algorithm~\ref{alg:sbv}, when
$\tilde c>36A\iota$,
\begin{equation}\label{eq:aggregate-improvement}
\begin{aligned}
\Gamma_{\tilde c}
&\le U_{\tilde c}(\widetilde E_i^t(s);s^t)\\
&\le U_{\tilde c}^{\mathrm{band}}
-\left(2\sqrt A-\frac3{\sqrt2}\right)
\sqrt{\frac{\iota}{\tilde c}}\\
&\quad-\frac{2H_t\eta}{\tilde c}
\left(\frac{\tilde c}{H_t^2}
-\widetilde E_i^t(s)\right)\\
&<U_{\tilde c}^{\mathrm{band}}
<6H_t\sqrt{\frac{A\iota}{\tilde c}},
\end{aligned}
\end{equation}
where $U_{\tilde c}^{\mathrm{band}}$
denotes the corresponding stage-wise upper bound.

\item \textbf{Improvement with aggregate-informed continuation prediction.} Suppose there exists a history-based
predictor
$\widehat v_j(s,a,d)\in[0,H_t-1]$
such that
$\big| \overline V_{t+1,i}^{\tilde l_j}(s') -\widehat v_j(s,a,\sigma(a_{-i})) \big| \le H_t\rho$, 
for all $a\in\mathcal A_i^t(s)$, $a_{-i}\in\mathcal A_{-i}^t(s)$, and $s'\in\mathcal S^{t+1}$, where $\rho\in[0,(H_t-1)/H_t]$ is known. 
For a residual-feedback variant of Algorithm~\ref{alg:sbv} constructed in Appendix~\ref{app:cce}, the following stage-wise Bellman error bound holds with probability at least $1-p/2$ for all stages with $\tilde{c}\ge\iota$: 
\begin{equation}\label{eq:residual-factor-three} 
\Gamma_{\tilde c}^{\mathrm{res}} \le 3H_t\sqrt{\frac{A\iota}{\tilde c}}, \qquad \text{if } A\ge13,\ \rho\le0.1. 
\end{equation} 
Furthermore, for any $A\ge2$, if
$\rho\le c_0/\sqrt A$ for a constant
$c_0$ independent of $A$, then
$\Gamma_{\tilde c}^{\mathrm{res}}
=O(H_t\sqrt{\iota/\tilde c})$.
 \end{enumerate} 
 \end{theorem}
The complete proof is deferred to Appendix~\ref{app:cce}.

Theorem~\ref{thm:pred-comparison}(i)
shows that the aggregation-predictive Exp3-IX
framework retains the worst-case stage-wise
bound $6H_t\sqrt{A\iota/\tilde c}$ for any
bounded, history-based forecast, matching the
stage-wise order of existing decentralized
CCE learning algorithms~\cite{mao2022improving}.
Algorithm~\ref{alg:sbv}, which uses an
empirical-aggregate forecast, inherits this
guarantee.
More importantly, Theorem~\ref{thm:pred-comparison}(ii) establishes a strictly
tighter stage-wise individual-gap bound for
Algorithm~\ref{alg:sbv} than that of standard
bandit Exp3-IX, providing a theoretical
justification for exploiting aggregate feedback
to improve learning efficiency. Theorem~\ref{thm:pred-comparison}(iii) further shows that, under an additional continuation-prediction condition, an aggregate-informed residual-feedback refinement can improve the bound to $O(H_t\sqrt{\iota/\tilde c})$.

These results highlight the value of aggregate
feedback in decentralized learning for AMGs.
While the aggregative reward structure simplifies
the strategic coupling in immediate rewards,
it does not necessarily extend to the value
functions or simplify the intertemporal coupling
induced by state transitions.
Exploiting aggregate feedback thus improves
stage-wise learning through richer information
about immediate strategic interactions, while
agents still need to learn the long-term effects
of their actions through repeated interactions
with the environment.
Further improvements are possible when aggregate
information can also support accurate
continuation prediction.

\subsubsection{Global CCE Sample Complexity}
We now turn to the global CCE guarantee. Following the certified-policy construction in Algorithm~\ref{alg:certify}, we establish confidence bounds for the corresponding value functions. 

For analysis, we introduce two auxiliary notations $\underline{V}$ and  $\underline{\hat{V}}$ as lower confidence bounds of the value estimates. Specifically, for any $i \in \mathcal{N},$ episode $ k \in [K],$  step $t \in [T+1],$  and state $ s^t \in \mathcal{S}^t$, we define $\underline{V}_{t,i}^k(s^t) = \underline{\hat{V}}_{t,i}^k(s^t) = 0$ if $t = T+1$ or $k$ is in the first stage of the $(t,s^t)$ pair, and 
\begin{align*}
\underline{\hat{V}}_{t,i}^k(s^t) &=\frac{1}{\tilde{c}}\sum_{j=1}^{\tilde{c}} \left(r_{i}^t(s^t,\bm{a}^{t,\tilde{l}_j}) + \underline{V}_{t+1,i}^{\tilde{l}_j}(s^{t+1,\tilde{l}_j}) \right)- b_{\tilde{c}},\\ &\text{ and } \underline{V}_{t,i}^k(s^t) = \max\left\{\underline{\hat{V}}_{t,i}^k(s^t), 0\right\}. 
\end{align*}

\begin{lemma}[Confidence bounds] \label{lemma:upperbound}
It holds with probability at least $1-p$ that for all agents $i \in \mathcal{N},$ episodes $k \in [K],$ steps $t \in \mathcal{T},$ and states $s^t \in \mathcal{S}^t$,
\begin{align}
 V_{t,i}^{\star,\bar{\pi}_{-i}^{t,k}}(s^t) \leq \overline{V}_{t,i}^k(s^t), \label{first inequality}\\
      V_{t,i}^{\bar{\bm\pi}^{t,k}}(s^t) \geq \underline{V}_{t,i}^k(s^t) . \label{second inequality}
\end{align}
\end{lemma}
The detailed proof is given in Appendix~\ref{app:cce}.

These bounds reduce the CCE gap analysis to bounding the difference between optimistic and pessimistic value estimates $\overline V-\underline V$. Combining these with the adaptive stage-length design, we present the following sample complexity guarantee for learning CCE in general-sum aggregative Markov games. 
\begin{theorem}[Sample complexity of learning CCE] \label{thm:cce} For any $\epsilon\in(0,T]$ and $p\in(0,1)$, let $\iota$ be defined in \eqref{eq:main-pred-parameters}. There exists an absolute constant $C>0$ such that, if the agents run Algorithm~\ref{alg:sbv} for \[ K\ge C\frac{SA_{\max}T^5\iota}{\epsilon^2} \] episodes, then, with probability at least $1-p$, the output policy $\bar{\bm\pi}$ of Algorithm~\ref{alg:certify} is an $\epsilon$-approximate CCE. 
\end{theorem}
\textit{Proof sketch:}

We provide a high-level overview of the proof and defer all technical details to Appendix~\ref{app:cce}.

The analysis proceeds in four main steps.

\textit{Step 1 (Stage-wise estimation).}
Proposition~\ref{prop:pred-regret}
establishes a high-probability
external-regret bound for the
aggregation-predictive hybrid-feedback
oracle.
Together with martingale concentration,
this yields the stage-wise Bellman
error bound in Lemma~\ref{lemma:bonus}.
Theorem~\ref{thm:pred-comparison}(i)
further guarantees
\[
b_{\tilde c}
\le
6H_t\sqrt{\frac{A_i\iota}{\tilde c}}.
\]

\textit{Step 2 (Confidence bounds).}
Using the certified-policy construction,
Lemma~\ref{lemma:upperbound}
establishes optimistic and pessimistic
bounds on the best-response and
certified-policy values.
Consequently, the CCE gap is bounded
by the difference
$\overline V-\underline V$.

\textit{Step 3 (Recursive error propagation).}
Let
\[
\delta^{t,k}
=
\overline V_{t,i}^k(s^{t,k})
-
\underline V_{t,i}^k(s^{t,k}).
\]
The stage-based value updates imply
\[
\delta^{t,k}
\le
\mathbf1\{\tilde c^{t,k}=0\}T
+
\frac1{\tilde c}
\sum_{j=1}^{\tilde c}
\delta^{t+1,\tilde l_j}
+
2b_{\tilde c},
\]
where terms involving an empty
previous stage are defined as zero.

The stage-length upper bound controls
the reuse of previous-stage samples,
while the cumulative stage regularity
controls the sum of the bonuses.
Summing over episodes and recursively
over steps gives
\[
\sum_{k=1}^K\delta^{1,k}
=
O\left(
ST^3+
\sqrt{SA_{\max}KT^5\iota}
\right).
\]

\textit{Step 4 (CCE guarantee).}
Averaging over episodes and using the definition of the output policy $\bar{\bm\pi}$, we obtain
\[
\begin{aligned}
V_{1,i}^{\star,\bar\pi_{-i}}(s^1)
-
V_{1,i}^{\bar{\bm\pi}}(s^1) \le
\frac1K\sum_{k=1}^K\delta^{1,k} =
O\left(
\sqrt{\frac{SA_{\max}T^5\iota}{K}}
\right),
\end{aligned}
\]
which establishes the claimed sample complexity result.
\hfill$\blacksquare$

\section{Decentralized CE Learning in AMGs}
\label{sec:ce}

The proposed framework can be further extended to decentralized correlated equilibrium (CE) learning. Unlike CCE, which requires small external regret, CE requires small swap regret. To achieve this, we apply an Ito-style reduction~\cite{ito2020tight} to the no-external-regret aggregation-predictive oracle in Algorithm~\ref{alg:sbv}. Specifically, each action is associated with a copy of the aggregation-predictive oracle, and the action-selection policy is obtained as a stationary distribution of the stochastic matrix formed by their output distributions. 

\subsection{Aggregation-Predictive No-Swap-Regret Learning}

Let $\mathcal A$ be an action set of size $A$. For each source action
$a\in\mathcal A$, maintain a row policy $q_j(\cdot\mid a)$, cumulative
estimated losses $Z_j(\cdot\mid a)$, a predictable forecast
$M_j(\cdot\mid a)$, and an update count $n_j(a)$. At round $j$, compute
the stationary distribution $p_j$ of the row-stochastic matrix $Q_j$:
\begin{equation}\label{eq:ce-stationary}
p_j(b)=\sum_{a\in\mathcal A}p_j(a)q_j(b\mid a),
\qquad b\in\mathcal A.
\end{equation}
The learner samples an internal index $J_j\sim p_j$, then plays
$A_j\sim q_j(\cdot\mid J_j)$. Since the rows of $Q_j$ are strictly
positive, this stationary distribution is unique. The marginal
distribution of the played action is also $p_j$.

The observed aggregate gives the full immediate-loss vector
$f_j\in[0,F]^A$, while the transition reveals a continuation-loss sample
$g_j\in[0,G]$ only for $A_j$, where $F+G\le1$. The selected row is updated
by
\begin{align}
Z_j(b\mid J_j)
&=Z_{j-1}(b\mid J_j)+f_j(b)
+\frac{\mathbf 1\{A_j=b\}g_j}
{q_j(b\mid J_j)+\beta},
\label{eq:ce-row-loss}\\
M_{j+1}(b\mid J_j)
&=\frac{n_j(J_j)M_j(b\mid J_j)+f_j(b)}
{n_j(J_j)+1},
\label{eq:ce-row-forecast}\\
q_{j+1}(b\mid J_j)
&\propto
\exp\{-\eta[Z_j(b\mid J_j)+M_{j+1}(b\mid J_j)]\}.
\label{eq:ce-row-policy}
\end{align}
Then increment $n_j(J_j)$. Other rows remain unchanged. At the start of
each stage, set $q_1(b\mid a)=1/A$, $Z_0(b\mid a)=0$,
$M_1(b\mid a)=F/2$, and $n_1(a)=0$ for all $a,b$.

Let $\ell_j(b)=\mathbb E[f_j(b)+g_j\mid
\mathcal F_{j-1},a_j=b]$ be the conditional expected normalized
loss of playing $b$. For a map $\phi:\mathcal A\to\mathcal A$, define
$\phi\diamond p$ by
$(\phi\diamond p)(b)=\sum_{a:\phi(a)=b}p(a)$. The pseudo swap regret
over $m$ rounds is
\begin{equation}\label{eq:ce-swap-regret}
\mathcal R_m^{\rm swap}
=
\max_{\phi:\mathcal A\to\mathcal A}
\sum_{j=1}^m
\left[
\langle p_j,\ell_j\rangle
-\sum_{a\in\mathcal A}p_j(a)\ell_j(\phi(a))
\right],
\end{equation}
where $\ell_j(b)$ is the conditional expected loss of playing $b$.
The stationary relation~\eqref{eq:ce-stationary} makes the first term
equal to the expected loss obtained by drawing a row from $p_j$ and then
an action from that row. This links swap regret to the external regret
of the selected row learners.

\begin{proposition}[Regret guarantee for aggregation-predictive no-swap learning]
\label{prop:ce-swap-regret}
Assume the hybrid-feedback conditions of
Proposition~\ref{prop:pred-regret}, with $f_j$ conditionally
mean-independent of the current internal-index and action draws, and
$M_j(\cdot\mid a)$ predictable for every row. Let every row use
\eqref{eq:ce-row-loss}--\eqref{eq:ce-row-policy} with common
$\eta>0$ and $\beta=2\eta G$, where $\eta F\le1$. For any
$m\ge1$ and $\delta\in(0,1)$, with probability at least $1-\delta$,
\begin{align}\label{eq:ce-swap-bound}
\begin{aligned}
\mathcal R_m^{\rm swap}\le{}&
\frac{A(\log A+L_\delta/4)}{\eta}
+2\eta E_m^{\rm CE}
+4\eta A mG^2\\
&+A L_\delta+2\sqrt{2AmL_\delta}\\
&+\sqrt{8m\left(A\log A+\log(2/\delta)\right)},
\end{aligned}
\end{align}
where
\[
L_\delta=\log\frac{8A^2m}{\delta},
\qquad
E_m^{\rm CE}
=\sum_{j=1}^m
\|f_j-M_j(\cdot\mid J_j)\|_\infty^2.
\]
\end{proposition}

The first terms in~\eqref{eq:ce-swap-bound} sum the external-regret
bounds for the $A$ row learners. The final term controls the deviation
between the realized row-index sequence and its conditional expectation.
The proof is given in Appendix~\ref{app:ce}.

\subsection{Decentralized CE Algorithm}

For a state $(t,s)$, write $A=|\mathcal A_i^t(s)|$,
$G_t=(H_t-1)/H_t$, and let
$m=\widetilde L^t(s)$ be the planned stage length. Define
\begin{equation}\label{eq:ce-pred-scales}
\iota_{\rm CE}=\log\frac{64NSA_{\max}^2TK^2}{p},
\qquad
C_A=\log A+\frac{\iota_{\rm CE}}4.
\end{equation}
At the start of each stage, set
\begin{equation}\label{eq:ce-pred-parameters}
\eta_{\rm CE}
=\min\left\{H_t,\,
\sqrt{\frac{A C_A}
{m(2/H_t^2+4A G_t^2)}}\right\},
~
\beta_{\rm CE}=2\eta_{\rm CE}G_t.
\end{equation}
For a completed stage, let
$E_m^{\rm CE}=\widetilde E_i^{t,\rm CE}(s)$ be its accumulated
row-prediction error, and define
\begin{align}
B_m^{\rm CE}:={}&
\frac{A C_A}{\eta_{\rm CE}}
+2\eta_{\rm CE}E_m^{\rm CE}
+4\eta_{\rm CE}A mG_t^2
+A\iota_{\rm CE}\notag\\
&+2\sqrt{2Am\iota_{\rm CE}}
+\sqrt{8m(A\log A+\iota_{\rm CE})}.
\label{eq:ce-swap-radius}
\end{align}
The stage-end bonus is
\begin{equation}\label{eq:ce-pred-bonus}
b_m^{\rm CE}
=\begin{cases}
\min\{H_t,H_t\sqrt{2\iota_{\rm CE}/m}\},&A=1,\\[1mm]
\min\{H_t,\frac{H_t}{m}
[B_m^{\rm CE}+\sqrt{2m\iota_{\rm CE}}]\},&A\ge2.
\end{cases}
\end{equation}
Since $E_m^{\rm CE}\le m/H_t^2$, the bonus has the worst-case order
$b_m^{\rm CE}=O(H_tA\sqrt{\iota_{\rm CE}/m})$.

Algorithm~\ref{alg:sbv-CE} retains the
two-timescale adaptive stage-based
V-learning framework of
Algorithm~\ref{alg:sbv}.
The optimistic value updates and
aggregate-based stage-length adaptation
remain unchanged.
The main difference lies in the
within-stage policy update: instead of
a single aggregation-predictive Exp3-IX
learner, the CE algorithm maintains
one learner per action and updates
only the sampled row.
At stage boundaries, all row learners
are reset, and the recorded stationary
marginal policies are used in the same
certified-policy construction.

\begin{algorithm*}[!tbp]
\caption{Adaptive Stage-Based V-Learning with Aggregation-Predictive No-Swap-Regret Exp3-IX (Agent $i$)}
\label{alg:sbv-CE}
\begin{algorithmic}[1]
\STATE \textbf{Initialize:} For every feasible $(t,s)$, set
$H_t=T-t+1$, $\overline V_{t,i}(s)=\widehat V_{t,i}(s)=H_t$,
$\widetilde L^t(s)=T$, $\widetilde C^t(s)=0$,
$\widetilde r_i^t(s)=\widetilde v_i^t(s)=0$,
$\mathcal D^t(s)=\emptyset$, and
$\widetilde E_i^{t,\rm CE}(s)=0$.
\STATE For all $a,b\in\mathcal A_i^t(s)$, initialize
$q_{t,i}^s(b\mid a)=1/A$, $Z_{t,i}^s(b\mid a)=0$,
$M_{t,i}^s(b\mid a)=1/(2H_t)$, and $n_{t,i}^s(a)=0$.
Set $\overline V_{T+1,i}=0$ and compute the stationary distribution
$\pi_i^t(\cdot\mid s)$ of the row policies.
\STATE Set $(\eta_{\rm CE},\beta_{\rm CE})$ by
\eqref{eq:ce-pred-parameters} using $m=\widetilde L^t(s)$.
\FOR{episode $k\gets1$ to $K$}
    \STATE Record
    $\pi_i^{t,k}(\cdot\mid s)
    \gets\pi_i^t(\cdot\mid s)$
    for all $t\in\mathcal T$
    and $s\in\mathcal S^t$.
    \STATE Receive initial state $s^1$.
    \FOR{step $t\gets1$ to $T$}
        \STATE $\widetilde c:=\widetilde C^t(s^t)
        \gets\widetilde C^t(s^t)+1$.
        \STATE Store the current stationary policy as
        $\pi_i^{t,k}(\cdot\mid s^t)\gets\pi_i^t(\cdot\mid s^t)$.
        Sample internal index
        $J^t\sim\pi_i^t(\cdot\mid s^t)$, then action
        $a_i^t\sim q_{t,i}^{s^t}(\cdot\mid J^t)$; observe
        $r_i^t,s^{t+1}$, obtain $d^t$, and recover
        $d_{-i}^t=\phi_i(a_i^t,d^t)$.
        \STATE Append $d^t$ to $\mathcal D^t(s^t)$ and accumulate
        $\widetilde r_i^t(s^t)\gets\widetilde r_i^t(s^t)+r_i^t$,
        $\widetilde v_i^t(s^t)\gets\widetilde v_i^t(s^t)
        +\overline V_{t+1,i}(s^{t+1})$.
        \STATE For all $a\in\mathcal A_i^t(s^t)$, compute
        $f(a)=[1-r_i^t(s^t,a,d_{-i}^t)]/H_t$; set
        $g=[H_t-1-\overline V_{t+1,i}(s^{t+1})]/H_t$.
        \STATE Accumulate the prediction error using the forecast
        before its update:
        $\widetilde E_i^{t,\rm CE}(s^t)\gets
        \widetilde E_i^{t,\rm CE}(s^t)
        +\|f-M_{t,i}^{s^t}(\cdot\mid J^t)\|_\infty^2$.
        \STATE Update only row $J^t$ of $Z$ by
        \eqref{eq:ce-row-loss}, using
        $a_j=a_i^t$, $g_j=g$, and $\beta=\beta_{\rm CE}$.
        \STATE Update only row $J^t$ of $M$ by
        \eqref{eq:ce-row-forecast}, then increment
        $n_{t,i}^{s^t}(J^t)$ and update that row policy by
        \eqref{eq:ce-row-policy} with $\eta=\eta_{\rm CE}$.
        \IF{$\widetilde C^t(s^t)=\widetilde L^t(s^t)$}
            \STATE Update $\widehat V_{t,i}(s^t)$ by
            \eqref{eq:description-value}, using
            $b_{\widetilde c}=b_{\widetilde c}^{\rm CE}$ from
            \eqref{eq:ce-pred-bonus}; then clip
            $\overline V_{t,i}(s^t)=\min\{\widehat V_{t,i}(s^t),H_t\}$.
            \STATE Set $\lambda(s^t)$ by the CV or MAD rule and update
            $\widetilde L^t(s^t)\gets
            \max\{\widetilde L^t(s^t)+1,
            \lfloor\lambda(s^t)(1+1/T)\widetilde L^t(s^t)\rfloor\}$.
            \STATE Reset $\widetilde C^t(s^t)$,
            $\widetilde r_i^t(s^t)$, $\widetilde v_i^t(s^t)$,
            $\widetilde E_i^{t,\rm CE}(s^t)$, and $\mathcal D^t(s^t)$
            to zero or $\emptyset$; reset all rows to their initial
            values and recompute $(\eta_{\rm CE},\beta_{\rm CE})$
            for the new planned stage length.
        \ENDIF
        \STATE Recompute $\pi_i^t(\cdot\mid s^t)$ as the stationary
        distribution of the updated row policies, for the next visit.
    \ENDFOR
\ENDFOR
\STATE \textbf{Output:}
The historical policy sequence
$\{\pi_i^{t,k}\}_{t\in\mathcal T,\,
k\in[K]}$.
\end{algorithmic}
\end{algorithm*}

\subsection{Sample Complexity of CE Learning}

The stage-wise swap-regret bound,
combined with Bellman sampling
concentration, yields the CE bonus
in~\eqref{eq:ce-pred-bonus}.
The subsequent confidence-bound
induction and recursive stage analysis
follow the CCE framework in
Appendix~\ref{app:cce},
with recommendation-dependent
swap deviations replacing
fixed-action deviations.

\begin{theorem}[Sample complexity of learning CE]
\label{thm:ce}
For any
$\epsilon\in(0,T]$ and $p\in(0,1)$, let $\iota_{\rm CE}$ be defined
in~\eqref{eq:ce-pred-scales}. There is an absolute constant $C>0$
such that, if the agents run Algorithm~\ref{alg:sbv-CE} for
\[
K\ge C\frac{SA_{\max}^2T^5\iota_{\rm CE}}{\epsilon^2},
\]
episodes, then, with probability at least $1-p$, the output policy generated by Algorithm~\ref{alg:certify} from the local histories of
Algorithm~\ref{alg:sbv-CE} is an $\epsilon$-approximate CE.
\end{theorem}
\textit{Proof sketch:}

The proposition in Section~\ref{sec:ce} controls the stage-wise
swap regret. Combining it with Bellman sampling gives
$b_m^{\rm CE}=O(H_tA\sqrt{\iota_{\rm CE}/m})$. The upper/lower
value induction and cumulative stage-length argument are the same as
in Appendix~\ref{app:cce}, with this CE bonus in place of the CCE bonus.
This yields a sufficient episode complexity of \(\widetilde O(SA_{\max}^2T^5/\epsilon^2)\). The complete proof is provided in Appendix~\ref{app:ce}.
\hfill$\blacksquare$

\section{Application to Smart Grids}\label{sec:simulations}\label{sec:smart-grid}

We apply the proposed framework to price-based
demand response in renewable-integrated smart
grids~\cite{maharjan2016demand,dai2025real},
a natural example of an AMG, where renewable
generation enters the system state and agents'
decisions are coupled through electricity
prices~\cite{zhang2026adaptive,etesami2018stochastic}.
Specifically, we consider a microgrid comprising multiple prosumers equipped with rooftop photovoltaic (PV) generation, a grid operator, and shared energy storage. 
In each period, prosumers independently determine their electricity consumption and accordingly purchase electricity from
the grid or sell surplus PV generation. The grid operator supplies power sequentially from public renewable generation, shared storage, and conventional generation, and sets a pricing
rule that depends on aggregate demand, renewable generation, and storage levels. The aggregate-dependent pricing couples prosumers' decisions, while shared storage introduces intertemporal coupling.

\subsection{Prosumer Aggregative Markov Game}

We consider $N=10$ prosumers over a horizon of
$T=5$ representative time periods.
The public renewable generation is modeled as a Markov process:
\begin{equation}
E_N^t=[\bar E_N^t+\xi^t]^+,
\qquad
\bar E_N=(14,17,23,16,11)\ \mathrm{kWh},
\end{equation}
where $\bar E_N^t$ is the forecast renewable
generation and $\xi^t\in\{-2,0,2\}$ is
a Markovian forecast error.
The initial distribution of $\xi^1$ is
$(0.25,0.50,0.25)$, and its transition
matrix is
\begin{equation}
\hat q_\xi=
\begin{pmatrix}
0.70&0.25&0.05\\
0.20&0.60&0.20\\
0.05&0.25&0.70
\end{pmatrix},
\end{equation}
with rows and columns ordered as
$(-2,0,2)$.
The shared energy storage has a capacity
of $B=12$ kWh, an initial level of
$b^1=6$ kWh, and discrete energy levels
$\mathcal B=\{0,1,\ldots,12\}$.

The interactions among prosumers are
formulated as an AMG with the following
components.

\textit{1) State space:}
At each period $t$, all prosumers observe
the common system state
$s^t=(E_N^t,b^t)$,
where $E_N^t$ and $b^t$ denote the
public renewable output and shared
storage level, respectively.
Each prosumer $i$ has local PV generation
$g_i^t=\theta_i E_N^t$, with
$\theta_i=0.09$ in this experiment.
Since $\theta_i$ is a fixed local
parameter, $g_i^t$ is determined by
the observed public state and does not need to 
introduce an additional state variable.

\textit{2) Action space:}
Each prosumer $i$ independently chooses
its electricity consumption level
$a_i^t=\ell_i^t\in\mathcal A_i
=\{2,3,4\}$ kWh.
Its electricity purchase and sale quantities
are determined by
\[
d_i^t=[\ell_i^t-g_i^t]^+,
\qquad
c_i^t=[g_i^t-\ell_i^t]^+,
\]
which satisfy $d_i^tc_i^t=0$.
Thus, a prosumer purchases electricity
when local PV generation is insufficient
and sells any surplus electricity to the grid.

\textit{3) Aggregative stage rewards:}
The stage reward of prosumer $i$ consists
of consumption utility, electricity sale
revenue, and electricity purchase cost:
\begin{equation}\label{eq:grid-reward}
\begin{aligned}
r_i^t= U_i(\ell_i^t)
+P_{\mathrm{sell}}(d^t,E_N^t,b^t)c_i^t-P(d^t,E_N^t,b^t)d_i^t,
\end{aligned}
\end{equation}
where $d^t=\sum_{i=1}^N d_i^t$
is the aggregate electricity purchase
demand, and $P$ and $P_{\mathrm{sell}}$
are the purchase and feed-in prices,
respectively.

The electricity prices (yuan/kWh)
are specified as
\begin{align}
D_{\mathrm{ref}}(E)
&=N[2-0.09E]^+,\notag\\
P(D,E,b)
&=\bigg[
0.35+0.90\frac{D-D_{\mathrm{ref}}(E)}{N}
\notag\\
&
+0.05\frac{16-E}{23}
+0.45\left(0.5-\frac b{12}\right)
\bigg],
\label{eq:grid-price}\\
P_{\mathrm{sell}}(D,E,b)
&=0.65P(D,E,b).
\notag
\end{align}
Here, $D_{\mathrm{ref}}(E)$ denotes
the aggregate purchase demand when
all prosumers choose their minimum
consumption level.
The pricing rule captures the effects
of aggregate electricity demand,
renewable generation, and storage
availability.

The consumption utility in \eqref{eq:grid-reward} is
\begin{equation}
U_i(\ell)=v_i\ell-\frac12m_i\ell^2.
\end{equation}
To capture heterogeneous preferences,
we consider two types of prosumers,
with five agents in each type:
\begin{equation}
(v_i,m_i)=
\begin{cases}
(1.6875,0.244375),&i=1,\ldots,5,\\
(1.9125,0.165625),&i=6,\ldots,10.
\end{cases}
\end{equation}
Type 1 prosumers have lower marginal
consumption utility, whereas Type 2
prosumers exhibit a stronger preference
for higher electricity consumption.
For learning, the rewards are normalized
as $\widetilde r_i^t=(r_i^t+2)/10
\in[0,1]$; operational results are
reported on the original scale.

Importantly, given $s^t$, the reward
in~\eqref{eq:grid-reward} depends only on
prosumer $i$'s own action and the aggregate
imports of the other prosumers:
\begin{equation}
r_i^t(s^t,\bm a^t)
=r_i^t(s^t,a_i^t,d_{-i}^t),
\qquad
d_{-i}^t=d^t-d_i^t.
\end{equation}
Thus, the game satisfies the
aggregative reward structure.

\textit{4) State transition:}
Renewable generation evolves according
to the Markov chain $\hat q_\xi$,
while shared storage evolves with
electricity trading and grid dispatch.
Let $C^t=\sum_i c_i^t$ denote
aggregate electricity sales, and
define the net electricity demand as
\begin{equation}
n^t=d^t-c^t-E_N^t.
\end{equation}
If $n^t\ge0$, the deficit is first covered by battery discharge, then by conventional thermal generation.
If $n^t<0$, surplus energy charges the battery, and any remaining surplus is curtailed.
With charging and discharging efficiencies
$\eta_c=\eta_d=0.9$, the storage
transition is
\begin{equation}\label{eq:grid-storage}
b^{t+1}=
\begin{cases}
\left\lfloor b^t-
\min\left\{b^t,\dfrac{n^t}{\eta_d}\right\}
\right\rfloor,
& n^t\ge0,\\[3mm]
\left\lfloor b^t+\eta_c
\min\left\{-n^t,
\dfrac{B-b^t}{\eta_c}\right\}
\right\rfloor,
& n^t<0.
\end{cases}
\end{equation}
where $\lfloor\cdot\rfloor$ discretizes
the storage level to the nearest
lower integer.

Let $\mathcal F(b^t,\bm a^t,E_N^t)$
denote the storage update in
\eqref{eq:grid-storage}. The resulting state transition kernel is
\begin{equation}
\begin{aligned}
q_t(s^{t+1}\mid s^t,\bm a^t)=
\hat q_\xi(\xi^{t+1}\mid\xi^t)
\mathbf1\left\{
b^{t+1}=
\mathcal F(b^t,\bm a^t,E_N^t)
\right\}. \notag
\end{aligned}
\end{equation}

\textit{Decentralized learning setting:}
Each prosumer aims to maximize its expected cumulative reward $V_{1,i}(\cdot)$~\eqref{eqn:value} over the $T$ periods via decentralized learning.
Each prosumer knows its own utility function and the pricing rules, while the state transition kernel and opponents' individual decisions remain unknown. Since the price is strictly increasing in $d^t$, each prosumer can recover $d^t$ from the observed price and state, and hence obtain $d_{-i}^t=d^t-d_i^t$.  This provides the aggregate feedback required by our algorithm.

\begin{figure*}[t]
    \centering
    \includegraphics[width=0.98\textwidth]{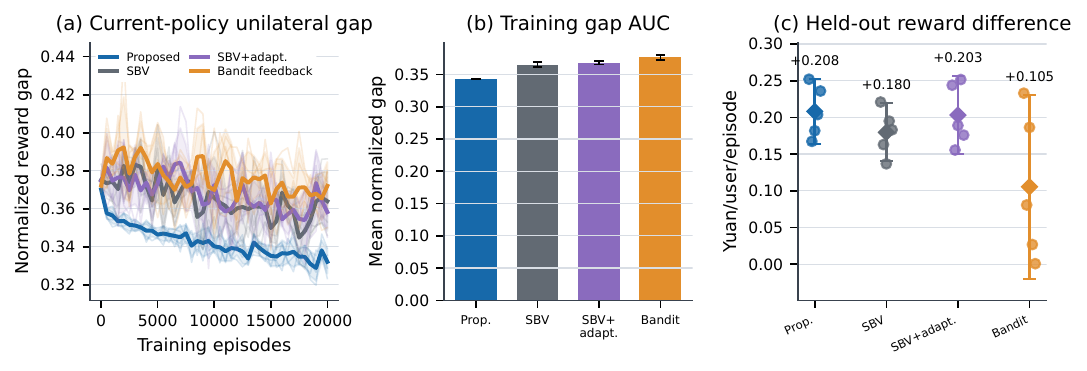}
    \caption{Learning performance over 20,000
    episodes. (a) Current-policy unilateral
    deviation gap during training.
    (b) Average gap over training, measured
    by the area under the gap curve (gap-AUC).
    (c) Final-policy reward difference relative
    to the Myopic price-taking rule.
    Results use five matched seeds;
    error bars in (b) show standard
    deviations, while (c) shows
    seed-level gains and 95\% paired
    confidence intervals.}
    \label{fig:grid-learning}
\end{figure*}

\begin{figure*}[t]
    \centering
    \includegraphics[width=0.86\textwidth]{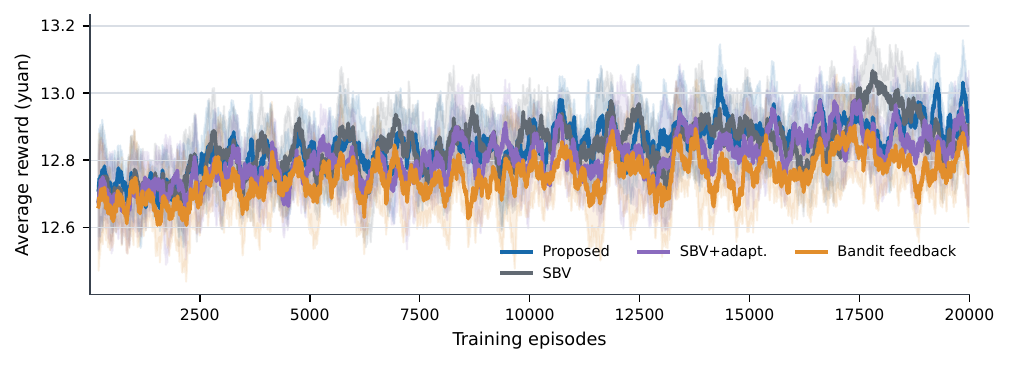}
    \caption{Mean realized reward per prosumer over each five-period episode during training, smoothed over 200 episodes. Shading shows $\pm$ one standard deviation across five seeds.}
    \label{fig:grid-training-reward}
\end{figure*}

\subsection{Simulation Results}

\subsubsection{Learning performance}
We apply Algorithm~\ref{alg:sbv} to the above AMG and first compare it with Stage-Based V-Learning for CCE (SBV)~\cite{mao2022improving} (a representative decentralized algorithm with CCE guarantees for general-sum MGs), its adaptive-stage variant (SBV+adapt.), and a bandit-feedback ablation (Bandit) of our algorithm without aggregate-based policy updates. All methods are trained for $20,000$ episodes using five matched random seeds. We evaluate equilibrium learning using the \emph{unilateral-deviation gap}, defined as the maximum gain any agent can obtain by switching to a best-response policy. 
A zero gap indicates an exact CCE, while smaller gaps indicate weaker incentives for unilateral deviations and hence greater strategic stability. As shown in Fig.~\ref{fig:grid-learning}, our algorithm achieves the smallest final gap ($0.3321$ versus $0.3640$ for SBV) and gap-AUC ($0.3425$). Furthermore, the inferior performance of Bandit and the limited improvement from stage adaptation alone indicate that exploiting aggregate feedback improves learning efficiency, consistent with Theorem~\ref{thm:pred-comparison}. Moreover, our algorithm achieves the highest held-out mean reward among the four stage-based methods (Fig.~\ref{fig:grid-training-reward} and Table~\ref{tab:grid-metrics}). Finally, the certified output policy constructed by Algorithm~\ref{alg:certify} achieves a small deviation gap of $0.13753$ in normalized cumulative payoff, consistent with the CCE guarantee in Theorem~\ref{thm:cce}.

\begin{figure}[t]
    \centering
    \includegraphics[width=\columnwidth]{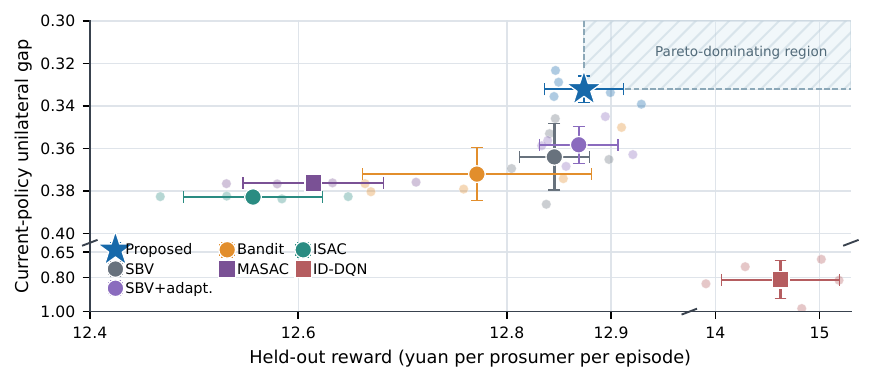}
    \caption{Held-out reward versus
    current-policy unilateral-deviation
    gap for the evaluated algorithms.
    Small markers represent individual
    seeds; large markers show means
    with standard deviations.
    The shaded region indicates
    strict Pareto dominance over
    the proposed algorithm.
    Rewards use original payoff units,
    while gaps are normalized.}
    \label{fig:grid-reward-gap}
\end{figure}

We further compare our algorithm with three representative MARL methods without established CCE guarantees: MASAC and ISAC~\cite{bettini2024benchmarl}, and ID-DQN~\cite{vanhasselt2016double}. As shown in Fig.~\ref{fig:grid-reward-gap}, our algorithm achieves the smallest deviation gap while maintaining competitive rewards, demonstrating greater strategic stability. Although ID-DQN achieves the highest mean reward, its substantially larger deviation gap indicates strong incentives for unilateral deviations. Moreover, no evaluated baseline simultaneously achieves a higher reward and a smaller gap than our algorithm. These comparisons highlight the favorable reward--stability trade-off achieved by our algorithm.

\subsubsection{Demand response and grid operation}

\begin{figure*}[t]
    \centering
    \includegraphics[width=0.95\textwidth]{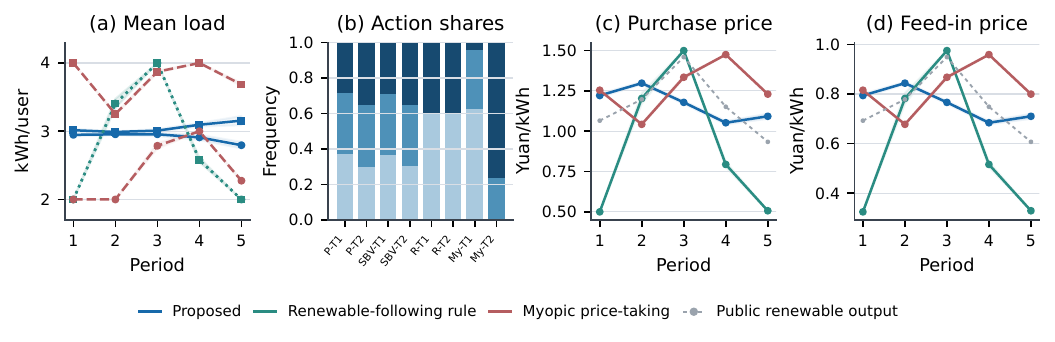}
    \caption{Prosumer consumption and
    electricity prices under the evaluated
    policies.
    (a) Mean consumption by user type;
    (b) action frequencies (2, 3, and
    4 kWh, bottom to top);
    (c) purchase price;
    (d) feed-in price.
    Circles and squares in (a) denote
    Types 1 and 2, respectively.
    Dashed curves in (c)--(d) show
    public renewable output.
    Shaded bands indicate
    $\pm$ one standard deviation
    across five seeds.}
    \label{fig:grid-users}
\end{figure*}

\begin{figure*}[t]
    \centering
    \includegraphics[width=0.98\textwidth]{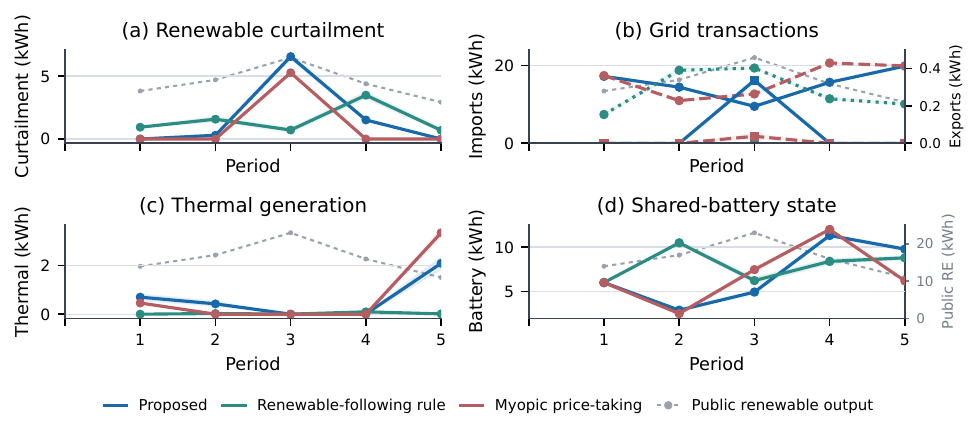}
    \caption{Grid operation under the evaluated policies: (a) renewable curtailment, (b) imports (circles) and exports (squares), (c) thermal generation, and (d) shared-battery state. Dashed curves show public renewable output; shading indicates $\pm$ one standard deviation across five seeds.}
    \label{fig:grid-operation}
\end{figure*}

\begin{table*}[htbp]
\centering
\caption{Summary of learning and
microgrid performance.
Results are means $\pm$ standard
deviations across five seeds.
Rewards are in yuan per prosumer
per episode, and deviation gaps
use normalized cumulative payoffs.
Thermal generation is in kWh
per episode.} \label{tab:grid-metrics}
\scriptsize
\setlength{\tabcolsep}{2pt}
\begin{tabular}{lcccccc}
\hline
Method & Reward & \shortstack{Current-policy\\gap (norm.)} & \shortstack{Certified-output gap$^{\dagger}$\\(norm.)} & Gap-AUC & \shortstack{Renewables\\used (\%)} & Thermal (kWh)\\
\hline
\multicolumn{7}{l}{\textit{Learning algorithms}}\\
Proposed & $12.874\pm0.038$ & $0.3321\pm0.0062$ & $0.13753\pm0.00009$ & $0.3425\pm0.0005$ & $94.55\pm0.16$ & $3.23\pm0.20$\\
SBV & $12.846\pm0.034$ & $0.3640\pm0.0156$ & -- & $0.3653\pm0.0041$ & $94.45\pm0.29$ & $3.34\pm0.10$\\
SBV+adapt. & $12.869\pm0.038$ & $0.3582\pm0.0087$ & -- & $0.3680\pm0.0026$ & $94.31\pm0.30$ & $3.33\pm0.33$\\
Bandit feedback & $12.771\pm0.110$ & $0.3721\pm0.0125$ & -- & $0.3765\pm0.0038$ & $94.57\pm0.19$ & $3.60\pm0.34$\\
MASAC (discrete) & $12.614\pm0.067$ & $0.3761\pm0.0005$ & -- & -- & $94.73\pm0.20$ & $3.92\pm0.27$\\
ISAC (discrete) & $12.557\pm0.066$ & $0.3828\pm0.0005$ & -- & -- & $94.91\pm0.20$ & $4.12\pm0.21$\\
Indep. Double DQN & $14.625\pm0.567$ & $0.8122\pm0.1113$ & -- & -- & $93.07\pm1.78$ & $1.11\pm0.67$\\
\hline
\multicolumn{7}{l}{\textit{Reference demand-response policies}}\\
Renewable-following rule & $13.002\pm0.034$ & $1.6776$ & -- & -- & $95.18\pm0.36$ & $0.14\pm0.02$\\
Myopic price-taking & $12.666\pm0.014$ & $1.0531$ & -- & -- & $96.57\pm0.04$ & $3.81\pm0.15$\\
\hline
\end{tabular}
\vspace{1mm}
\parbox{\textwidth}{\footnotesize $^{\dagger}$Gap of Algorithm~\ref{alg:certify}'s certified output policy, evaluated using its shared historical index.}
\end{table*}

We further examine prosumer consumption and grid operation under the learned policy and two representative demand-response rules. The \emph{renewable-following} rule selects 2 kWh when public renewable output is at most 17 kWh and 4 kWh otherwise, representing an idealized strategy that adjusts consumption directly to renewable availability without considering electricity prices or storage conditions. The \emph{myopic price-taking} (Myopic) rule maximizes each prosumer's current-period reward at a reference price computed when all prosumers consume 3 kWh, ignoring future storage effects. All policies are evaluated over 300 independent test episodes per seed. As shown in Fig.~\ref{fig:grid-users}, under the learned policy, the mean consumption remains stable at around 3 kWh across periods, while retaining the demand difference between the two prosumer types. The learned policy also exhibits more moderate electricity-price variations, with period-mean purchase prices ranging from 1.053 to 1.298 yuan/kWh, compared with 1.043--1.475 under Myopic and 0.50--1.50 under renewable-following. Such stable consumption patterns and moderate price variations can facilitate demand-side management and supply--demand balancing.

Figure~\ref{fig:grid-operation} presents grid-operation profiles, including renewable curtailment, thermal generation, and shared storage dynamics, while Table~\ref{tab:grid-metrics} summarizes the economic, operational, and equilibrium metrics. All three policies maintain high renewable utilization above 94\%. Compared with Myopic, our policy achieves higher user rewards and lower thermal generation, while Myopic exhibits larger interperiod consumption variations, reflecting the limitations of short-sighted decisions in a multistage setting. Compared with renewable-following, which achieves substantially lower thermal generation by directly aligning consumption with renewable availability, our policy attains a much smaller unilateral-deviation gap (0.332 versus 1.678). This indicates that demand-response rules designed for good operational performance may fail to match the incentives of self-interested prosumers. These findings illustrate the potential of decentralized CCE learning to support strategically stable demand response in renewable-integrated smart grids.

\section{Conclusion}\label{sec:conclusions}
This paper investigates decentralized equilibrium learning in aggregative Markov games (AMGs). By incorporating aggregate feedback into an adaptive stage-based V-learning framework with an aggregation-predictive hybrid-feedback Exp3-IX oracle, we prove a high-probability no-regret guarantee for the proposed oracle and derive tighter stage-wise individual-gap bounds than prior methods. We further derive an episode-complexity bound of $\widetilde O(SA_{\max}T^5/\epsilon^2)$ for approximate coarse correlated equilibrium (CCE) learning under the prescribed adaptive stage-length schedule, together with an extension to correlated equilibrium (CE) learning. We apply the algorithm to a renewable-rich smart grid with strategic prosumers. Numerical experiments show that our algorithm achieves smaller unilateral-deviation gaps and stronger strategic stability compared with existing baselines, while attaining competitive user payoffs and favorable grid operational performance. In the future, we plan to tighten sample-complexity bounds by better exploiting the aggregative structure, study decentralized Nash equilibrium learning under additional structural assumptions, and extend the framework to other classes of structured Markov games.

\bibliographystyle{IEEEtran}
\bibliography{refs}
\appendices

\section{Proofs for Section~\ref{sec:cce_ce}}\label{app:cce}

\subsection{Stage-Oracle Parameters}\label{app:pred-parameters}
Let the remaining horizon and confidence logarithm be
\begin{equation}\label{eq:pred-scales}
 H_t=T-t+1,\qquad
 \iota=\log\frac{16NSA_{\max}TK}{p}.
\end{equation}
For a stage of length $\tilde c$ with $A=|\mathcal A_i^t(s)|$,
let $C_A=\log A+\iota/4$ and set
\begin{equation}\label{eq:pred-parameters}
 \begin{aligned}
 \eta&=\min\left\{H_t,\,
 H_t\sqrt{\frac{C_A}{\tilde c[2+4A(H_t-1)^2]}}\right\},\\
 \beta&=\frac{2\eta(H_t-1)}{H_t}.
 \end{aligned}
\end{equation}
In Algorithm~\ref{alg:sbv}, $(\eta,\beta)$ denotes a separately stored
pair for each $(t,s)$, fixed within its current stage.
Evaluate \eqref{eq:pred-parameters} at its planned length $\widetilde L^t(s)$.
The learning rate
balances the regret terms, while $\beta$ controls implicit exploration.

For a completed stage, define the cumulative prediction error
$E_{\tilde c}
:=
\sum_{j=1}^{\tilde c}
\|f_j-M_j\|_\infty^2$,
where each $M_j$ is the forecast used before observing
the $j$th sample.
For the running-mean forecast $M_{t,i}(\cdot)$ (line-\ref{line:forecast-update}) in Algorithm~\ref{alg:sbv},
this quantity is exactly the prediction-error accumulator
maintained by the algorithm:
$E_{\tilde c}=\widetilde E_i^t(s)$ (line-\ref{pred-error-update}).

For $A\ge2$, define
 \begin{align*}
 U_{\tilde c}(x;s^t):=\frac{H_t}{\tilde c}\Bigg[
 &\frac{C_A}{\eta}+2\eta x+
 \frac{4\eta A\tilde c (H_t-1)^2}{H_t^2}\\
 &+\iota+(1+\frac1{2H_t})\sqrt{2\tilde c\iota}\Bigg].
 \end{align*}
Restating the main-text definition for reference, the stage-end bonus is
\begin{equation}\label{eq:app-pred-bonus}
 b^{\mathrm{pred}}_{t,i}(\tilde c,x;s^t)=
 \begin{cases}
 \min\{H_t,H_t\sqrt{2\iota/\tilde c}\},&A=1,\\
 \min\{H_t,U_{\tilde c}(x;s^t)\},&A\ge2.
 \end{cases}
\end{equation}
For $A=1$, only the Bellman sampling error remains.

The stage length is initialized at $T$ and updated by
\[
\widetilde L_{\mathrm{new}}
=\max\left\{\widetilde L+1,
\left\lfloor\lambda\left(1+\frac1T\right)\widetilde L\right\rfloor\right\},~
\lambda\in\left[\frac{T+1/2}{T+1},1\right].
\]

\subsection{Estimation and CCE guarantees}

Throughout this appendix, agents know their reward maps and, after each joint action, obtain the common aggregate and recover the opponents' aggregate as specified in the main text. Conditional on the pre-action joint history, agents sample actions independently. Rewards lie in $[0,1]$, and the remaining-horizon value functions lie in $[0,H_t]$.

We first introduce a few notations to facilitate the analysis. For a step $t\in \mathcal{T}$ of an episode $k\in[K]$, we denote by $s^{t,k}$ the state that the agents observe at this time step. For any state $s^t\in\mathcal{S}^t$, we let $\pi_{i}^{t,k}(\cdot \mid s^t)\in\Delta(\mathcal{A}_i^t(s^t))$ be the distribution prescribed by Algorithm~\ref{alg:sbv} to agent $i$ at this step. Notice that such notations are well-defined for every $s^t \in\mathcal{S}^t$ even if $s^t$ might not be the state $s^{t,k}$ that is actually visited at the given step. We further let $\pi_{i}^{t,k} = \{\pi_{i}^{t,k}(\cdot\mid s^t):s^t \in\mathcal{S}^t\}$, and let $a_{i}^{t,k}\in\mathcal{A}_i^t(s^t)$ be the actual action taken by agent $i$. For any $s^t \in\mathcal{S}^t$, let $\tilde{C}^{t,k}(s^t)$ denotes the value of $\tilde{C}^{t}(s^t)$ at the \emph{beginning} of the $k$-th episode. The analysis uses a coefficient construction with common stage boundaries, so the same visit-count and predecessor-stage notation applies across agents. We also use $\overline{V}_{t,i}^{k}(s^t)$ and $\hat{V}_{t,i}^{k}(s^t)$ to denote the values of $\overline{V}_{t,i}(s^t)$ and $\hat{V}_{t,i}(s^t)$, respectively, at the beginning of the $k$-th episode from agent $i$'s perspective. 

Further, for a state $s^{t,k}$, let $\tilde{c}^{t,k}$ denote the number of times that state $s^{t,k}$ has been visited (at the $t$-th step) in the stage right before the current stage, and let $\tilde{l}_{j}^{t,k}$ denote the index of the episode that this state was visited the $j$-th time among the $\tilde{c}^{t,k}$ times. For notational convenience, we use $\tilde{c}$ to denote $\tilde{c}^{t,k}$, and $\tilde{l}_j$ to denote $\tilde{l}_{j}^{t,k}$, whenever $t$ and $k$ are clear from the context. With the new notations, the update rule in Line~\ref{line:14} of Algorithm~\ref{alg:sbv} can be equivalently expressed as
\begin{equation}\label{eqn:update_rule_new}
\hat{V}_{t,i}(s^t) \gets \frac{1}{\tilde{c}}\sum_{j=1}^{\tilde{c}} \left(r_{i}^t(s^t,\bm{a}^{t,\tilde{l}_j}) + \overline{V}_{t+1,i}^{\tilde{l}_j}(s^{t+1,\tilde{l}_j}) \right)+ b_{\tilde{c}}. 
\end{equation}

To streamline the Bellman equations for the value and Q-functions, we introduce two auxiliary operators. First, for any stage-$t$ state-action pair $(s^t,\bm{a}^t)$ and any value function $V$ defined on the stage-$t+1$ state space $\mathcal{S}^{t+1}$, the state transition operator is
$$
\mathbb{P}_{t} V(s^t, \bm{a}^t) := \mathbb{E}_{s^{t+1} \sim p^{t}(\cdot \mid s^t, \bm{a}^t)} V\left(s^{t+1}\right).
$$
Second, for a joint policy profile $\bm\pi = (\pi_i^t)_{i\in\mathcal{N},t\in\mathcal{T}}$, we define the stage-$t$ marginal joint decision rule as $\boldsymbol{\pi}^t := (\pi_1^t, \pi_2^t, \dots, \pi_N^t)$, which is the collection of stage-$t$ decision rules of all agents extracted from the full joint Markov policy $\bm\pi$. For any stage-$t$ state $s^t$ and any Q-function $Q$ defined on $(s^t,\mathcal{A}^t(s^t))$, the policy expectation operator is
$$
\mathbb{D}_{\boldsymbol{\pi}^t} Q(s^t) := \mathbb{E}_{\bm{a}^t \sim \boldsymbol{\pi}^t(\cdot \mid s^t)} Q(s^t, \bm{a}^t),
$$
where $\bm{a}^t \sim \boldsymbol{\pi}^t(\cdot \mid s^t)$ means the joint action $\bm{a}^t$ is sampled from the product distribution induced by $\boldsymbol{\pi}^t$ at state $s^t$.

With these notations and definitions, the Bellman equations for the joint value and Q-functions can be written as
\begin{align*}
Q_{t}^{\bm\pi}(s^t, \bm{a}^t) &= \left(r_{t} + \mathbb{P}_{t} V_{t+1}^{\bm\pi}\right)(s^t, \bm{a}^t),  \\
V_{t}^{\bm\pi}(s^t) &= \left(\mathbb{D}_{\boldsymbol{\pi}^t} Q_{t}^{\bm\pi}\right)(s^t) 
\end{align*}
for all $t \in \mathcal{T}$, $s^t \in \mathcal{S}^t$, and $\bm{a}^t\in \mathcal{A}^t(s^t)$.

The certified output policy first samples an episode index uniformly from $[K]$. At each step, if the current state has a nonempty predecessor stage, it samples uniformly one of that stage's visit indices, sets the active episode index to the selected episode, and then executes the corresponding joint policy. If the predecessor stage is empty, it uses the fixed uniform fallback for the remainder of the episode. Thus the policy indexed by $k$ at step $t$ uses the sampled predecessor-stage index before selecting its action.

In the following proof, we assume without loss of generality that the initial state $s^1$ is fixed (i.e., the initial state distribution $\rho$ is a point mass at $s^1$). Our proof can be straightforwardly generalized to the case where $s^1$ is drawn from an arbitrary fixed distribution $\rho\in\Delta(\mathcal{S}^1)$.

We first restate and prove the Stage-wise individual gap bound.
Write $A=|\mathcal A_i^t(s)|$,
and $d_j=d_{-i}^{t,\tilde l_j}$; $b_{\tilde c}$ is the main-text
bonus in \eqref{eq:pred-bonus}, restated as
\eqref{eq:app-pred-bonus} for this appendix.

\begin{restatement}{Stage-wise individual gap}
With probability at least
$1-p/2$, it holds for all agents $i \in \mathcal{N},$ episodes $k \in [K],$  steps $t \in \mathcal{T},$  and states $s^t\in\mathcal S^t$ with $\tilde c>0$,
\begin{align}
 &\Gamma_{\tilde c} := \max_{\pi_i^t}\frac1{\tilde c}\sum_{j=1}^{\tilde c}
 \mathbb D_{\pi_i^t\times\pi_{-i}^{t,\tilde l_j}}
 (r_i^t+\mathbb P_t\overline V_{t+1,i}^{\tilde l_j})(s)
 \notag\\
 &\quad-\frac1{\tilde c}\sum_{j=1}^{\tilde c}
 \left[r_i^t(s,\bm a^{t,\tilde l_j})
 +\overline V_{t+1,i}^{\tilde l_j}(s^{t+1,\tilde l_j})\right]
 \le b_{\tilde c}.\label{eq:app-pred-lemma-bound}
\end{align}
\end{restatement}

\emph{Proof:}
The proof proceeds in three steps.
Fix an agent $i$, a step-state pair $(t,s)$,
and an initiated stage of planned length
$\tilde c$.
All probability bounds below are applied
conditionally on stage initiation, when
the stage length and oracle parameters
are fixed.
Set $\delta_0=p/(4NSTK)$.

\textbf{Step 1: Return sampling error.}
Let $\mathcal F_{j-1}$ denote the joint
history immediately before the $j$th action
at $(t,s)$ within the stage, including
intervening episodes.
Define
\[
Y_j=r_i^t(s,\bm a^{t,\tilde l_j})
+\overline V_{t+1,i}^{\tilde l_j}
(s^{t+1,\tilde l_j}).
\]
By the definition of the state transition
kernel and the joint policy, we have
\[
\mathbb E[Y_j\mid\mathcal F_{j-1}]
=
\mathbb D_{\bm\pi^{t,\tilde l_j}}
(r_i^t+\mathbb P_t
\overline V_{t+1,i}^{\tilde l_j})(s).
\]
Thus,
$\{Y_j-\mathbb E[Y_j\mid\mathcal F_{j-1}]\}_j$
forms a martingale difference sequence.
Since $Y_j\in[0,H_t]$, the conditional range
of each martingale difference has width
at most $H_t$.
Applying conditional Hoeffding's inequality,
with probability at least $1-\delta_0$,
we have
\begin{align}\label{eqn:martingale difference}
&\frac1{\tilde c}\sum_{j=1}^{\tilde c}
\mathbb D_{\bm\pi^{t,\tilde l_j}}
(r_i^t+\mathbb P_t
\overline V_{t+1,i}^{\tilde l_j})(s)
\notag\\
&\quad-\frac1{\tilde c}
\sum_{j=1}^{\tilde c}
\left[
r_i^t(s,\bm a^{t,\tilde l_j})
+\overline V_{t+1,i}^{\tilde l_j}
(s^{t+1,\tilde l_j})
\right]
\le H_t\sqrt{\frac{\iota}{2\tilde c}}.
\end{align}

\textbf{Step 2: Within-stage external regret.}
It remains to bound
\begin{align*}
R_{\tilde c}^\star:={}&
\max_{\pi_i^t}\frac1{\tilde c}
\sum_{j=1}^{\tilde c}
\mathbb D_{\pi_i^t\times
\pi_{-i}^{t,\tilde l_j}}
(r_i^t+\mathbb P_t
\overline V_{t+1,i}^{\tilde l_j})(s)
\notag\\
&-\frac1{\tilde c}\sum_{j=1}^{\tilde c}
\mathbb D_{\bm\pi^{t,\tilde l_j}}
(r_i^t+\mathbb P_t
\overline V_{t+1,i}^{\tilde l_j})(s).
\end{align*}
This is the average external regret of
the agent's within-stage policies relative
to the best fixed action in hindsight.

Let $p_j=\pi_i^{t,\tilde l_j}(\cdot\mid s)$
denote the agent's action distribution at
the $j$th visit, and define the normalized
expected loss of action $a$ as
\begin{equation}\label{eq:stage-expected-loss}
\ell_j(a)=
1-\frac1{H_t}
\mathbb E_{a_{-i}\sim
\pi_{-i}^{t,\tilde l_j}(\cdot\mid s)}
\left[
(r_i^t+\mathbb P_t
\overline V_{t+1,i}^{\tilde l_j})
(s,a,a_{-i})
\right].
\end{equation}
By linearity in the comparison policy,
\begin{equation}\label{eq:regret-equivalence}
R_{\tilde c}^\star=
\frac{H_t}{\tilde c}\left[
\sum_{j=1}^{\tilde c}\langle p_j,\ell_j\rangle
-\min_a\sum_{j=1}^{\tilde c}\ell_j(a)
\right].
\end{equation}

Thus, minimizing $R_{\tilde c}^\star$~\eqref{eq:regret-equivalence} reduces
to an adversarial multi-armed bandit problem
with hybrid feedback: the observed opponents'
aggregate enables evaluation of the immediate
loss of every action, whereas the continuation
loss is observed only for the selected action.

Algorithm~\ref{alg:sbv} addresses this problem
using the aggregation-predictive Exp3-IX oracle,
whose high-probability regret guarantee is
established in the following proposition.

Since $A=|\mathcal A_i^t(s)|\le A_{\max}$,
we have
$L_{\delta_0}=\log(16ANSTK/p)\le\iota$
and $\log A+L_{\delta_0}/4\le C_A$.
Applying the regret guarantee stated below,
with probability at least $1-\delta_0$,
for $A\ge2$,
\begin{equation}\label{eqn:no-regret}
\begin{aligned}
R_{\tilde c}^\star\le
\frac{H_t}{\tilde c}\Bigg[
&\frac{C_A}{\eta}
+2\eta E_{\tilde c}
+4\eta A\tilde c G^2\\
&+\iota+
\left(\frac12+\frac1{2H_t}\right)
\sqrt{2\tilde c\iota}
\Bigg],
\end{aligned}
\end{equation}
where $F=1/H_t$, $G=(H_t-1)/H_t$, and
$E_{\tilde c}=\sum_{j=1}^{\tilde c}
\|f_j-M_j\|_\infty^2$.
For $A=1$, we have $R_{\tilde c}^\star=0$.

\textbf{Step 3: Combining the bounds.}
By the definition of $\Gamma_{\tilde c}$,
adding \eqref{eqn:martingale difference}
and \eqref{eqn:no-regret} yields, for $A\ge2$,
\begin{align*}
\Gamma_{\tilde c}
\le\frac{H_t}{\tilde c}\Bigg[
&\frac{C_A}{\eta}
+2\eta E_{\tilde c}
+4\eta A\tilde c G^2\\
&+\iota+
\left(1+\frac1{2H_t}\right)
\sqrt{2\tilde c\iota}
\Bigg].
\end{align*}
For Algorithm~\ref{alg:sbv},
$E_{\tilde c}=\widetilde E_i^t(s)$.
For $A=1$, \eqref{eqn:martingale difference}
directly gives
$\Gamma_{\tilde c}
\le H_t\sqrt{\iota/(2\tilde c)}$.
Combining these bounds with the deterministic
bound $\Gamma_{\tilde c}\le H_t$ yields
$\Gamma_{\tilde c}\le b_{\tilde c}$
as defined in \eqref{eq:app-pred-bonus}.

The probability bounds apply to each
initiated stage, without conditioning on
its eventual completion.
A union bound over at most $NSTK$ initiated
stages yields a total failure probability
of at most $2NSTK\delta_0=p/2$.
Consequently, the claimed bound~\eqref{eq:app-pred-lemma-bound} holds
simultaneously for all completed stages.
\hfill$\blacksquare$

To establish the regret guarantee used in
the stage-wise Bellman bound above, we analyze the
aggregation-predictive Exp3-IX update in a
general adversarial bandit setting.

Consider $A$ actions and $\tilde c$ rounds.
At round $j$, the learner selects $a_j\sim p_j$
and gets the immediate-loss vector
$f_j\in[0,F]^A$ for all actions, together with
a continuation loss $g_j\in[0,G]$ only for the
selected action, where $F+G\le1$.
The immediate-loss feedback satisfies
\[
\mathbb E[f_j(a)\mid\mathcal F_{j-1},a_j=b]
=
\mathbb E[f_j(a)\mid\mathcal F_{j-1}]
\]
for all actions $a,b$, where $\mathcal F_{j-1}$
is the pre-round history.
The forecast $M_j\in[0,F]^A$ is
$\mathcal F_{j-1}$-measurable.

The learner updates
\begin{align}
p_j(a)&\propto
\exp\{-\eta[Z_{j-1}(a)+M_j(a)]\},
\label{eq:app-pred-oracle-policy}\\
Z_j(a)&=Z_{j-1}(a)+f_j(a)
+\frac{\mathbf1\{a_j=a\}g_j}
{p_j(a)+\beta},
\label{eq:app-pred-oracle-loss}
\end{align}
where $Z_0=0$, $\eta>0$, $\eta F\le1$,
and $\beta=2\eta G$.
Define the conditional expected loss by
\begin{align*}
\ell_j(a)=
\mathbb E[f_j(a)+g_j
\mid\mathcal F_{j-1},a_j=a].
\end{align*}
For the stage-wise Markov-game application,
this is the normalized Bellman loss in
\eqref{eq:stage-expected-loss}.

\begin{restatement}{Regret guarantee for aggregation-predictive oracle}
For the hybrid-feedback adversarial
bandit problem described above,
suppose the learner follows
\eqref{eq:app-pred-oracle-policy}--
\eqref{eq:app-pred-oracle-loss}. Then, for any $\tilde c\ge1$ and
$\delta\in(0,1)$, with probability at least
$1-\delta$,
\begin{equation}\label{eq:app-pred-regret-prop}
\begin{aligned}
\mathcal R_{\tilde c}
:={}&
\sum_{j=1}^{\tilde c}\langle p_j,\ell_j\rangle
-\min_a\sum_{j=1}^{\tilde c}\ell_j(a)\\
\le{}&
\frac{\log A+L_\delta/4}{\eta}
+2\eta E_{\tilde c}
+4\eta A\tilde c G^2\\
&+L_\delta
+ \frac{2F+G}{2} \sqrt{2\tilde c L_\delta},
\end{aligned}
\end{equation}
where $L_\delta=\log(4A/\delta)$ and
$E_{\tilde c}=\sum_{j=1}^{\tilde c}
\|f_j-M_j\|_\infty^2$. For fixed $A$, $F$, $G$, and $\delta$,
the bound further implies
$\mathcal R_{\tilde c}=O(\sqrt{\tilde c})$
when $\eta=\Theta(\tilde c^{-1/2})$.
\end{restatement}

\emph{Proof:}
The objective is to bound the regret
with respect to the conditional expected
losses $\ell_j$. Since the exponential-weights
policy is based on the accumulated loss
estimates $Z_j$, we first control the
regret for these estimates and then
account for the estimation errors.

Consider
$\widehat\ell_j(a)
:=Z_j(a)-Z_{j-1}(a)
=f_j(a)+\widetilde g_j(a)$,
where
$\widetilde g_j(a)
=\frac{\mathbf1\{a_j=a\}g_j}
{p_j(a)+\beta}$.
Thus $\widehat\ell_j$ is the loss estimate
actually accumulated by the learner,
and $\widetilde g_j$ is its IX continuation
component.

\textbf{Step 1: Regret for the estimated losses.}
We first bound
$\sum_{j=1}^{\tilde c}
[\langle p_j,\widehat\ell_j\rangle
-\widehat\ell_j(a)]$
for any fixed comparison action $a$.
To analyze the exponential-weights update,
we consider the change in the potential
$\sum_b e^{-\eta Z_j(b)}$, whose logarithm
telescopes across rounds.

By $Z_j=Z_{j-1}+\widehat\ell_j$ and
the definition of $p_j$ \eqref{eq:app-pred-oracle-policy},
\begin{align*}
&\frac{\sum_b e^{-\eta Z_j(b)}}
{\sum_b e^{-\eta Z_{j-1}(b)}}\\
&=
\frac{\sum_b
e^{-\eta[Z_{j-1}(b)+M_j(b)]}}
{\sum_b e^{-\eta Z_{j-1}(b)}}
\sum_b p_j(b)
e^{-\eta[\widehat\ell_j(b)-M_j(b)]}.
\end{align*}
Taking logarithms, the first factor
satisfies
\begin{align*}
\log
\frac{\sum_b
e^{-\eta[Z_{j-1}(b)+M_j(b)]}}
{\sum_b e^{-\eta Z_{j-1}(b)}}
\le-\eta\langle p_j,M_j\rangle,
\end{align*}
by the nonnegativity of the KL divergence
between $p_j$ and the normalized weights
proportional to $e^{-\eta Z_{j-1}}$.

For the second factor, note that
$\widehat\ell_j(b)\ge0$ and
$M_j(b)\le F$. Hence
$\eta[\widehat\ell_j(b)-M_j(b)]
\ge-\eta F\ge-1$.
Using $e^{-x}\le1-x+x^2$ for $x\ge-1$
and $\log(1+x)\le x$, we obtain
\begin{align*}
&\log\sum_b p_j(b)
e^{-\eta[\widehat\ell_j(b)-M_j(b)]}\\
&\le
-\eta\langle p_j,\widehat\ell_j-M_j\rangle
+\eta^2
\langle p_j,(\widehat\ell_j-M_j)^2\rangle.
\end{align*}
Combining the two factors gives
\begin{align}\label{eq:pred-log-potential}
\log\frac{\sum_b e^{-\eta Z_j(b)}}
{\sum_b e^{-\eta Z_{j-1}(b)}} 
\le-\eta\langle p_j,\widehat\ell_j\rangle
+\eta^2
\langle p_j,(\widehat\ell_j-M_j)^2\rangle.
\end{align}

Summing \eqref{eq:pred-log-potential}
over $j=1,\ldots,\tilde c$ gives
\begin{align}
&\log\sum_b e^{-\eta Z_{\tilde c}(b)}
-\log\sum_b e^{-\eta Z_{0}(b)} \notag \\
&= \log\sum_b e^{-\eta Z_{\tilde c}(b)}
-\log A \notag \\ 
&\le
-\eta\sum_{j=1}^{\tilde c}
\langle p_j,\widehat\ell_j\rangle +\eta^2\sum_{j=1}^{\tilde c}
\langle p_j,(\widehat\ell_j-M_j)^2\rangle, \label{eq1}
\end{align}
where $Z_0=0$.

Moreover, since the sum contains the
term corresponding to action $a$,
\begin{align}\label{eq2}
\log\sum_b e^{-\eta Z_{\tilde c}(b)}
\ge-\eta Z_{\tilde c}(a)
=-\eta\sum_{j=1}^{\tilde c}\widehat\ell_j(a).
\end{align}
Combining inequalities \eqref{eq1}--\eqref{eq2} and
rearranging yields
\begin{align}\label{eq:pred-optimistic}
&\sum_{j=1}^{\tilde c}
[\langle p_j,\widehat\ell_j\rangle
-\widehat\ell_j(a)] \le\frac{\log A}{\eta}
+\eta\sum_{j=1}^{\tilde c}
\langle p_j,(\widehat\ell_j-M_j)^2\rangle.
\end{align}

Next, using
$\widehat\ell_j=f_j+\widetilde g_j$
and $(x+y)^2\le2x^2+2y^2$,
\begin{align*}
&\langle p_j,(\widehat\ell_j-M_j)^2\rangle\\
&\le2\langle p_j,(f_j-M_j)^2\rangle
+2\langle p_j,\widetilde g_j^2\rangle\\
&\le2\|f_j-M_j\|_\infty^2
+2\langle p_j,\widetilde g_j^2\rangle,
\end{align*}
where the last inequality uses
$\sum_a p_j(a)=1$.

Furthermore, only the selected action
can have a nonzero IX estimate. Thus,
\begin{align*}
\langle p_j,\widetilde g_j^2\rangle
&=\frac{p_j(a_j)g_j^2}
{[p_j(a_j)+\beta]^2}\\
&\le G\frac{g_j}{p_j(a_j)+\beta}
=G\sum_b\widetilde g_j(b),
\end{align*}
where we used $g_j\le G$ and
$p_j(a_j)/(p_j(a_j)+\beta)\le1$.

Substituting these bounds into
\eqref{eq:pred-optimistic}, we obtain
\begin{align}
&\sum_{j=1}^{\tilde c}
[\langle p_j,\widehat\ell_j\rangle
-\widehat\ell_j(a)]\notag\\
&\quad\le\frac{\log A}{\eta}
+2\eta E_{\tilde c}
+2\eta G\sum_{j=1}^{\tilde c}
\sum_b\widetilde g_j(b).
\label{eq:pred-deterministic}
\end{align}

\textbf{Step 2: Control of IX estimation errors.}
The bound in \eqref{eq:pred-deterministic}
still contains the random IX estimates.
We now control their cumulative size
and the estimation error for each comparison action.

Let $\bar g_j(a)
=\mathbb E[g_j\mid
\mathcal F_{j-1},a_j=a]$
denote the expected continuation loss
when action $a$ is selected.

We use the implicit-exploration
concentration argument
\cite{neu2015explore}.
For $p>0$, $g\in[0,G]$, and any
$w\ge0$ satisfying $wG\le2\beta$,
\[
\frac{wg}{p+\beta}
\le\frac{wg/p}{1+wg/(2p)}
\le\log(1+wg/p).
\]
Consequently, exponentiating gives
\[
e^{wg/(p+\beta)}
\le1+\frac{wg}{p}.
\]

Applying this inequality to the
selected action and conditioning on
$\mathcal F_{j-1}$ yields
\begin{align}\label{eq3}
&\mathbb E\left[
e^{w\sum_b\widetilde g_j(b)}
\mid\mathcal F_{j-1}\right] \notag \\
&=\sum_b p_j(b)
\mathbb E\left[
e^{wg_j/[p_j(b)+\beta]}
\mid\mathcal F_{j-1},a_j=b
\right] \notag \\
&\le
\sum_b p_j(b)
\left[1+\frac{w\bar g_j(b)}{p_j(b)}\right] \notag \\
&=1+w\sum_b\bar g_j(b) \notag\\
&\le e^{w\sum_b\bar g_j(b)}.
\end{align}
Similarly, for each fixed action $a$,
\[
\mathbb E\left[
e^{w\widetilde g_j(a)}
\mid\mathcal F_{j-1}\right]
\le e^{w\bar g_j(a)}.
\]

Since $\bar g_j(b)$ is
$\mathcal F_{j-1}$-measurable, \eqref{eq3}
implies
\[
\mathbb E \left[
\exp\left(
w\sum_b[\widetilde g_j(b)-\bar g_j(b)]
\right)
\middle|\mathcal F_{j-1}
\right] \le1.
\]
By successive conditioning over the rounds,
\[
\mathbb E\left[
\exp\left(
w\sum_{j=1}^{\tilde c}\sum_b
[\widetilde g_j(b)-\bar g_j(b)]
\right)\right]\le1.
\]
The same argument, using the corresponding
one-round inequality for each fixed action $a$,
gives
\[
\mathbb E\left[
\exp\left(
w\sum_{j=1}^{\tilde c}
[\widetilde g_j(a)-\bar g_j(a)]
\right)\right]\le1.
\] 

Therefore, by Markov's inequality,
for any $\rho\in(0,1)$,
\[
\begin{aligned}
&\Pr\left(
w\sum_{j=1}^{\tilde c}\sum_b
[\widetilde g_j(b)-\bar g_j(b)]
>\log(1/\rho)
\right)\\
&\quad\le
\rho\,
\mathbb E\left[
e^{w\sum_{j,b}
[\widetilde g_j(b)-\bar g_j(b)]}
\right]\le\rho.
\end{aligned}
\]
The same probability bound holds for
the corresponding sum over a fixed
action $a$.

First, take $w=4\eta G$, which satisfies
$wG\le2\beta$ since $G\le1$ and
$\beta=2\eta G$.
Setting $\rho=\delta/4$ gives
\[
4\eta G\sum_{j,b}
[\widetilde g_j(b)-\bar g_j(b)]
\le\log(4/\delta)
\]
with probability at least $1-\delta/4$.
Since $\bar g_j(b)\le G$,
\begin{equation}\label{eq:pred-ix-total}
4\eta G\sum_{j,b}\widetilde g_j(b)
\le4\eta A\tilde c G^2
+\log(4/\delta).
\end{equation}

Next, take $w=4\eta$, for which
$wG=2\beta$.
Applying the fixed-action bound with
$\rho=\delta/(4A)$ and taking a union
bound over all actions gives
\begin{equation}\label{eq:pred-ix-comparator}
\sum_{j=1}^{\tilde c}
[\widetilde g_j(a)-\bar g_j(a)]
\le\frac{L_\delta}{4\eta},
\qquad\forall a,
\end{equation}
with probability at least $1-\delta/4$.

\textbf{Step 3: Regret for conditional expected losses.}
We finally transfer the estimated-loss
bound to the target regret in
\eqref{eq:app-pred-regret-prop}.

For each fixed comparison action $a$,
the target regret admits the decomposition
\begin{equation}\label{eq:decomposition}
\begin{aligned}
&\sum_{j=1}^{\tilde c}
[\langle p_j,\ell_j\rangle-\ell_j(a)]\\
&=\sum_{j=1}^{\tilde c}
[\langle p_j,f_j\rangle+g_j
-f_j(a)-\bar g_j(a)]  \\
&\quad+\sum_{j=1}^{\tilde c}
\Big[
\langle p_j,\ell_j-f_j\rangle-g_j
+f_j(a)
-\mathbb E[f_j(a)\mid\mathcal F_{j-1}]
\Big].
\end{aligned}
\end{equation}
The second sum is the sampling error
between the observed feedback and its
conditional expectation.

The conditional-mean independence condition
on the immediate-loss vector $f_j$ gives
$\ell_j(a)=
\mathbb E[f_j(a)\mid\mathcal F_{j-1}]
+\bar g_j(a)$.
Moreover,
$\mathbb E[g_j\mid\mathcal F_{j-1}]
=\sum_b p_j(b)\bar g_j(b)$.
It follows that
\begin{align*}
&\mathbb E\left[
\langle p_j,\ell_j-f_j\rangle-g_j
\mid\mathcal F_{j-1}\right]\\
&=
\sum_b p_j(b)\bar g_j(b)
-\sum_b p_j(b)\bar g_j(b)=0,
\end{align*}
and
\[
\mathbb E\left[
f_j(a)-\mathbb E[f_j(a)\mid
\mathcal F_{j-1}]
\mid\mathcal F_{j-1}\right]=0.
\]
For each fixed action $a$, define
\[
X_j(a)=
\langle p_j,\ell_j-f_j\rangle-g_j
+f_j(a)
-\mathbb E[f_j(a)\mid\mathcal F_{j-1}].
\]
By conditional-mean independence,
$\mathbb E[X_j(a)\mid\mathcal F_{j-1}]=0$.
Moreover, we can write
$X_j(a)=C_j(a)+f_j(a)-\langle p_j,f_j\rangle-g_j$,
where
$C_j(a)=\langle p_j,\ell_j\rangle
-\mathbb E[f_j(a)\mid\mathcal F_{j-1}]$
is $\mathcal F_{j-1}$-measurable.
Since
$f_j(a),\langle p_j,f_j\rangle\in[0,F]$
and $g_j\in[0,G]$, we have
$C_j(a)-F-G\le X_j(a)\le C_j(a)+F$.
Thus, the conditional range of $X_j(a)$
has width at most $2F+G$.

Applying the conditional Hoeffding inequality
and taking a union bound over all actions,
with probability at least $1-\delta/2$,
simultaneously for all $a$,
\begin{equation}\label{eq:pred-mean-concentration}
\sum_{j=1}^{\tilde c}X_j(a)
\le
\frac{2F+G}{2}
\sqrt{2\tilde c L_\delta}.
\end{equation}

Next, we bound the first sum in \eqref{eq:decomposition}. Recall the IX estimate \(\widetilde{g}_j(a) = \frac{\mathbf{1}\{a_j=a\} g_j}{p_j(a)+\beta}\), we have
\begin{align}\label{eq5}
\langle p_j,\widetilde g_j\rangle
&=\frac{p_j(a_j)g_j}
{p_j(a_j)+\beta} =g_j-\beta
\frac{g_j}{p_j(a_j)+\beta}\notag  \\
&=g_j-\beta\sum_b\widetilde g_j(b).
\end{align}
Substituting
$\widehat\ell_j=f_j+\widetilde g_j$
and \eqref{eq5} into the summand yields
\begin{align}\label{eq6}
&\langle p_j,f_j\rangle+g_j
-f_j(a)-\bar g_j(a) \notag \\
&=\langle p_j,\widehat\ell_j\rangle
-\widehat\ell_j(a)
+\beta\sum_b\widetilde g_j(b) +\widetilde g_j(a)-\bar g_j(a).
\end{align}

Using \eqref{eq6} to rewrite the first sum in
\eqref{eq:decomposition}, and applying
\eqref{eq:pred-mean-concentration} to the second sum,
we obtain, with probability at least $1-\delta/2$,
simultaneously for all comparison actions $a$,
\begin{align}
&\sum_{j=1}^{\tilde c}
[\langle p_j,\ell_j\rangle-\ell_j(a)]
\notag\\
&\le
\sum_{j=1}^{\tilde c}
[\langle p_j,\widehat\ell_j\rangle
-\widehat\ell_j(a)]
+\beta\sum_{j,b}\widetilde g_j(b)
\notag\\
&\quad+
\sum_{j=1}^{\tilde c}
[\widetilde g_j(a)-\bar g_j(a)]
+\frac{2F+G}{2} \sqrt{2\tilde c L_\delta}.
\label{eq:pred-path}
\end{align}

We now combine the bounds established in the three steps.
Substituting \eqref{eq:pred-deterministic}
into \eqref{eq:pred-path} and using
$\beta=2\eta G$ yields
\begin{align*}
&\sum_{j=1}^{\tilde c}
[\langle p_j,\ell_j\rangle-\ell_j(a)] \notag \\
&\le\frac{\log A}{\eta}
+2\eta E_{\tilde c}
+4\eta G\sum_{j,b}\widetilde g_j(b) \notag \\
&\quad+
\sum_{j=1}^{\tilde c}
[\widetilde g_j(a)-\bar g_j(a)]
+\frac{2F+G}{2} \sqrt{2\tilde c L_\delta}.
\end{align*}
Next, applying the two concentration bounds
\eqref{eq:pred-ix-total} and
\eqref{eq:pred-ix-comparator} from Step 2,
we obtain
\begin{align*}
&\sum_{j=1}^{\tilde c}
[\langle p_j,\ell_j\rangle-\ell_j(a)]\\
&\le
\frac{\log A}{\eta}
+2\eta E_{\tilde c}
+4\eta A\tilde c G^2
+L_\delta +\frac{L_\delta}{4\eta}
+\frac{2F+G}{2} \sqrt{2\tilde c L_\delta}.
\end{align*}
By the union bound over the three high-probability
events used in Steps 2 and 3, the above bound holds
with probability at least $1-\delta$,
simultaneously for all comparison actions $a$.
Taking the maximum over $a$ proves
\eqref{eq:app-pred-regret-prop}.

When $G=0$, we have $g_j=0$ and
$\widetilde g_j=0$.
The algorithm reduces to optimistic
full-information exponential weights,
and the same conclusion follows from
Step 1 and the immediate-loss
concentration argument in Step 3.

Finally, since $E_{\tilde c}\le\tilde c F^2$,
choosing $\eta=\Theta(\tilde c^{-1/2})$
yields an $O(\sqrt{\tilde c})$ regret upper bound
for fixed $A$, $F$, $G$, and $\delta$.
\hfill$\blacksquare$

The stage-wise individual gap bound proved above depends explicitly on
the cumulative prediction error $E_{\tilde c}$,
which varies with the choice of prediction rule
$M_j$.
The following result establishes a worst-case
guarantee, quantifies the improvement over
standard bandit Exp3-IX, and shows how further
improvements can be achieved through accurate
continuation prediction.

\begin{restatement}{Prediction-dependent bounds on the stage-wise individual gap}
Consider the stage-based learning framework
in Algorithm~\ref{alg:sbv}.
For an agent $i$ and a completed stage
at $(t,s)$ of length $\tilde c$, let
$A=|\mathcal A_i^t(s)|\ge2$ and
$H_t=T-t+1$.
For any $p\in(0,1)$, let $\iota$ be defined
in \eqref{eq:pred-scales}.
The following results hold.
\begin{enumerate}
\renewcommand{\labelenumi}{\roman{enumi}).}
\item \textbf{Worst-case guarantee.} For any bounded, history-based prediction
rule $M_j\in[0,1/H_t]^A$ used in the
aggregation-predictive oracle, let
$\Gamma_{\tilde c}^{M}$ denote the
corresponding stage-wise individual gap,
defined as in Lemma~\ref{lemma:bonus}.
Then, with probability at least $1-p/2$,
\begin{equation}\label{eq:app-pred-worst-gap}
\begin{aligned}
\Gamma_{\tilde c}^{M}
\le
\min\left\{
H_t,\,
U_{\tilde c}(E_{\tilde c}^{M};s^t)
\right\} \le
6H_t\sqrt{\frac{A\iota}{\tilde c}},
\end{aligned}
\end{equation}
where $E_{\tilde c}^{M} =\sum_{j=1}^{\tilde c} \|f_j-M_j\|_\infty^2$. In particular, Algorithm~\ref{alg:sbv} satisfies this bound with $E_{\tilde c}^{M}=\widetilde E_i^t(s)$ and $\Gamma_{\tilde c}^{M}=\Gamma_{\tilde c}$ in Lemma~\ref{lemma:bonus}.  

\item \textbf{Improvement over standard bandit Exp3-IX.}
Replacing the proposed oracle with standard
bandit Exp3-IX yields a larger stage-wise
upper bound. Specifically, for
Algorithm~\ref{alg:sbv}, when
$\tilde c>36A\iota$,
\begin{equation}\label{eq:app-aggregate-improvement}
\begin{aligned}
\Gamma_{\tilde c}
&\le U_{\tilde c}(\widetilde E_i^t(s);s^t)\\
&\le U_{\tilde c}^{\mathrm{band}}
-\left(2\sqrt A-\frac3{\sqrt2}\right)
\sqrt{\frac{\iota}{\tilde c}}\\
&\quad-\frac{2H_t\eta}{\tilde c}
\left(\frac{\tilde c}{H_t^2}
-\widetilde E_i^t(s)\right)\\
&<U_{\tilde c}^{\mathrm{band}}
<6H_t\sqrt{\frac{A\iota}{\tilde c}},
\end{aligned}
\end{equation}
where $U_{\tilde c}^{\mathrm{band}}$
denotes the corresponding stage-wise upper bound.

\item \textbf{Improvement with aggregate-informed continuation prediction.} Suppose there exists a history-based
predictor
$\widehat v_j(s,a,d)\in[0,H_t-1]$
such that
$\big| \overline V_{t+1,i}^{\tilde l_j}(s') -\widehat v_j(s,a,\sigma(a_{-i})) \big| \le H_t\rho$, 
for all $a\in\mathcal A_i^t(s)$, $a_{-i}\in\mathcal A_{-i}^t(s)$, and $s'\in\mathcal S^{t+1}$, where $\rho\in[0,(H_t-1)/H_t]$ is known. 
Using the residual-feedback variant of
the aggregation-predictive Exp3-IX oracle
constructed in the proof below, the resulting
stage-wise individual gap
$\Gamma_{\tilde c}^{\mathrm{res}}$
satisfies, with probability at least
$1-p/2$, simultaneously over all completed
stages with $\tilde c\ge\iota$,
\begin{equation}\label{eq:app-residual-factor-three}
\Gamma_{\tilde c}^{\mathrm{res}}
\le
3H_t\sqrt{\frac{A\iota}{\tilde c}},
\qquad
\text{if } A\ge13,\ \rho\le0.1.
\end{equation}
Furthermore, for any $A\ge2$, if
$\rho\le c_0/\sqrt A$ for a constant
$c_0$ independent of $A$, then
\begin{equation}\label{eq:app-residual-action-rate}
\Gamma_{\tilde c}^{\mathrm{res}}
=
O\left(
H_t\sqrt{\frac{\iota}{\tilde c}}
\right).
\end{equation}
\end{enumerate}
\end{restatement}

\emph{Proof:}
The proof proceeds in three parts.

\noindent
i). The proof of the stage-wise individual gap above
applies to any bounded, history-based
prediction rule $M_j$, since the regret
proposition above does not require a
specific prediction rule.
Thus, with probability at least $1-p/2$,
\[
\Gamma_{\tilde c}^{M}
\le
\min\left\{
H_t,\,
U_{\tilde c}(E_{\tilde c}^{M};s^t)
\right\}.
\]
Since $f_j,M_j\in[0,1/H_t]^A$, we have
\[
E_{\tilde c}^{M}
\le\frac{\tilde c}{H_t^2}.
\]

Let $D_A:=2+4A(H_t-1)^2$.
Recall that
\[
\eta
=
\min\left\{
H_t,\,
H_t\sqrt{\frac{C_A}{\tilde cD_A}}
\right\},
\qquad
C_A=\log A+\frac{\iota}{4}.
\]

If $\tilde c\le36A\iota$, then
\[
\min\left\{
H_t,\,
U_{\tilde c}(E_{\tilde c}^{M};s^t)
\right\}
\le H_t
\le6H_t\sqrt{\frac{A\iota}{\tilde c}}.
\]

Otherwise, $\tilde c>36A\iota$.
Since $C_A\le5\iota/4$ and $D_A\ge2$,
the learning-rate cap is inactive, giving
\[
\eta=H_t\sqrt{\frac{C_A}{\tilde cD_A}}.
\]
As $U_{\tilde c}(x;s^t)$ is increasing in $x$,
we obtain
\begin{align}
U_{\tilde c}(E_{\tilde c}^{M};s^t)
&\le
U_{\tilde c}\left(
\frac{\tilde c}{H_t^2};s^t\right)
\notag\\
&=
2\sqrt{\frac{C_AD_A}{\tilde c}}
+\frac{H_t\iota}{\tilde c}
+\left(1+\frac1{2H_t}\right)
H_t\sqrt{\frac{2\iota}{\tilde c}}.
\label{eq:pred-worst-case-substitution}
\end{align}

Using $D_A\le4AH_t^2$,
$H_t\ge1$, and $\tilde c>36A\iota$,
we have
\begin{align}
\frac{
U_{\tilde c}(\tilde c/H_t^2;s^t)
}{
H_t\sqrt{A\iota/\tilde c}
}
&\le
4\sqrt{\frac{C_A}{\iota}}
+\frac3{\sqrt{2A}}
+\sqrt{\frac{\iota}{A\tilde c}}
\notag\\
&\le
4\sqrt{\frac{C_A}{\iota}}
+\frac3{\sqrt{2A}}
+\frac1{6A}.
\label{eq:pred-constant-bound}
\end{align}

Since $\iota\ge\log(16A)$, we have
\[
\frac{C_A}{\iota}
=
\frac14+\frac{\log A}{\iota}
\le
\begin{cases}
0.45, & A=2,\\
1.25, & A\ge3.
\end{cases}
\]
Consequently, the right-hand side of
\eqref{eq:pred-constant-bound}
is bounded above by
\[
\begin{cases}
4\sqrt{0.45}+\dfrac32+\dfrac1{12}<6,
& A=2,\\[1ex]
4\sqrt{1.25}+\dfrac3{\sqrt6}+\dfrac1{18}<6,
& A\ge3.
\end{cases}
\]
This proves \eqref{eq:app-pred-worst-gap}.

\medskip
\noindent
ii). Standard bandit Exp3-IX observes only
the selected action's normalized loss,
corresponding to $F=0$, $G=1$, and $M_j=0$
in the regret proposition above.
Under the same regret and concentration
analysis, its stage-wise upper bound
before truncation at $H_t$ is
\[
U_{\tilde c}^{\mathrm{band}}
=
\frac{H_t}{\tilde c}
\left[
\frac{C_A}{\eta_{\mathrm{band}}}
+4\eta_{\mathrm{band}}A\tilde c
+\iota+\sqrt{2\tilde c\iota}
\right],
\]
where
\[
\eta_{\mathrm{band}}
=
\min\left\{
1,\sqrt{\frac{C_A}{4A\tilde c}}
\right\}.
\]

For $\tilde c>36A\iota$, the learning-rate
caps of both oracles are inactive.
Thus, by \eqref{eq:pred-worst-case-substitution},
\[
\begin{aligned}
U_{\tilde c}
\left(\frac{\tilde c}{H_t^2};s^t\right)
={}&
2\sqrt{\frac{C_AD_A}{\tilde c}}
+\frac{H_t\iota}{\tilde c}\\
&+\left(1+\frac1{2H_t}\right)
H_t\sqrt{\frac{2\iota}{\tilde c}},\\
U_{\tilde c}^{\mathrm{band}}
={}&
4H_t\sqrt{\frac{AC_A}{\tilde c}}
+\frac{H_t\iota}{\tilde c}
+H_t\sqrt{\frac{2\iota}{\tilde c}}.
\end{aligned}
\]

Using
$\sqrt{D_A}\le2(H_t-1)\sqrt A+\sqrt2$
and $C_A\ge\iota/4$, we obtain
\[
\begin{aligned}
&U_{\tilde c}^{\mathrm{band}}
-U_{\tilde c}
\left(\frac{\tilde c}{H_t^2};s^t\right)\\
&=
2\sqrt{\frac{C_A}{\tilde c}}
\left(2H_t\sqrt A-\sqrt{D_A}\right)
-\sqrt{\frac{\iota}{2\tilde c}}\\
&\ge
\left(2\sqrt A-\frac3{\sqrt2}\right)
\sqrt{\frac{\iota}{\tilde c}}
>0,
\end{aligned}
\]
where the last inequality follows from
$2\sqrt A-3/\sqrt2\ge1/\sqrt2$ for $A\ge2$.

Moreover, since $C_A\le5\iota/4$
and $\tilde c>36A\iota$,
\[
\begin{aligned}
\frac{U_{\tilde c}^{\mathrm{band}}}
{H_t\sqrt{A\iota/\tilde c}}
&=
4\sqrt{\frac{C_A}{\iota}}
+\sqrt{\frac2A}
+\sqrt{\frac{\iota}{A\tilde c}}\\
&<
4\sqrt{\frac54}+1+\frac1{12}
<6.
\end{aligned}
\]

Finally, Algorithm~\ref{alg:sbv} satisfies
$\widetilde E_i^t(s)\le\tilde c/H_t^2$.
By the definition of $U_{\tilde c}$,
\[
\begin{aligned}
&U_{\tilde c}
\left(\frac{\tilde c}{H_t^2};s^t\right)
-U_{\tilde c}(\widetilde E_i^t(s);s^t)\\
&\qquad=
\frac{2H_t\eta}{\tilde c}
\left(
\frac{\tilde c}{H_t^2}
-\widetilde E_i^t(s)
\right)
\ge0.
\end{aligned}
\]
Combining these inequalities with
Lemma~\ref{lemma:bonus} gives
\[
\begin{aligned}
\Gamma_{\tilde c}
&\le U_{\tilde c}(\widetilde E_i^t(s);s^t)\\
&\le U_{\tilde c}^{\mathrm{band}}
-\left(2\sqrt A-\frac3{\sqrt2}\right)
\sqrt{\frac{\iota}{\tilde c}}\\
&\quad-\frac{2H_t\eta}{\tilde c}
\left(
\frac{\tilde c}{H_t^2}
-\widetilde E_i^t(s)
\right)\\
&<U_{\tilde c}^{\mathrm{band}}
<6H_t\sqrt{\frac{A\iota}{\tilde c}}.
\end{aligned}
\]
This proves \eqref{eq:app-aggregate-improvement}
on the high-probability event of part (i)
for Algorithm~\ref{alg:sbv}.

\medskip
\noindent
iii). We construct a residual-feedback
oracle and establish its stage-wise
individual-gap bound.

For a fixed stage at $(t,s)$, the immediate
and continuation losses remain as defined
in Algorithm~\ref{alg:sbv}.
Define the predicted continuation loss
and normalization constant by
\[
q_j(a,d):=
\frac{H_t-1-\widehat v_j(s,a,d)}{H_t},
\qquad
C_\rho:=1+2\rho.
\]

Using the observed opponents' aggregate
$d_j$ and its history-based empirical
distribution $\widehat\mu_j$,
we replace the original prediction
and feedback by
\begin{align*}
M_j^{\mathrm{res}}(a)
&=
\frac1{C_\rho}
\mathbb E_{d\sim\widehat\mu_j}
\left[
f_j(s,a,d)+q_j(a,d)
\right],\\
h_j(a)
&=
\frac{
f_j(s,a,d_j)+q_j(a,d_j)
}{C_\rho},\\
z_j
&=
\frac{
g_j-q_j(a_j,d_j)+\rho
}{C_\rho}.
\end{align*}

The policy and cumulative-loss updates
retain the form of
\eqref{eq:app-pred-oracle-policy}--
\eqref{eq:app-pred-oracle-loss},
with $(M_j,f_j,g_j)$ replaced by
$(M_j^{\mathrm{res}},h_j,z_j)$,
respectively, and with
\begin{align*}
\eta_{\tilde c}^{\mathrm{res}}
&=
C_\rho
\min\left\{
1,\,
\sqrt{
\frac{C_A}{
\tilde c(2+16A\rho^2)}
}
\right\},\\
\beta_{\tilde c}^{\mathrm{res}}
&=
\frac{4\eta_{\tilde c}^{\mathrm{res}}\rho}
{C_\rho}.
\end{align*}
As in Algorithm~\ref{alg:sbv}, these
parameters are fixed at stage initiation
using the planned stage length.

For the stage-end update, replace the
original prediction-error accumulator
and bonus by
\begin{align}
\widetilde E_i^{t,\mathrm{res}}(s)
&=
\sum_{j=1}^{\tilde c}
\|h_j-M_j^{\mathrm{res}}\|_\infty^2,\\
b_{\tilde c}^{\mathrm{res}}
&=
\min\Bigg\{
H_t,\,
\frac{H_tC_\rho}{\tilde c}
\Bigg[
\frac{C_A}{\eta_{\tilde c}^{\mathrm{res}}}
+2\eta_{\tilde c}^{\mathrm{res}}
\widetilde E_i^{t,\mathrm{res}}(s)
\notag\\
&\qquad
+\frac{
16\eta_{\tilde c}^{\mathrm{res}}
A\tilde c\rho^2}{C_\rho^2}
+\iota
+\frac{3/2+\rho}{C_\rho}
\sqrt{2\tilde c\iota}
\Bigg]
\Bigg\}.
\label{eq:residual-stage-bonus}
\end{align}

By the residual-prediction condition,
$|g_j-q_j(a_j,d_j)|\le\rho$.
Moreover, since
$f_j(s,a,d)\in[0,1/H_t]$ and
$q_j(a,d)\in[0,(H_t-1)/H_t]$, we have
\[
z_j\in\left[0,\frac{2\rho}{C_\rho}\right],
\qquad
h_j,M_j^{\mathrm{res}}
\in[0,1/C_\rho]^A.
\]

Since $\widehat v_j$ and $\widehat\mu_j$
are history-based and the opponents'
aggregate is conditionally independent
of the current action draw,
$M_j^{\mathrm{res}}$ is predictable
and $h_j$ satisfies the conditional
mean-independence condition of the regret
proposition above.

By construction,
\[
h_j(a_j)+z_j
=
\frac{f_j(a_j)+g_j+\rho}{C_\rho}.
\]
Taking conditional expectations given
$\mathcal F_{j-1}$ and $a_j=a$ yields
\[
\ell_j^{\mathrm{res}}(a)
=
\frac{\ell_j(a)+\rho}{C_\rho}.
\]
Consequently, for the loss sequence and
policies generated by the residual oracle,
\[
\mathcal R_{\tilde c}
=
C_\rho\mathcal R_{\tilde c}^{\mathrm{res}}.
\]

We apply the regret proposition above with
\[
F=\frac1{C_\rho},
\qquad
G=\frac{2\rho}{C_\rho},
\qquad
\eta=\eta_{\tilde c}^{\mathrm{res}},
\qquad
\beta=\beta_{\tilde c}^{\mathrm{res}},
\]
and $\delta_0=p/(4NSTK)$.
These parameters satisfy
$F+G=1$, $\eta F\le1$, and $\beta=2\eta G$.
Since $A\le A_{\max}$, we have
$L_{\delta_0}\le\iota$ and
$\log A+L_{\delta_0}/4\le C_A$.
Thus, with probability at least $1-\delta_0$,
\begin{align}
\mathcal R_{\tilde c}^{\mathrm{res}}
\le{}&
\frac{C_A}{\eta_{\tilde c}^{\mathrm{res}}}
+2\eta_{\tilde c}^{\mathrm{res}}
\widetilde E_i^{t,\mathrm{res}}(s)
\notag\\
&+\frac{
16\eta_{\tilde c}^{\mathrm{res}}
A\tilde c\rho^2}{C_\rho^2}
+\iota
+\frac{1+\rho}{C_\rho}
\sqrt{2\tilde c\iota}.
\label{eq:residual-regret-bound}
\end{align}

Using the return sampling bound in
Lemma~\ref{lemma:bonus}, with probability
at least $1-\delta_0$, we obtain
\[
\begin{aligned}
\Gamma_{\tilde c}^{\mathrm{res}}
&\le
\frac{H_t}{\tilde c}
\mathcal R_{\tilde c}
+H_t\sqrt{\frac{\iota}{2\tilde c}}\\
&=
\frac{H_tC_\rho}{\tilde c}
\left[
\mathcal R_{\tilde c}^{\mathrm{res}}
+\frac1{2C_\rho}\sqrt{2\tilde c\iota}
\right].
\end{aligned}
\]
Combining this with
\eqref{eq:residual-regret-bound}
and the deterministic bound
$\Gamma_{\tilde c}^{\mathrm{res}}\le H_t$
gives
\[
\Gamma_{\tilde c}^{\mathrm{res}}
\le b_{\tilde c}^{\mathrm{res}}
\]
with probability at least $1-2\delta_0$.

It remains to derive the two stated bounds.
From the ranges of $h_j$ and
$M_j^{\mathrm{res}}$, we have
\[
\widetilde E_i^{t,\mathrm{res}}(s)
\le\frac{\tilde c}{C_\rho^2}.
\]

Let $D_\rho:=2+16A\rho^2$.
Since $\tilde c\ge\iota$,
$C_A\le5\iota/4$, and $D_\rho\ge2$,
the learning-rate cap is inactive, giving
\[
\eta_{\tilde c}^{\mathrm{res}}
=
C_\rho\sqrt{
\frac{C_A}{\tilde cD_\rho}
}.
\]

Substituting this and
$\widetilde E_i^{t,\mathrm{res}}(s)
\le\tilde c/C_\rho^2$
into \eqref{eq:residual-stage-bonus},
we obtain
\begin{align*}
\Gamma_{\tilde c}^{\mathrm{res}}
\le{}&
2\sqrt{\frac{C_AD_\rho}{\tilde c}}
+\frac{H_tC_\rho\iota}{\tilde c}\\
&+\left(\frac32+\rho\right)
H_t\sqrt{\frac{2\iota}{\tilde c}}.
\end{align*}
In the first term, the normalization
factor cancels with that in the learning
rate, while the original return scale
$H_t$ is retained.
Hence the bound is
\begin{align}
\Gamma_{\tilde c}^{\mathrm{res}}
\le H_t\sqrt{\frac{\iota}{\tilde c}}
\Bigg[
&\sqrt{5(2+16A\rho^2)}
+\left(\frac32+\rho\right)\sqrt2
\notag\\
&+(1+2\rho)
\sqrt{\frac{\iota}{\tilde c}}
\Bigg].
\label{eq:residual-intermediate-bound}
\end{align}

If $A\ge13$ and $\rho\le0.1$,
then $\sqrt{\iota/\tilde c}\le1$, and
\[
\begin{aligned}
\frac{\Gamma_{\tilde c}^{\mathrm{res}}}
{H_t\sqrt{A\iota/\tilde c}}
&\le
\sqrt{0.8+\frac{10}{A}}
+\frac{1.6\sqrt2+1.2}{\sqrt A}\\
&\le
\sqrt{0.8+\frac{10}{13}}
+\frac{1.6\sqrt2+1.2}{\sqrt{13}}
<3.
\end{aligned}
\]
This proves \eqref{eq:app-residual-factor-three}.

Furthermore, if $\rho\le c_0/\sqrt A$
for a constant $c_0$ independent of $A$,
the expression in square brackets in
\eqref{eq:residual-intermediate-bound}
is bounded by a constant independent of $A$.
Consequently, we obtain an
$O(H_t\sqrt{\iota/\tilde c})$ upper bound
on $\Gamma_{\tilde c}^{\mathrm{res}}$,
proving \eqref{eq:app-residual-action-rate}.

The probability bounds are applied
conditionally on stage initiation.
As in Lemma~\ref{lemma:bonus}, a union
bound over at most $NSTK$ initiated stages
gives a total failure probability of
at most $2NSTK\delta_0=p/2$.
\hfill$\blacksquare$

These results show that the aggregation-predictive Exp3-IX framework retains the worst-case stage-wise bound \(6H_t\sqrt{A\iota/\tilde c}\) for any bounded, history-based forecast, matching the stage-wise order of prior decentralized CCE learning algorithms~\cite{mao2022improving}. Algorithm~\ref{alg:sbv} equipped with empirical-aggregate forecasts inherits this guarantee. They further establish a strictly tighter stage-wise individual-gap bound for Algorithm~\ref{alg:sbv} compared with standard bandit Exp3-IX, theoretically justifying the use of aggregate feedback to enhance learning efficiency. Under an extra continuation-prediction condition, an aggregate-informed residual-feedback refinement can improve the bound to \(O(H_t\sqrt{\iota/\tilde c})\).

\begin{algorithm}[!b]
\caption{Construction of the Correlated Policy $\bar{\bm{\pi}}^{t,k}$}\label{alg:certifyhk}
\begin{algorithmic}[1]
    \STATE \textbf{Input:} The distribution trajectory $\{\pi_{i}^{t,k}:i\in\mathcal{N},t\in \mathcal{T},k\in[K]\}$ specified by Algorithm~\ref{alg:sbv}.
    
    \STATE \textbf{Initialize:} $k'\gets k$.
    
    \FOR{step $t'\gets t$ to $T$}
        \STATE Receive $s^{t'}$;
        \IF{$\tilde c^{t',k'}(s^{t'})=0$}
            \STATE Execute fixed uniform local policies for the remaining episode and stop this construction.
        \ENDIF
        \STATE Uniformly sample $j$ from $\{1,2,\dots, \tilde{c}^{t', k'}(s^{t'})\}$;
        \STATE Set $k' \gets \tilde{l}_{j}^{t',k'}$; \COMMENT{$\tilde{l}_{j}^{t', k'}$ is the index of the episode that this state was visited the $j$-th time (among the total $\tilde{c}^{t',k'}$ times) in the last stage}
        \STATE Take joint action $\bm{a}^{t'}\sim \times_{i=1}^N\pi_{i}^{t', k'}(\cdot \mid s^{t'})$;
    \ENDFOR
\end{algorithmic}
\end{algorithm}

We now turn to the global CCE guarantee. Based on the trajectory of the distributions $\{\pi_{i}^{t,k}:i\in\mathcal{N},t\in \mathcal{T},k\in[K]\}$ specified by Algorithm~\ref{alg:sbv}, we construct a correlated policy $\bar{\bm\pi}^{t,k}$ for each $(t,k)\in \mathcal{T}\times [K]$. Our construction of the correlated policies, largely inspired by the ``certified policies'' \cite{bai2020near} for learning in two-player zero-sum games, is formally presented  in Algorithm~\ref{alg:certifyhk}. We further define an output policy $\bar{\bm\pi}$: first uniformly sample an index $k$ from $[K]$, then execute the  policy $\bar{\bm\pi}^{1,k}$. A formal description of $\bar{\bm\pi}$ is given in Algorithm~\ref{alg:certify}. By construction of the correlated policies $\bar{\bm\pi}^{t,k}$, we know that for any $i \in \mathcal{N},$ episode $ k \in [K],$  step $t \in [T+1],$  and state $ s^t \in \mathcal{S}^t$, the corresponding value function can be written recursively as follows:
\[
V_{t,i}^{\bar{\bm\pi}^{t,k}}(s^t) = \frac{1}{\tilde{c}}\sum_{j=1}^{\tilde{c}} \mathbb{D}_{\bm\pi^{t,\tilde{l}_j}}\bigg(r_{i}^t+\mathbb{P}_{t} V_{t+1,i}^{\bar{\bm\pi}^{t+1,\tilde{l}_j}}\bigg)(s^t), 
\]
This recursion applies when the previous completed stage is nonempty. At $t=T+1$ the value is zero. In a first-stage case the certified policy uses the fixed fallback in Algorithm~\ref{alg:certifyhk}; its true value lies in $[0,T-t+1]$ and is not set to zero. We also immediately obtain that
\[
V_{1,i}^{\bar{\bm\pi}}(s^1) = \frac{1}{K}\sum_{k=1}^K V_{1,i}^{\bar{\bm\pi}^{1,k}}(s^1).
\]

For analytical purposes, we introduce two auxiliary notations $\underline{V}$ and  $\underline{\hat{V}}$ as lower confidence bounds of the value estimates. Specifically, for any $i \in \mathcal{N},$ episode $ k \in [K],$  step $t \in [T+1],$  and state $ s^t \in \mathcal{S}^t$, we define $\underline{V}_{t,i}^k(s^t) = \underline{\hat{V}}_{t,i}^k(s^t) = 0$ if $t = T+1$ or $k$ is in the first stage of the $(t,s^t)$ pair, and 
\begin{align*}
\underline{\hat{V}}_{t,i}^k(s^t) &=\frac{1}{\tilde{c}}\sum_{j=1}^{\tilde{c}} \left(r_{i}^t(s^t,\bm{a}^{t,\tilde{l}_j}) + \underline{V}_{t+1,i}^{\tilde{l}_j}(s^{t+1,\tilde{l}_j}) \right)- b_{\tilde{c}},\\ &\text{ and } \underline{V}_{t,i}^k(s^t) = \max\left\{\underline{\hat{V}}_{t,i}^k(s^t), 0\right\}. 
\end{align*}
Note that these notations are only for analysis and agents do not need to maintain them explicitly during learning. Further, recall that $V_{t,i}^{\star,\bar{\pi}_{-i}^{t,k}}(s^t)$ is agent $i$'s best response value against its opponents' policy $\bar{\pi}_{-i}^{t,k}$. Our next lemma shows that $\overline{V}_{t,i}^k(s^t)$ and $\underline{V}_{t,i}^k(s^t)$ are indeed valid upper and lower bounds of $V_{t,i}^{\star,\bar{\pi}_{-i}^{t,k} }(s^t)$ and $V_{t,i}^{\bar{\bm\pi}^{t,k}}(s^t)$, respectively. 

\begin{restatement}{Confidence bounds}
With probability at least $1-p$, simultaneously for all agents $i \in \mathcal{N}$, episodes $k \in [K]$, steps $t \in \mathcal{T}$, and states $s^t \in \mathcal{S}^t$,
\begin{align}
 V_{t,i}^{\star,\bar{\pi}_{-i}^{t,k}}(s^t) &\leq \overline{V}_{t,i}^k(s^t), \label{eq:app-upper-confidence}\\
 V_{t,i}^{\bar{\bm\pi}^{t,k}}(s^t) &\geq \underline{V}_{t,i}^k(s^t). \label{eq:app-lower-confidence}
\end{align}
\end{restatement}
\emph{Proof:}
Fix an agent $i\in \mathcal{N}$, episode$ k \in [K],$  step $t \in \mathcal{T},$  and state $ s^t \in \mathcal{S}^t$. The desired result clearly holds for any state $s^t$ that is in its first stage, due to our initialization of $\overline{V}_{t,i}^k(s^t)$ and  $\underline{V}_{t,i}^k(s^t)$ for this special case. In the following, we only need to focus on the case where $\overline{V}^k_{t,i}(s^t)$ and  $\underline{V}_{t,i}^k(s^t)$ have been updated at least once at the given state $s^t$ before the $k$-th episode. 

	We first prove the first inequality \eqref{eq:app-upper-confidence}. It suffices to show that $\hat{V}_{t,i}^k(s^t)\geq V_{t,i}^{\star,\bar{\pi}_{-i}^{t,k}}(s^t)$, since $\overline{V}_{t,i}^k(s^t)=\min\{\hat{V}_{t,i}^k(s^t), H_t\}$, and $V_{t,i}^{\star,\bar{\pi}_{-i}^{t,k}}(s^t)$ is always less than or equal to $H_t$. Our proof relies on induction on $k\in [K]$. The base case $k=1$ holds by initialization logic. For the inductive step, consider two cases for $t \in \mathcal{T}$ and $s^t \in \mathcal{S}^t$:
	
	\textbf{Case 1 (for \eqref{eq:app-upper-confidence}):} $\hat{V}_{t,i}(s^t)$ has just been updated in (the end of) episode $k-1$. In this case, by definition of stage-based updates:	
    	\begin{equation}
	\hat{V}^k_{t,i}(s^t) = \frac{1}{\tilde{c}}\sum_{j=1}^{\tilde{c}} \left( r_{i}^t(s^t,\bm{a}^{t,\tilde{l}_j}) + \overline{V}_{t+1,i}^{\tilde{l}_j}(s^{t+1,\tilde{l}_j}) \right) + b_{\tilde{c}} .\nonumber
	\end{equation}
	And by the definition of $V_t^{\star, \bar{\pi}_{-i}^{t,k}}(s^t)$, it holds with probability at least $1-\frac{p}{4NSKT}$ that
    \begin{align}
	V_{t,i}^{\star,\bar{\pi}_{-i}^{t,k}}(s^t)&\leq  \max_{\pi_{i}^t} \frac{1}{\tilde{c}}\sum_{j=1}^{\tilde{c}} \mathbb{D}_{\pi_{i}^t\times \pi_{-i}^{t,\tilde{l}_{j}}} \left( r_{i}^t + \mathbb{P}_tV_{t+1,i}^{\star, \bar{\pi}_{-i}^{t+1,\tilde{l}_{j}}}\right)  (s^t)\nonumber\\
	&\leq\max_{\pi_{i}^t} \frac{1}{\tilde{c}}\sum_{j=1}^{\tilde{c}} \mathbb{D}_{\pi_{i}^t\times \pi_{-i}^{t,\tilde{l}_{j}}} \left( r_{i}^t + \mathbb{P}_t\overline{V}_{t+1,i}^{\tilde{l}_j}\right)  (s^t)\nonumber\\
	&\leq  \frac{1}{\tilde{c}} \sum_{j=1}^{\tilde{c}} \left( r_{i}^t(s^t,\bm{a}^{t,\tilde{l}_j}) + \overline{V}_{t+1,i}^{\tilde{l}_j}(s^{t+1,\tilde{l}_j}) \right) + b_{\tilde c}\nonumber\\
	&=  \hat{V}^k_{t,i}(s^t),\nonumber
	\end{align}
	where the second step is by the induction hypothesis, the third step holds due to the stage-wise Bellman bound above, and the last step is by the definition of $b_{\tilde{c}}$. 
	
	\textbf{Case 2 (for \eqref{eq:app-upper-confidence}): } $\hat{V}_{t,i}(s^t)$ was not updated in (the end of) episode $k-1$.  Since we have excluded the case that $\hat{V}_{t,i}$ has never been updated, we are guaranteed that there exists an episode $j$ such that $\hat{V}_{t,i}(s^t)$ has been updated in the end of episode $j-1$ most recently. In this case, $\hat{V}_{t,i}^k(s^t) = \hat{V}_{t,i}^{k-1}(s^t) = \dots = \hat{V}_{t,i}^{j}(s^t) \geq V_{t,i}^{\star,\bar{\pi}_{-i}^{t,j}}(s^t)$, where the last step is by the induction hypothesis. Finally, by the construction of stage-based policies (Algorithm~\ref{alg:certify}), $\bar{\pi}_{-i}^{t,j}$ is constant within the same stage (i.e., unchanged for episodes in that stage), so $V_{t,i}^{\star,\bar{\pi}_{-i}^{t,j}}(s^t)$ is also constant for all episodes $j$ in the same stage. Since we know that episode $j$ and episode $k$ lie in the same stage, we can conclude that $V_{t,i}^{\star,\bar{\pi}_{-i}^{t,k}}(s^t) = V_{t,i}^{\star,\bar{\pi}_{-i}^{t,j}}(s^t) \leq \hat{V}_{t,i}^k(s^t)$. 
	
	Combining the two cases and applying a union bound over all agents $i \in \mathcal{N},$ episodes $k \in [K],$  steps $t \in \mathcal{T},$  and states $ s^t \in \mathcal{S}^t$, the first inequality holds with probability at least $1-\frac{p}{2}$. 
	
	Next, we prove the second inequality \eqref{eq:app-lower-confidence} in the statement of the lemma. Notice that it suffices to show $\underline{\hat{V}}_{t,i}^k(s^t)\leq V_{t,i}^{\bar{\bm\pi}^{t,k}}(s^t)$ because $\underline{V}_{t,i}^k(s^t) = \max\{\underline{\hat{V}}_{t,i}^k(s^t),0\}$. Our proof again relies on induction on $k\in[K]$. Similar to the proof of the first inequality, the claim apparently holds for $k=1$, and we consider the following two cases for each step $t\in \mathcal{T}$ and $s^t \in\mathcal{S}^t$.	
	
    \textbf{Case 1 (for \eqref{eq:app-lower-confidence}):} The value of $\underline{\hat{V}}_{t,i}(s^t)$ has just changed in (the end of) episode $k-1$. In this case, 
	\begin{equation}
	\underline{\hat{V}}^k_{t,i}(s^t) = \frac{1}{\tilde{c}}\sum_{j=1}^{\tilde{c}} \left( r_{i}^t(s^t,\bm{a}^{t,\tilde{l}_j}) + \underline{V}_{t+1,i}^{\tilde{l}_j}(s^{t+1,\tilde{l}_j}) \right)- b_{\tilde{c}} .\nonumber
	\end{equation}
	By the definition of $V_{t,i}^{\bar{\bm\pi}^{t,k}}(s^t)$, it holds with probability at least $1-\frac{p}{4NSKT}$ that
	\begin{align}
	V_{t,i}^{\bar{\bm\pi}^{t,k}}(s^t)=  &\frac{1}{\tilde{c}}\sum_{j=1}^{\tilde{c}} \mathbb{D}_{\bm{\pi}^{t,\tilde{l}_j}} \left( r_{i}^t+\mathbb{P}_{t} V_{t+1,i}^{\bar{\bm\pi}^{t+1, \tilde{l}_j}}\right)(s^t)\nonumber\\
	\geq&\frac{1}{\tilde{c}}\sum_{j=1}^{\tilde{c}} \mathbb{D}_{\bm{\pi}^{t,\tilde{l}_j}} \left( r_{i}^t+\mathbb{P}_{t} \underline{\hat{V}}_{t+1,i}^{\tilde{l}_j}\right)(s^t)\nonumber\\
	\geq  \frac{1}{\tilde{c}} \sum_{j=1}^{\tilde{c}}  \Big( r_{i}^t(s^t,\bm{a}^{t,\tilde{l}_j})& +\underline{V}_{t+1,i}^{\tilde{l}_j}(s^{t+1,\tilde{l}_j}) \Big) - \min\{H_t,H_t\sqrt{2\iota/\tilde c}\} \nonumber\\
	\geq & \underline{\hat{V}}^k_{t,i}(s^t),\nonumber
	\end{align}
	where the second step is by the induction hypothesis, the third step holds due to the Azuma-Hoeffding inequality, and the last step is by the definition of $b_{\tilde{c}}$. 
	
	\textbf{Case 2 (for \eqref{eq:app-lower-confidence}): } The value of $\underline{\hat{V}}_{t,i}(s^t)$ has not changed in (the end of) episode $k-1$.  Since we have excluded the case that $\underline{\hat{V}}_{t,i}$ has never been updated, we are guaranteed that there exists an episode $j$ such that $\underline{\hat{V}}_{t,i}(s^t)$ has changed in the end of episode $j-1$ most recently. In this case, we know that indices $j$ and $k$ belong to the same stage, and $\underline{\hat{V}}_{t,i}^k(s^t) = \underline{\hat{V}}_{t,i}^{k-1}(s^t) = \dots = \underline{\hat{V}}_{t,i}^{j}(s^t) \leq V_{t,i}^{\bar{\bm\pi}^{t,j}}(s^t)$, where the last step is by the induction hypothesis. Finally, by stage-based policy construction (Algorithm~\ref{alg:certify}), $\bar{\bm\pi}^{t,j}$ is constant within the same stage, so $V_{t,i}^{\bar{\bm\pi}^{t,j}}(s^t)$ is constant for all episodes $j$ in that stage. Since $j$ and $k$ lie in the same stage, we can conclude  that $V_{t,i}^{\bar{\bm\pi}^{t,k}}(s^t)= V_{t,i}^{\bar{\bm\pi}^{t,j}}(s^t) \geq \underline{\hat{V}}_{t,i}^k(s^t)$. 
	
	Again, combining the two cases and applying a union bound over all agents
$i \in \mathcal{N}$, episodes $k \in [K]$, steps $t \in \mathcal{T}$, and states
$s^t \in \mathcal{S}^t$ , the second inequality holds with probability at least $1-\frac{p}{2}$.

By the union bound over both inequalities, the lemma holds with probability at least $1-p$. 
\hfill \ensuremath{\blacksquare}

We conclude by restating the global sample-complexity result and proving it from the confidence bounds and stage-wise guarantees above.

\begin{restatement}{Sample complexity of learning CCE}
For any $\epsilon\in(0,T]$ and $p\in(0,1)$, let $\iota$ be defined in \eqref{eq:main-pred-parameters}. There exists an absolute constant $C>0$ such that, if the agents run Algorithm~\ref{alg:sbv} for $K\ge C\frac{SA_{\max}T^5\iota}{\epsilon^2}$ episodes, then, with probability at least $1-p$, the output policy $\bar{\bm\pi}$ of Algorithm~\ref{alg:certify} is an $\epsilon$-approximate CCE. 
\end{restatement}

\emph{Proof:}
We first recall the definitions of several notations and define a few new ones. In the sums below, terms involving an empty predecessor stage are defined as zero, including its average and bonus; the initialized interval is accounted for separately. For a state $s^{t,k}$, recall that $\tilde{c}^{t,k}$ denotes the number of visits to the state $s^{t,k}$ (at the $t$-th step)  in the stage right before the current stage, and $\tilde{l}_{j}^{t,k}$ denotes the $j$-th episode among the $\tilde{c}^{t,k}$ episodes. When $t$ and $k$ are clear from context, we abbreviate $\tilde{l}_{j}^{t,k}$ as $\tilde{l}_j$ and $\tilde{c}^{t,k}$ as $\tilde{c}$.
    
    On the event of the confidence bounds above and by the construction of the output policy $\bar{\bm\pi}$ in Algorithm~\ref{alg:certify}, we know that
	\[
	\begin{aligned}
	V_{1,i}^{\star,\bar{\pi}_{-i}}(s^1) - V_{1,i}^{\bar{\bm\pi}}(s^1) \leq & \frac{1}{K}\sum_{k=1}^K  \left( V_{1,i}^{\star,\bar{\pi}_{-i}^{1,k}}(s^1) - V_{1,i}^{\bar{\bm\pi}^{1,k}} (s^1) \right)\\
	\leq & \frac{1}{K}\sum_{k=1}^K  \left( \overline{V}_{1,i}^{k}(s^1) - \underline{V}_{1,i}^{k}(s^1)\right). 
	\end{aligned}
	\]
	Thus, it suffices to upper bound $\frac{1}{K}\sum_{k=1}^K ( \overline{V}_{1,i}^{k}(s^1) - \underline{V}_{1,i}^{k}(s^1))$. For a fixed agent $i\in\mathcal{N}$, we define the following notation:
	$$
	\delta^{t,k} := \overline{V}_{t,i}^{k}(s^{t,k})- \underline{V}_{t,i}^{k}(s^{t,k}).
	$$ 
	The key idea of the subsequent proof is to upper bound $\sum_{k=1}^K \delta^{t,k}$ by the next step $\sum_{k=1}^K \delta^{t+1,k}$, and then obtain a recursive formula. From the update rule of $\overline{V}_{t,i}^k(s^{t,k})$ in \eqref{eqn:update_rule_new}, we have:
    \begin{align*}
	\overline{V}_{t,i}^k(s^{t,k}) \leq & \mathbb{I}[c^{t,k} = 0]T  \\ & + \frac{1}{\tilde{c}}\sum_{j=1}^{\tilde{c}}  \left( r_{i}^t(s^t,\bm{a}^{t,\tilde{l}_j}) + \overline{V}_{t+1,i}^{\tilde{l}_j}(s^{t+1,\tilde{l}_j}) \right)+ b_{\tilde{c}},
	\end{align*}
	where the $\mathbb{I}[c^{t,k}=0]$ term counts for the event that the optimistic value function has never been updated for the given state.
	
	Combining this with the definition of $\underline{V}_{t,i}^{k}(s^{t,k})$, we have
	\begin{align}
	\delta^{t,k} \leq & \mathbb{I}[c^{t,k} = 0]T \notag \\
    &+ \frac{1}{\tilde{c}}\sum_{j=1}^{\tilde{c}}  \left( \overline{V}_{t+1,i}^{\tilde{l}_j}(s^{t+1,\tilde{l}_j}) - \underline{V}_{t+1,i}^{\tilde{l}_j}(s^{t+1,\tilde{l}_j}) \right)+2 b_{\tilde{c}}\nonumber\\
	=&
\underbrace{\mathbb{I}[\tilde c^{t,k}=0]T}
_{\text{Term 1}}
+
\underbrace{\frac{1}{\tilde c}
\sum_{j=1}^{\tilde c}\delta^{t+1,\tilde l_j}}
_{\text{Term 2}}
+
\underbrace{2b_{\tilde c}}
_{\text{Term 3}}. \label{eqn:delta}
	\end{align}
	To find an upper bound of $\sum_{k=1}^K \delta^{t,k}$, we proceed to upper bound each term on the RHS of \eqref{eqn:delta} separately:
    
\textbf{Term 1:} $\sum_{k=1}^K \mathbb{I}[c^{t,k}=0]T$.
For a fixed step $t$, each state contributes at most $T$ visits before its first stage completes. Thus there are at most $ST$ such visits, and $\sum_{k=1}^K\mathbb I[c^{t,k}=0]T\le ST^2$.

\textbf{Term 2:} $\sum_{k=1}^K \frac{1}{\tilde{c}^{t,k}}\sum_{j=1}^{\tilde{c}^{t,k}} \delta^{t+1,\tilde{l}_{j}^{t,k}}$.
Switching the order of summation, we rewrite this term as:
	\begin{align}
	\sum_{k=1}^K \frac{1}{\tilde{c}^{t,k}}\sum_{j=1}^{\tilde{c}^{t,k}} \delta^{t+1,\tilde{l}_{j}^{t,k}} =& \sum_{k=1}^K \sum_{k'=1}^K \frac{1}{\tilde{c}^{t,k}} \delta^{t+1,k'} \sum_{j=1}^{\tilde{c}^{t,k}} \mathbb{I}\left[\tilde{l}_{j}^{t,k} = k'\right]\nonumber\\
	=&\sum_{k'=1}^K  \delta^{t+1,k'} \sum_{k=1}^K  \frac{1}{\tilde{c}^{t,k}} \sum_{j=1}^{\tilde{c}^{t,k}} \mathbb{I}\left[\tilde{l}_{j}^{t,k} = k'\right].\label{eqn:d1}
	\end{align}
	For a fixed episode $k'$, notice that $\sum_{j=1}^{\tilde{c}^{t,k}} \mathbb{I}[\tilde{l}_{j}^{t,k} = k']\leq 1$, and that $\sum_{j=1}^{\tilde{c}^{t,k}} \mathbb{I}[\tilde{l}_{j}^{t,k} = k']= 1$ happens if and only if $s^{t,k} = s^{t,k'}$ and $(t,k')$ lies in the previous stage of $(t,k)$ with respect to the step-state pair $(t,s^{t,k})$. 
    Define $\mathcal{K}_{k'}:= \{ k\in[K]: \sum_{j=1}^{\tilde{c}^{t,k}} \mathbb{I}[\tilde{l}_{j}^{t,k} = k']= 1 \}$; this set consists of all episodes $k$ for which $k'$ is a pre-stage episode of $(t,s^{t,k})$. Then we know that all episode indices $k\in \mathcal{K}_{k'}$ belong to the same stage, and hence these episodes have the same value of $\tilde{c}^{t,k}$. That is, there  exists an integer $N_{k'}>0$, such that $\tilde{c}^{t,k} = N_{k'},\forall k \in \mathcal{K}_{k'}$. Further, since the stages are partitioned in a way such that each stage is at most $(1+\frac{1}{T})$ times longer than the previous stage (Line~\ref{line:17} of Algorithm~\ref{alg:sbv}), we know that $|\mathcal{K}_{k'}|\leq (1+\frac{1}{T})N_{k'}$. Therefore, for every $k'$, it holds that
	\begin{equation}\label{eqn:d2}
	\sum_{k=1}^K  \frac{1}{\tilde{c}^{t,k}} \sum_{j=1}^{\tilde{c}^{t,k}} \mathbb{I}\left[\tilde{l}_{j}^{t,k} = k'\right] \leq 1+ \frac{1}{T}.
	\end{equation} 
	Combining \eqref{eqn:d1} and \eqref{eqn:d2} leads to the following upper bound of the second term in \eqref{eqn:delta}:	
	\begin{equation}
	\sum_{k=1}^K \frac{1}{\tilde{c}^{t,k}}\sum_{j=1}^{\tilde{c}^{t,k}} \delta^{t+1,\tilde{l}_{j}^{t,k}}   \leq (1+\frac{1}{T}) \sum_{k=1}^{K} \delta^{t+1,k}. \nonumber
	\end{equation}
	So far, we have obtained the following upper bound:
	\[
	\sum_{k=1}^K\delta^{t,k} \leq ST^2 + (1+\frac{1}{T})  \sum_{k=1}^K \delta^{t+1,k} + 2\sum_{k=1}^K b_{\tilde{c}^{t,k}}. 
	\]
	
	Iterating the above inequality over $t = T, T-1, \dots, 1$ leads to
	\begin{align*}
	\sum_{k=1}^{K}\delta^{1,k} &\leq \left( \sum_{t=1}^T (1+\frac{1}{T})^{t-1} \right) ST^2+ 2\sum_{t=1}^T \sum_{k=1}^K (1+\frac{1}{T})^{t-1}b_{\tilde{c}^{t,k}} \notag \\ &=  \left((1+\frac{1}{T})^T-1 \right) ST^3 + 2\sum_{t=1}^T \sum_{k=1}^K (1+\frac{1}{T})^{t-1}b_{\tilde{c}^{t,k}}  \notag \\&\leq O\left( ST^3 + \sum_{t=1}^T \sum_{k=1}^K (1+\frac{1}{T})^{t-1}b_{\tilde{c}^{t,k}} \right),
	\end{align*}
	where we used the fact that $(1+\frac{1}{T})^T \leq e$. 

    \textbf{Term 3: $b_{\tilde c^{t,k}}$.}
By the worst-case part of the stage-wise bound above,
$b_{\tilde c}\le6T\sqrt{A_i\iota/\tilde c}$.
For a fixed $(t,s)$, let $n_t(s)$ be its total number of visits,
and let $e_1,\ldots,e_J$ be the completed predecessor-stage lengths
that are used by subsequent stages.
Write $w_j$ for the number of actual visits in the stage following
the predecessor of length $e_j$.
The final following stage can be truncated, so only the upper bound
$0\le w_j\le(1+1/T)e_j$ is used.
The adaptive update gives $e_{j+1}\le(1+1/T)e_j$ and
$\sum_{j=1}^J e_j\le n_t(s)$.
Let $S_j=\sum_{h=1}^j e_h$ and $S_0=0$.
By the stage-length update and
$\lambda_{\min}\ge (T+1/2)/(T+1)$, each successive planned stage
length satisfies
\begin{align*}
    & e_{j+1}\ge e_j+\max\left\{1,\left\lfloor\frac{e_j}{2T}\right\rfloor\right\}
 \ge\left(1+\frac{1}{4T}\right)e_j, \\
& e_{j+1}\le\left(1+\frac1T\right)e_j,
\end{align*}
where $e_1=T$ and we used
$\max\{1,\lfloor x\rfloor\}\ge x/2$ for $x\ge0$.
Consequently, every completed prefix obeys
\[
 S_j=\sum_{h=1}^j e_h
 \le e_j\sum_{r=0}^{j-1}\left(1+\frac{1}{4T}\right)^{-r}
 \le (1+4T)e_j\le5T e_j.
\]
Therefore,
\[
 \sqrt{S_j}-\sqrt{S_{j-1}}
 =\frac{e_j}{\sqrt{S_j}+\sqrt{S_{j-1}}}
 \ge\frac{\sqrt{e_j}}{2\sqrt{5T}}.
\]
Summing and using $w_j\le(1+1/T)e_j$ yields
\[
 \sum_j\frac{w_j}{\sqrt{e_j}}
 \le2(1+1/T)\sqrt{5Tn_t(s)}
 =O(\sqrt{Tn_t(s)}).
\]
Since $\sum_s n_t(s)=K$, Cauchy--Schwarz and
$(1+1/T)^{t-1}\le e$ give
\begin{align}
 &\sum_{t=1}^T\sum_{k=1}^K
 (1+1/T)^{t-1}b_{\tilde c^{t,k}}\notag\\
 &\quad\le O\left(
 T\sqrt{A_i\iota}\sum_{t=1}^T\sum_s\sqrt{Tn_t(s)}
 \right)
 \le O\left(\sqrt{SA_iKT^5\iota}\right).\nonumber
\end{align}
This uses only completed-stage bonuses and also covers a last
partially visited stage.
Summarizing the results above leads to
	\[
	\sum_{k=1}^K \delta_1^k \leq O\left( ST^3 + \sqrt{SA_i KT^5\iota} \right). 
	\]
	In the case when $K$ is large enough, such that $K \geq \frac{ST}{A_{\max}\iota}$, use $A_i\le A_{\max}$ to absorb the initialization term and obtain:	
	\[
	V_{1,i}^{\star,\bar{\pi}_{-i}}(s^1) - V_{1,i}^{\bar{\bm\pi}}(s^1) \leq \frac{1}{K}\sum_{k=1}^{K}\delta_1^k \leq O\left(\sqrt{SA_{\max} T^5\iota/K} \right) .
	\]
	The finite-sample bound above holds when $K\ge ST/(A_{\max}\iota)$. Under the sample-size condition in the main-text theorem and $\epsilon\le T$, this requirement is satisfied. Choosing the absolute constant in that condition sufficiently large makes the finite-sample gap at most $\epsilon$ for every agent. The confidence event is uniform over initial states, so the conclusion also holds for any fixed initial-state distribution. Thus the output policy is an $\epsilon$-approximate CCE with probability at least $1-p$.
\hfill$\blacksquare$

\section{Proofs for Section~\ref{sec:ce}}
\label{app:ce}

This appendix proves the no-swap-regret
guarantee, establishes the stage-wise CE
Bellman-error bound, and completes the
proof of Theorem~\ref{thm:ce}.
We use the notation and certified-policy
construction introduced in
Appendix~\ref{app:cce}.
Conditional on the pre-action joint
history, agents randomize independently,
and the CE learner's internal index is
independent of the opponents' current
actions.

\subsection{No-Swap-Regret Guarantee}

We first restate the regret guarantee
of Proposition~\ref{prop:ce-swap-regret}.

\begin{restatement}
{Regret guarantee for aggregation-predictive
no-swap learning}
Under the hybrid-feedback assumptions
of Proposition~\ref{prop:pred-regret},
suppose that every row learner follows
\eqref{eq:ce-row-loss}--
\eqref{eq:ce-row-policy}
with a common $\eta>0$,
$\eta F\le1$, and $\beta=2\eta G$.
For any $m\ge1$ and $\delta\in(0,1)$,
with probability at least $1-\delta$,
\begin{align*}
\mathcal R_m^{\rm swap}
\le{}&
\frac{A(\log A+L_\delta/4)}{\eta}
+2\eta E_m^{\rm CE}
+4\eta AmG^2\\
&+AL_\delta
+2\sqrt{2AmL_\delta}\\
&+\sqrt{
8m\left(
A\log A+\log(2/\delta)
\right)},
\end{align*}
where
\[
L_\delta=\log\frac{8A^2m}{\delta},
\qquad
E_m^{\rm CE}
=\sum_{j=1}^m
\|f_j-M_j(\cdot\mid J_j)\|_\infty^2.
\]
\end{restatement}
\emph{Proof:}
For each row $a\in\mathcal A$,
define its selected-round set and
number of updates by
\[
\mathcal I_a
=\{j\in[m]:J_j=a\},
\qquad
N_a=|\mathcal I_a|.
\]
The conditional-expected external
regret of row $a$ is
\begin{equation}
\begin{aligned}
\mathcal R_a^{\rm ext}
:={}
\sum_{j\in\mathcal I_a}
\langle q_j(\cdot\mid a),\ell_j\rangle-\min_{b\in\mathcal A}
\sum_{j\in\mathcal I_a}\ell_j(b), \nonumber
\end{aligned}
\end{equation}
with $\mathcal R_a^{\rm ext}=0$
when $N_a=0$.

On its selected subsequence,
row $a$ follows the hybrid-feedback
Exp3-IX update of
Proposition~\ref{prop:pred-regret}.
The conditional mean-independence
condition for the immediate-loss
vector is preserved on this
subsequence: conditional on the
pre-round history, $J_j$ is sampled
independently of the opponents'
current actions.

Apply Proposition~\ref{prop:pred-regret}
to every row and every possible
selected-subsequence prefix length
$n\in[m]$, using failure probability
$\delta/(2Am)$.
A union bound implies that,
with probability at least
$1-\delta/2$, simultaneously
for all rows,
\begin{equation}\label{eq:app-ce-row-bound}
\begin{aligned}
\mathcal R_a^{\rm ext}
\le{}&
\frac{\log A+L_\delta/4}{\eta}
+2\eta E_a\\
&+4\eta A N_aG^2
+L_\delta
+2\sqrt{2N_aL_\delta},
\end{aligned}
\end{equation}
where
\[
E_a
=\sum_{j\in\mathcal I_a}
\|f_j-M_j(\cdot\mid a)\|_\infty^2,
\qquad
L_\delta=\log\frac{8A^2m}{\delta}.
\]
The inequality is also valid
for $N_a=0$.

Since
\[
\sum_aN_a=m,
\qquad
\sum_aE_a=E_m^{\rm CE},
\qquad
\sum_a\sqrt{N_a}\le\sqrt{Am},
\]
summing \eqref{eq:app-ce-row-bound}
over all rows gives
\begin{equation}\label{eq:app-ce-sum-row}
\begin{aligned}
\sum_a\mathcal R_a^{\rm ext}
\le{}&
\frac{A(\log A+L_\delta/4)}{\eta}
+2\eta E_m^{\rm CE}\\
&+4\eta AmG^2
+AL_\delta
+2\sqrt{2AmL_\delta}.
\end{aligned}
\end{equation}

We next relate the row-wise
external regret to swap regret.
For a fixed mapping
$\phi:\mathcal A\to\mathcal A$,
define
\[
X_j^\phi
=
\langle q_j(\cdot\mid J_j),\ell_j\rangle
-\ell_j(\phi(J_j)).
\]
By the stationary relation
\eqref{eq:ce-stationary},
\begin{equation}\label{eq:app-ce-swap-mean}
\begin{aligned}
\mathbb E[X_j^\phi
\mid\mathcal F_{j-1}]
={}&
\langle p_j,\ell_j\rangle\\
&-\sum_{a\in\mathcal A}
p_j(a)\ell_j(\phi(a)).
\end{aligned}
\end{equation}
Because $\ell_j(a)\in[0,1]$,
we have $|X_j^\phi|\le1$.
Consequently, Azuma--Hoeffding
and a union bound over the
$A^A$ possible mappings yield,
with probability at least
$1-\delta/2$, simultaneously
for all $\phi$,
\begin{equation}\label{eq:app-ce-index-azuma}
\begin{aligned}
&\sum_{j=1}^m
\mathbb E[X_j^\phi
\mid\mathcal F_{j-1}]
-\sum_{j=1}^mX_j^\phi\\
&\qquad\le
\sqrt{
8m\left(
A\log A+\log\frac2\delta
\right)}.
\end{aligned}
\end{equation}

Furthermore, for every mapping
$\phi$,
\begin{align}
\sum_{j=1}^mX_j^\phi
={}&
\sum_{a\in\mathcal A}
\sum_{j\in\mathcal I_a}
\left[
\langle q_j(\cdot\mid a),\ell_j\rangle
-\ell_j(\phi(a))
\right]\notag\\
\le{}&
\sum_{a\in\mathcal A}
\mathcal R_a^{\rm ext}.
\label{eq:app-ce-swap-decomp}
\end{align}

Combining
\eqref{eq:app-ce-swap-mean}--
\eqref{eq:app-ce-swap-decomp}
and maximizing over $\phi$ gives
\[
\mathcal R_m^{\rm swap}
\le
\sum_a\mathcal R_a^{\rm ext}
+
\sqrt{
8m\left(
A\log A+\log\frac2\delta
\right)}.
\]
Substituting
\eqref{eq:app-ce-sum-row}
proves Proposition~
\ref{prop:ce-swap-regret}
with probability at least
$1-\delta$.
\hfill \ensuremath{\blacksquare}

\subsection{Stage-Wise Bellman Error}

Fix an agent $i$, a step $t$,
a state $s\in\mathcal S^t$,
and a completed stage with
$m=\tilde c>0$ visits.
Let $\tilde l_j$ denote the
episode index of its $j$th visit,
and let $p_j$ be the stationary
marginal policy used at that visit.
Let $A=|\mathcal A_i^t(s)|$.

Define the normalized expected
Bellman loss by
\begin{equation}\label{eq:app-ce-bellman-loss}
\ell_j(a)
=
1-\frac1{H_t}
\mathbb E_{a_{-i}\sim\pi_{-i}^{t,\tilde l_j}(\cdot\mid s)}
\Big[
(r_i^t+\mathbb P_t\overline V_{t+1,i}^{\tilde l_j})
(s,a,a_{-i})
\Big].
\end{equation}
In Algorithm~\ref{alg:sbv-CE},
$f_j(a)=[1-r_i^t(s,a,d_{-i}^{t,\tilde l_j})]/H_t$ and
$g_j=[H_t-1-\overline V_{t+1,i}^{\tilde l_j}
(s^{t+1,\tilde l_j})]/H_t$.
Thus, conditional mean independence of the aggregate feedback and
the current action draw gives
\[
\ell_j(a)
=\mathbb E[f_j(a)+g_j\mid\mathcal F_{j-1},a_j=a],
\]
so this is exactly the conditional expected loss used in
Proposition~\ref{prop:pred-regret}.

\begin{restatement}
{Stage-wise CE Bellman error}
With probability at least $1-p/2$,
simultaneously for all agents, steps, states, and
completed stages with $m>0$,
\begin{align*}
\Gamma_m^{\rm CE}
:={}&
\max_{\phi:\mathcal A_i^t(s)
\to\mathcal A_i^t(s)}
\frac1m\sum_{j=1}^m
\mathbb D_{
(\phi\diamond p_j)
\times\pi_{-i}^{t,\tilde l_j}}
\left(
r_i^t+
\mathbb P_t\overline V_{t+1,i}^{\tilde l_j}
\right)(s)\\
&-\frac1m\sum_{j=1}^m
\left[
r_i^t(s,\bm a^{t,\tilde l_j})
+
\overline V_{t+1,i}^{\tilde l_j}
(s^{t+1,\tilde l_j})
\right]\\
\le{}& b_m^{\rm CE},
\end{align*}
where $b_m^{\rm CE}$ is defined
in \eqref{eq:ce-pred-bonus}.
\end{restatement}

\emph{Proof:}
Define the observed Bellman sample
and its conditional expectation by
\begin{align*}
Y_j
&=
r_i^t(s,\bm a^{t,\tilde l_j})
+
\overline V_{t+1,i}^{\tilde l_j}
(s^{t+1,\tilde l_j}),\\
\mu_j
&=
\mathbb D_{
p_j\times\pi_{-i}^{t,\tilde l_j}}
\left(
r_i^t+
\mathbb P_t\overline V_{t+1,i}^{\tilde l_j}
\right)(s).
\end{align*}

By \eqref{eq:app-ce-bellman-loss},
the Bellman-error expression
decomposes exactly as
\begin{equation}\label{eq:app-ce-bellman-decomp}
\Gamma_m^{\rm CE}
=
\frac{H_t}{m}\mathcal R_m^{\rm swap}
+
\frac1m\sum_{j=1}^m(\mu_j-Y_j).
\end{equation}

We first bound the swap-regret term.
For each initiated stage, allocate
failure probability
\[
\delta_0=\frac{p}{4NSTK}.
\]
Since $A\le A_{\max}$ and
$m\le K$, the logarithmic factors
in Proposition~\ref{prop:ce-swap-regret}
satisfy
\begin{align*}
L_{\delta_0}
&=
\log\frac{8A^2m}{\delta_0}\\
&\le
\log\frac{
32NSA_{\max}^2TK^2}{p}
\le\iota_{\rm CE},
\\
\log\frac2{\delta_0}
&\le\iota_{\rm CE}.
\end{align*}
Hence, for $A\ge2$,
Proposition~\ref{prop:ce-swap-regret}
and the definition
\eqref{eq:ce-swap-radius}
give
\begin{equation}\label{eq:app-ce-stage-regret}
\mathcal R_m^{\rm swap}
\le B_m^{\rm CE}
\end{equation}
with failure probability at most
$\delta_0$.
When $A=1$, the swap regret
is identically zero.

Next, conditional on the
pre-action history,
\[
\mathbb E[Y_j\mid\mathcal F_{j-1}]
=\mu_j,
\qquad
0\le Y_j\le H_t.
\]
Thus $\{\mu_j-Y_j\}$ is a
bounded martingale-difference
sequence. Azuma--Hoeffding gives
\begin{equation}\label{eq:app-ce-bellman-azuma}
\frac1m\sum_{j=1}^m(\mu_j-Y_j)
\le
H_t\sqrt{\frac{2\iota_{\rm CE}}m}
\end{equation}
with failure probability at most
$\delta_0$.

Combining
\eqref{eq:app-ce-bellman-decomp}--
\eqref{eq:app-ce-bellman-azuma}
yields, for $A\ge2$,
\[
\Gamma_m^{\rm CE}
\le
\frac{H_t}{m}
\left[
B_m^{\rm CE}
+\sqrt{2m\iota_{\rm CE}}
\right].
\]
For $A=1$, only the sampling
term remains.

Moreover, the Bellman return
belongs to $[0,H_t]$, so
$\Gamma_m^{\rm CE}\le H_t$
deterministically.
Taking the minimum of the two
bounds gives exactly
$b_m^{\rm CE}$ in
\eqref{eq:ce-pred-bonus}.

The planned stage length is fixed
at its start. Therefore, the
preceding concentration bounds
can be applied conditionally on
the stage-start history.
Unfinished stages may be extended
by zero-loss virtual rounds;
the event for a completed stage
is unaffected by this extension.
There are at most $NSTK$
initiated agent-state stages.
A union bound over the regret
and sampling events therefore
gives total failure probability
at most
$2NSTK\delta_0=p/2$.
\hfill \ensuremath{\blacksquare}

We also record a deterministic
envelope for the CE bonus,
which will be used in the
global analysis.

\begin{lemma}
\label{lemma:app-ce-bonus-envelope}
For every completed stage with
$m\ge1$ and
$A=|\mathcal A_i^t(s)|$,
the bonus in
\eqref{eq:ce-pred-bonus}
satisfies
\[
b_m^{\rm CE}
=
O\left(
H_tA\sqrt{
\frac{\iota_{\rm CE}}m}
\right).
\]
\end{lemma}

\emph{Proof:}
The case $A=1$ follows directly
from \eqref{eq:ce-pred-bonus}.
Consider $A\ge2$ and write
\[
F=\frac1{H_t},\qquad
G=\frac{H_t-1}{H_t},\qquad
D=2F^2+4AG^2.
\]
Since both $f_j$ and the running-mean forecasts
$M_j(\cdot\mid J_j)$ lie in $[0,F]^A$,
$E_m^{\rm CE}\le mF^2$.
Thus the first three terms in
$B_m^{\rm CE}$ are bounded by
\begin{equation}\label{eq:app-ce-eta-terms}
\frac{AC_A}{\eta_{\rm CE}}
+\eta_{\rm CE}mD.
\end{equation}

If
\[
\eta_{\rm CE}
=
\sqrt{\frac{AC_A}{mD}}
\le H_t,
\]
then \eqref{eq:app-ce-eta-terms}
equals
\[
2\sqrt{AC_A mD}.
\]
Using $D\le5A$ and
$C_A=O(\iota_{\rm CE})$,
this is
$O(A\sqrt{m\iota_{\rm CE}})$.

For $m\ge\iota_{\rm CE}$,
the remaining terms in
\eqref{eq:ce-swap-radius},
together with the Bellman
sampling term, have the same
order. Therefore,
\[
b_m^{\rm CE}
=
O\left(
H_tA\sqrt{\frac{\iota_{\rm CE}}m}
\right).
\]

If $m<\iota_{\rm CE}$,
the same bound follows from
$b_m^{\rm CE}\le H_t$.

Finally, if the learning rate
is capped at $\eta_{\rm CE}=H_t$,
then
\[
m\le
\frac{AC_A}{H_t^2D}
\le\frac{AC_A}{2}
=O(A\iota_{\rm CE}).
\]
The trivial bound
$b_m^{\rm CE}\le H_t$
again implies the stated
envelope. This completes
the proof.
\hfill \ensuremath{\blacksquare}

\subsection{CE Value-Confidence Bounds}

We next extend the confidence-bound
induction of Appendix~\ref{app:cce}
to recommendation-dependent
deviations.

As in the CE definition,
$\Psi_i$ consists of admissible
action-modification rules
$\psi_i^t(s,\cdot)$ based on
the current state and the
recommended action.
These rules do not observe the
internal row index $J_j$ or
the hidden historical index
used by the certified-policy
construction.
Consequently, at a fixed
$(t,s)$, the same action map
is applied across the historical
policies in the certified mixture.

For the analysis, define
$\underline V_{T+1,i}^k=0$.
At a state with a nonempty
completed predecessor stage,
let
\begin{align}
\underline{\widehat V}_{t,i}^k(s)
={}&
\frac1m\sum_{j=1}^m
\Big[
r_i^t(s,\bm a^{t,\tilde l_j})
+\underline V_{t+1,i}^{\tilde l_j}
(s^{t+1,\tilde l_j})
\Big]
-b_m^{\rm CE}, \notag \\
\underline V_{t,i}^k(s)
={}&
\max\left\{
0,\underline{\widehat V}_{t,i}^k(s)
\right\}.
\label{eq:app-ce-lower-clip}
\end{align}
Before the first completed stage,
set $\underline V_{t,i}^k(s)=0$.
The upper estimates
$\overline V_{t,i}^k(s)$
are those maintained by
Algorithm~\ref{alg:sbv-CE}.

\begin{restatement}
{CE value-confidence bounds}
With probability at least $1-p$,
simultaneously for all agents $i$,
episodes $k$, steps $t$, and
states $s\in\mathcal S^t$,
\[
\max_{\psi_i\in\Psi_i}
V_{t,i}^{\psi_i\diamond
\bar{\bm\pi}^{t,k}}(s)
\le
\overline V_{t,i}^k(s),
\qquad
V_{t,i}^{\bar{\bm\pi}^{t,k}}(s)
\ge
\underline V_{t,i}^k(s).
\]
\end{restatement}

\emph{Proof:}
We proceed by backward induction
on $t$.
Both inequalities hold at
$t=T+1$.
Before the first completed stage,
they follow from
\[
0\le V_{t,i}\le H_t,
\qquad
\overline V_{t,i}^k(s)=H_t,
\qquad
\underline V_{t,i}^k(s)=0.
\]

Consider a state with a nonempty
completed predecessor stage.
The certified-policy construction
selects a historical index
$\tilde l_j$ uniformly from this
stage and uses the corresponding
historical marginal policies.
For any admissible deviation
$\psi_i$, let
$\phi=\psi_i^t(s,\cdot)$
be its current action-modification
map.

By the induction hypothesis
at step $t+1$,
\begin{align}
&V_{t,i}^{\psi_i\diamond
\bar{\bm\pi}^{t,k}}(s)
\notag\\
&\quad\le
\frac1m\sum_{j=1}^m
\mathbb D_{
(\phi\diamond p_j)
\times\pi_{-i}^{t,\tilde l_j}}
\left(
r_i^t+
\mathbb P_t\overline V_{t+1,i}^{\tilde l_j}
\right)(s).
\label{eq:app-ce-upper-induction}
\end{align}
The same map $\phi$ appears
in every summand because the
deviation cannot condition
on the hidden historical index.

Maximizing
\eqref{eq:app-ce-upper-induction}
over admissible deviations
and applying the stage-wise
CE Bellman-error bound gives
\begin{align*}
&\max_{\psi_i\in\Psi_i}
V_{t,i}^{\psi_i\diamond
\bar{\bm\pi}^{t,k}}(s)\\
&\quad\le
\max_\phi\frac1m\sum_{j=1}^m
\mathbb D_{
(\phi\diamond p_j)
\times\pi_{-i}^{t,\tilde l_j}}
\left(
r_i^t+
\mathbb P_t\overline V_{t+1,i}^{\tilde l_j}
\right)(s)\\
&\quad\le
\frac1m\sum_{j=1}^m
\left[
r_i^t(s,\bm a^{t,\tilde l_j})
+\overline V_{t+1,i}^{\tilde l_j}
(s^{t+1,\tilde l_j})
\right]
+b_m^{\rm CE}\\
&\quad=
\widehat V_{t,i}^k(s).
\end{align*}
Since all values are at most
$H_t$, the upper clipping
$\overline V_{t,i}^k(s)
=\min\{\widehat V_{t,i}^k(s),H_t\}$
preserves this inequality.

For the lower bound, the
certified-policy recursion
and the induction hypothesis
imply
\begin{align}
V_{t,i}^{\bar{\bm\pi}^{t,k}}(s)
\ge
\frac1m\sum_{j=1}^m
\mathbb D_{
p_j\times\pi_{-i}^{t,\tilde l_j}}
\left(
r_i^t+
\mathbb P_t\underline V_{t+1,i}^{\tilde l_j}
\right)(s).
\label{eq:app-ce-lower-induction}
\end{align}

To compare this expectation
with the empirical lower
estimate, define
\[
Y_j^-
=
r_i^t(s,\bm a^{t,\tilde l_j})
+
\underline V_{t+1,i}^{\tilde l_j}
(s^{t+1,\tilde l_j}).
\]
The lower continuation estimate
$\underline V_{t+1,i}^{\tilde l_j}$
is determined by the history
available before the corresponding
action is drawn. Hence
\[
\mathbb E[Y_j^-
\mid\mathcal F_{j-1}]
=
\mathbb D_{
p_j\times\pi_{-i}^{t,\tilde l_j}}
\left(
r_i^t+
\mathbb P_t\underline V_{t+1,i}^{\tilde l_j}
\right)(s).
\]
Since $0\le Y_j^-\le H_t$, for each
initiated stage allocate failure probability
$p/(2NSTK)$. Since
$\log(2NSTK/p)\le\iota_{\rm CE}$,
Azuma--Hoeffding and a union bound yield, uniformly over all completed
stages and with total failure probability at most $p/2$,
\begin{align}
&\frac1m\sum_{j=1}^m
\mathbb D_{
p_j\times\pi_{-i}^{t,\tilde l_j}}
\left(
r_i^t+
\mathbb P_t\underline V_{t+1,i}^{\tilde l_j}
\right)(s)
\notag\\
&\quad\ge
\frac1m\sum_{j=1}^mY_j^-
-
H_t\sqrt{
\frac{2\iota_{\rm CE}}m}.
\label{eq:app-ce-lower-azuma}
\end{align}
The uniform event follows
by conditioning on stage-start
histories and taking a union
bound over initiated stages,
as in the preceding subsection.

If the sampling radius is
at most $H_t$, the CE bonus
is no smaller than this radius.
Combining
\eqref{eq:app-ce-lower-induction}
and \eqref{eq:app-ce-lower-azuma}
therefore gives
\[
V_{t,i}^{\bar{\bm\pi}^{t,k}}(s)
\ge
\underline{\widehat V}_{t,i}^k(s).
\]
If the sampling radius exceeds
$H_t$, the bonus equals $H_t$,
and the lower estimate is zero;
the inequality is then trivial.
The clipping in
\eqref{eq:app-ce-lower-clip}
preserves the lower bound
because values are nonnegative.

Between stage boundaries,
the value estimates and
the corresponding certified-policy
construction remain unchanged,
so both inequalities persist.

The stage-wise CE upper event
holds with probability at least
$1-p/2$, and the additional
lower sampling event holds with
probability at least $1-p/2$.
Their intersection has probability
at least $1-p$, completing
the induction.
\hfill \ensuremath{\blacksquare}

\subsection{Proof of the CE Sample
Complexity Guarantee}

We now prove Theorem~\ref{thm:ce}.
Define the CE gap of the output
policy by
\[
\operatorname{CEGap}(\bar{\bm\pi})
:=
\max_{i\in\mathcal N}
\sup_{\psi_i\in\Psi_i}
\left[
V_{1,i}^{\psi_i\diamond
\bar{\bm\pi}}(s^1)
-
V_{1,i}^{\bar{\bm\pi}}(s^1)
\right].
\]

\emph{Proof:}[Proof of
Theorem~\ref{thm:ce}]
Work on the simultaneous
confidence event established
above.
For each agent $i$, define
\[
\delta_i^{t,k}
=
\overline V_{t,i}^k(s^{t,k})
-
\underline V_{t,i}^k(s^{t,k}).
\]

The output certified policy
is the uniform mixture of
the $K$ episode-indexed
certified policies.
The confidence bounds imply,
for each agent $i$,
\begin{align}
\sup_{\psi_i\in\Psi_i}
\left[
V_{1,i}^{\psi_i\diamond
\bar{\bm\pi}}(s^1)
-
V_{1,i}^{\bar{\bm\pi}}(s^1)
\right]
\le
\frac1K\sum_{k=1}^K
\delta_i^{1,k}.
\label{eq:app-ce-gap-to-delta}
\end{align}

We next bound the cumulative
value-estimation gap.
At a state with a nonempty
completed predecessor stage,
the upper and lower value
updates give
\begin{align}
\delta_i^{t,k}
\le{}&
\frac1m\sum_{j=1}^m
\delta_i^{t+1,\tilde l_j}
+2b_m^{\rm CE}.
\notag
\end{align}
Before the first completed
stage, the gap is at most $T$.
Thus, using the predecessor-stage
notation of Appendix~\ref{app:cce},
\begin{equation}\label{eq:app-ce-delta-full}
\delta_i^{t,k}
\le
T\mathbf1\{\tilde c^{t,k}=0\}
+
\mathbf1\{\tilde c^{t,k}>0\}
\Big[
\frac1{\tilde c^{t,k}}
\sum_{j=1}^{\tilde c^{t,k}}
\delta_i^{t+1,\tilde l_j}
+
2b_{\tilde c^{t,k}}^{\rm CE}
\Big].
\end{equation}

As in Appendix~\ref{app:cce},
the first-stage contribution
is at most $ST^2$ at each step.
Furthermore, the stage-length
upper bound ensures that
each sample from a completed
stage is reused with total
weight at most $1+1/T$
in the following stage.
Summing
\eqref{eq:app-ce-delta-full}
over episodes therefore gives
\begin{equation}\label{eq:app-ce-summed-recursion}
\begin{aligned}
\sum_{k=1}^K\delta_i^{t,k}
\le{}&
ST^2+
\left(1+\frac1T\right)
\sum_{k=1}^K\delta_i^{t+1,k}\\
&+
2\sum_{\substack{k\in[K]:\\
\tilde c^{t,k}>0}}
b_{\tilde c^{t,k}}^{\rm CE}.
\end{aligned}
\end{equation}

Since $\delta_i^{T+1,k}=0$,
iterating
\eqref{eq:app-ce-summed-recursion}
over $t$ yields
\begin{align}
\sum_{k=1}^K\delta_i^{1,k}
={}&
O\left(
ST^3+
\sum_{t=1}^T
\left(1+\frac1T\right)^{t-1}
\sum_{\substack{k\in[K]:\\
\tilde c^{t,k}>0}}
b_{\tilde c^{t,k}}^{\rm CE}
\right).
\label{eq:app-ce-unrolled}
\end{align}

It remains to control the
cumulative bonuses.
By Lemma~\ref{lemma:app-ce-bonus-envelope},
\[
b_m^{\rm CE}
\le
C_bH_tA_{\max}
\sqrt{\frac{\iota_{\rm CE}}m}
\]
for an absolute constant $C_b$.

For a fixed $(t,s)$,
let $e_1,\ldots,e_J$
be its completed stage lengths.
The adaptive stage-length rule,
together with the coefficient
bounds in Algorithms~\ref{alg:fluctMAD}, ensures
\begin{equation}\label{eq:app-ce-stage-growth}
\left(1+\frac1{4T}\right)e_j
\le e_{j+1}
\le
\left(1+\frac1T\right)e_j.
\end{equation}
As shown in Appendix~\ref{app:cce},
these inequalities also hold
with the integer stage-length
update used in the algorithm.

Each visit in the stage
following a completed stage
of length $e_j$ uses the bonus
$b_{e_j}^{\rm CE}$.
The number of such visits is
at most $(1+1/T)e_j$.
Consequently, the total bonus
contribution at $(t,s)$ is
bounded by
\begin{equation}\label{eq:app-ce-state-bonus}
\begin{aligned}
\sum_{\substack{k:\,
s^{t,k}=s\\
\tilde c^{t,k}>0}}
b_{\tilde c^{t,k}}^{\rm CE}
&\le
C H_tA_{\max}
\sqrt{\iota_{\rm CE}}
\sum_{j=1}^J\sqrt{e_j},
\end{aligned}
\end{equation}
where $C$ is an absolute
constant.

Let
$r=1+1/(4T)$.
The lower growth bound in
\eqref{eq:app-ce-stage-growth}
implies $e_h\le e_jr^{-(j-h)}$
whenever $h<j$.
Therefore,
\begin{align}
\left(\sum_{j=1}^J\sqrt{e_j}\right)^2
&=
\sum_{j=1}^Je_j
+
2\sum_{h<j}\sqrt{e_he_j}
\notag\\
&\le
\left(
1+2\sum_{d=1}^{\infty}
r^{-d/2}
\right)
\sum_{j=1}^Je_j
\notag\\
&=
O\left(
T\sum_{j=1}^Je_j
\right).
\label{eq:app-ce-stage-sqrt-sum}
\end{align}

Let $K_{t,s}$ be the number
of visits to state $s$ at
step $t$ during the $K$
episodes.
Since
$\sum_j e_j\le K_{t,s}$,
\eqref{eq:app-ce-state-bonus}
and
\eqref{eq:app-ce-stage-sqrt-sum}
give
\[
\sum_{\substack{k:\,
s^{t,k}=s\\
\tilde c^{t,k}>0}}
b_{\tilde c^{t,k}}^{\rm CE}
=
O\left(
H_tA_{\max}
\sqrt{T K_{t,s}\iota_{\rm CE}}
\right).
\]

Using
$\sum_sK_{t,s}=K$
and Cauchy--Schwarz,
we obtain
\begin{align}
\sum_{\substack{k\in[K]:\\
\tilde c^{t,k}>0}}
b_{\tilde c^{t,k}}^{\rm CE}
&=
O\left(
H_tA_{\max}
\sqrt{STK\iota_{\rm CE}}
\right).
\label{eq:app-ce-step-bonus}
\end{align}

Finally,
$(1+1/T)^{t-1}\le e$
and
$\sum_{t=1}^T H_t=O(T^2)$.
It follows from
\eqref{eq:app-ce-step-bonus}
that
\begin{equation}\label{eq:app-ce-total-bonus}
\begin{aligned}
&\sum_{t=1}^T
\left(1+\frac1T\right)^{t-1}
\sum_{\substack{k\in[K]:\\
\tilde c^{t,k}>0}}
b_{\tilde c^{t,k}}^{\rm CE}\\
&\qquad=
O\left(
\sqrt{
SA_{\max}^2KT^5
\iota_{\rm CE}}
\right).
\end{aligned}
\end{equation}

Substituting
\eqref{eq:app-ce-total-bonus}
into \eqref{eq:app-ce-unrolled}
gives, simultaneously for
every agent $i$,
\begin{equation}\label{eq:app-ce-global-delta}
\sum_{k=1}^K\delta_i^{1,k}
=
O\left(
ST^3+
\sqrt{
SA_{\max}^2KT^5
\iota_{\rm CE}}
\right).
\end{equation}

Combining
\eqref{eq:app-ce-gap-to-delta}
and
\eqref{eq:app-ce-global-delta},
and then taking the maximum
over agents, yields
\begin{equation}\label{eq:app-ce-final-gap}
\operatorname{CEGap}(\bar{\bm\pi})
=
O\left(
\frac{ST^3}{K}
+
\sqrt{
\frac{
SA_{\max}^2T^5
\iota_{\rm CE}}{K}}
\right).
\end{equation}

For
\[
K\ge
C\frac{
SA_{\max}^2T^5
\iota_{\rm CE}}{\epsilon^2},
\]
with a sufficiently large
absolute constant $C$,
the first term in
\eqref{eq:app-ce-final-gap}
is absorbed into the second,
and the CE gap is at most
$\epsilon$.
The simultaneous confidence
event has probability at least
$1-p$.

Therefore, the output policy
$\bar{\bm\pi}$ is an
$\epsilon$-approximate CE
with probability at least
$1-p$, completing the proof
of Theorem~\ref{thm:ce}.
\hfill \ensuremath{\blacksquare}

\end{document}